\documentclass{pinchcr}

\usepackage{amssymb}
\usepackage[dvips]{graphicx}
\usepackage{graphicx}
\usepackage{amsmath}

\title{Introduction to Metal-Insulator Transitions}

\author{V.~Dobrosavljevi\'c}
\affiliation{Department of Physics and National High Magnetic Field Laboratory\\
Florida State University,Tallahassee, FL 32306, USA}

\begin{document}

\maketitle

\maintext

\chapter{Introduction to Metal-Insulator Transitions}

\section{Why study metal-insulator transitions?}

The metal-insulator transition (MIT) is one of the oldest, yet one
of the fundamentally least understood problems in condensed matter
physics. Materials which we understand well include good insulators
such as silicon and germanium, and good metals such as silver and
gold. Remarkably simple theories \cite{Ashcroft} have been successful
in describing these limiting situations: in both cases low temperature
dynamics can be well described through a dilute set of elementary
excitations. Unfortunately, this simplicity comes at a price: the
physical properties of such materials are extremely \textit{stable}.
They prove to be very difficult to manipulate or modify in order to
the meet the needs modern technology, or simply to explore novel and
interesting phenomena.
\begin{figure}[h]
\begin{center}
\includegraphics[width=0.8\textwidth]{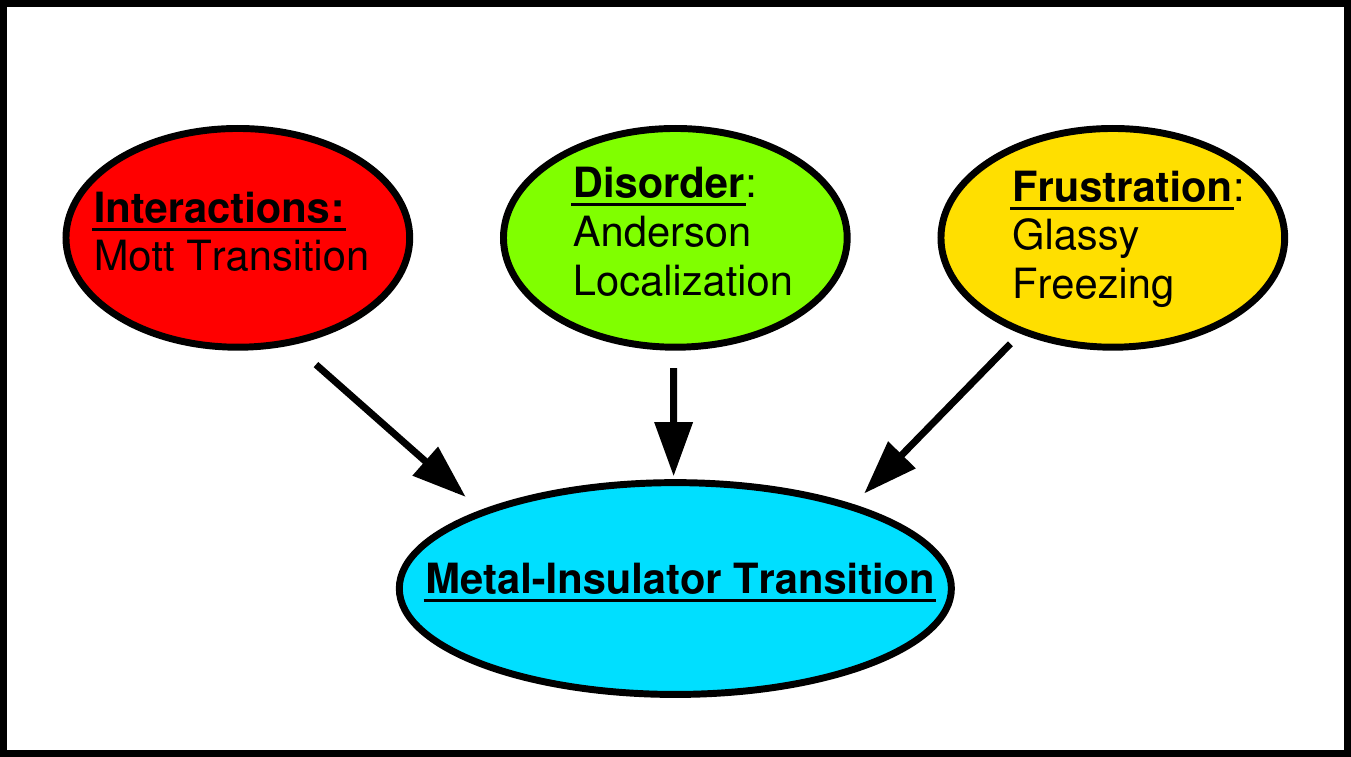}
\end{center}
\caption{Three basic mechanism for electron localization.}
\label{routes} 
\end{figure}

The situation is more promising in more complicated materials, where
relatively few charge carriers are introduced in an otherwise insulating
host. Several such systems have been fabricated even years ago, and
some have, in fact, served as basic building blocks of modern information
technology. The most familiar are, of course, the doped semiconductors
which led to the discovery of the transistor. More recent efforts
drifted to structures of reduced dimensionality and devices such as
silicon MOSFETs (metal-oxide-semiconductor field-effect transistors),
which can be found in any integrated circuit. 
\begin{figure}[h]
\begin{center}
\includegraphics[width=0.9\textwidth]{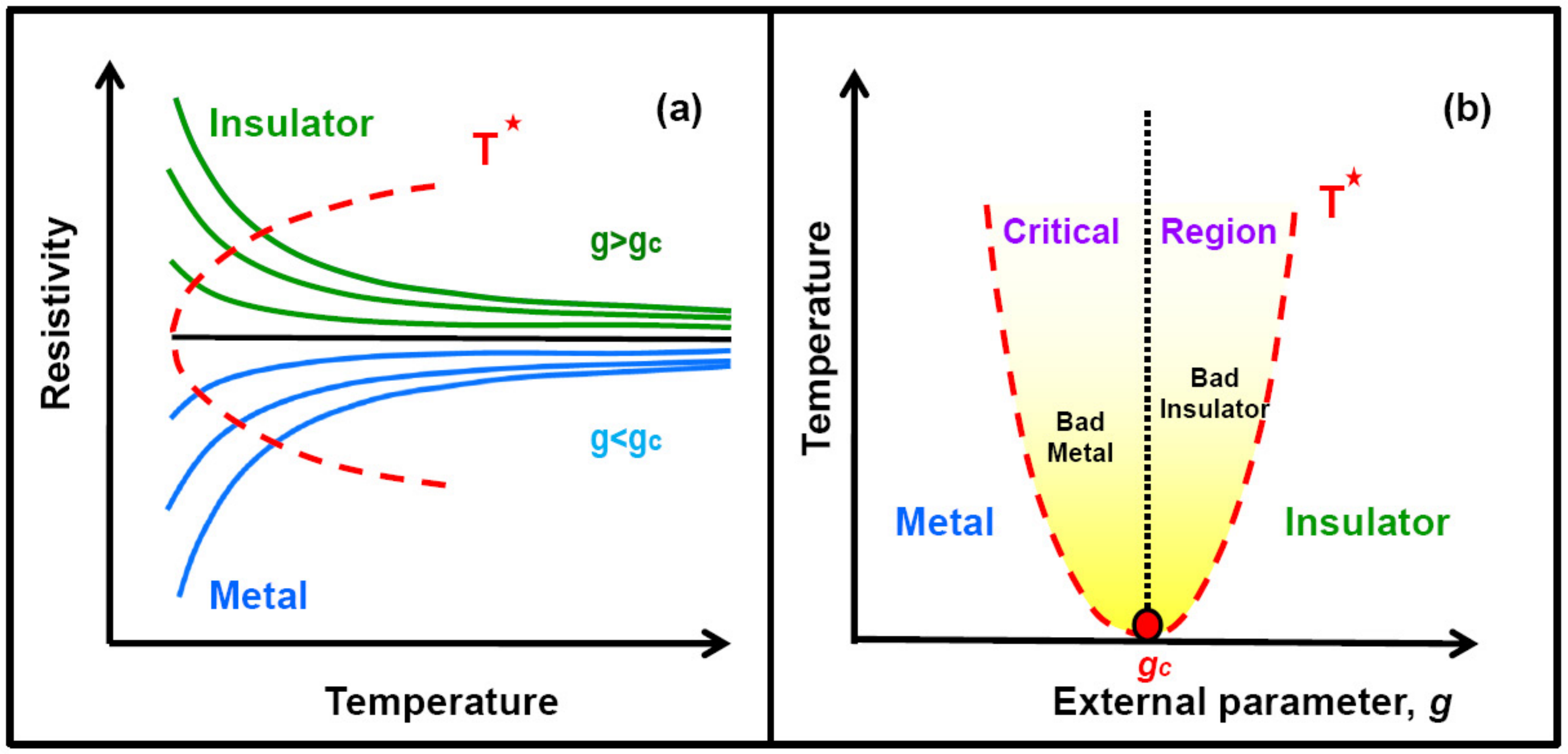}
\end{center}

\caption{Quantum critical behavior near a metal-insulator transition. 
Temperature dependence
of the resistance for different carrier concentrations is shown schematically
in (a).Well defined metallic or insulating behavior is observed only at temperatures
lower than a characteristic temperature $T<T^{*}$ that vanishes at
the transition. At $T<T^{*}$, the system is in the ``quantum
critical region'', as shown in (b). 
 As the system crosses over from metal to insulator, the temperature
dependence of the resistivity changes slope from positive to negative.}

\label{quantumcritical} 
\end{figure}

\subsection{Why is the MIT an important problem? }

In contrast to elemental materials, in systems close to the MIT the
physical properties change dramatically with the variation of control
parameters such as the carrier concentration, the temperature, or
the external magnetic field. Such sensitivity to small changes is, indeed,
quite common in any material close to a phase transition. In doped  
insulators this sensitivity follows from the vicinity to the metal-insulator
transition. The sharp critical behavior is seen here only at the lowest
accessible temperatures, because a qualitative distinction between
a metal and an insulator exists only at $T=0$ (Fig. \ref{quantumcritical}).
Since the basic degrees of freedom controlling the electrical transport
properties are electrons, and the transition is found at $T=0$, quantum
fluctuations dominate the critical behavior. The metal-insulator transition
should therefore be viewed as perhaps the best example of a quantum
critical point (QCP), a subject that has attracted much of the physicist's
fancy and imagination in recent years \cite{sachdevbook}. As near
other QCPs, one expects the qualitative behavior here to display a
degree of universality, allowing an understanding based on simple
yet fundamental physical pictures and concepts. Before we understand
the basic mechanisms and process that control this regime (Fig. 1.1) one can
hardly hope to have control over material properties even in very
simple situations. 

\subsection{Why is the MIT a difficult problem?}

From the theoretical point of view the problem at hand is extremely
difficult for reasons that are easy to guess. The two limits, that
of a good metal and that of a good insulator, are very different physical
systems, which can be characterized by very different elementary excitations.
For metals, these are fermionic quasiparticles corresponding to electrons
excited above the Fermi sea. For insulators, in contrast, these are
long-lived bosonic (collective) excitations such as phonons and spin
waves. In the intermediate regime of the metal-insulator transition,
both types of excitations coexist, and simple theoretical tools prove
of little help. Some of these conceptual difficulties are a general
feature of quantum critical points. For QCPs involving spin or charge
ordering, the critical behavior can be described by examining the
order parameter fluctuations associated with an appropriate symmetry
breaking. In contrast, the MIT is more appropriately described as
a dynamical transition, and an obvious order parameter theory is not
available. For all these reasons, the intermediate regime between
the metal and the insulator has remained difficult to understand both
from the practical and the conceptual point of view.

\subsection{Disorder and complexity}

\begin{figure*}[h]
\begin{center}
\includegraphics[width=0.9\textwidth]{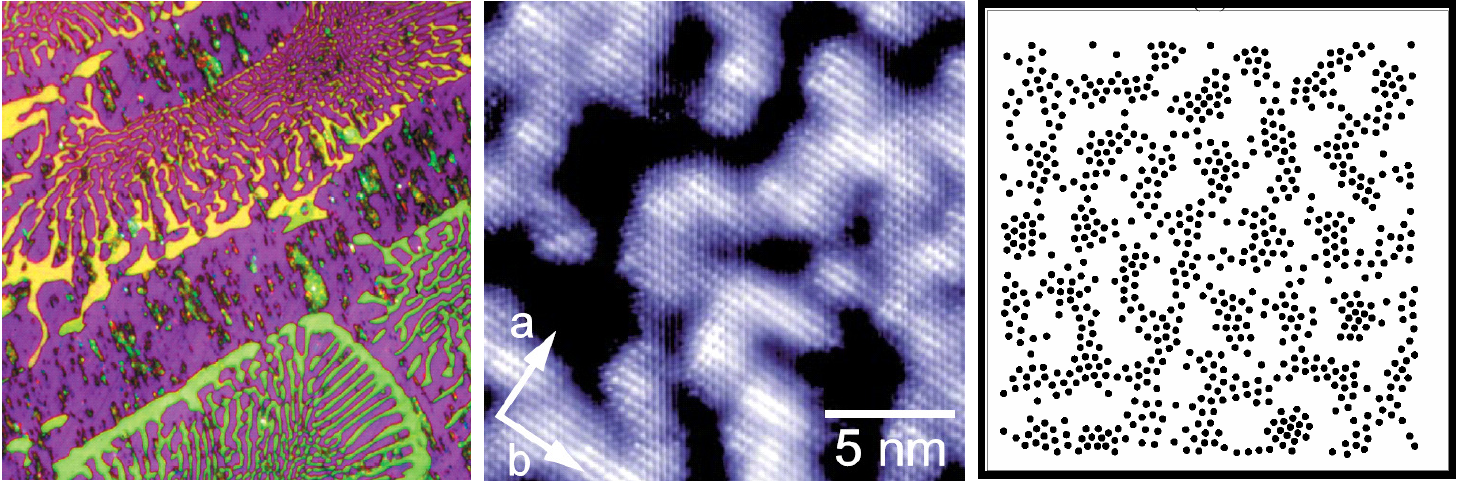}
\end{center}
\caption{In the last few years, fascinating examples of complex ordering around
the metal-insulator transition are starting to emerge, due to advances
of both the experimental probes and the theoretical tools available.
Left panel: Percolative conduction in the half-metallic ferromagnetic
and ferroelectric mixture $(La,Sr)MnO_{3}$ \protect\cite{cheong-prl04}
. Central panel: Inhomogeneous charge distribution revealed by scanning tunneling microscope (STM) 
spectroscopy  \protect\cite{takagi-prl04} on underdoped cuprate$Ca_{2-x}Na_{x}CuO_{2}Cl_{2}$.
Right panel: Strikingly similar ``stripe glass''
ordering is observed in a computer simulation of an appropriate model
\protect\cite{reichhardt-lanl05}. }

\label{complex} 
\end{figure*}

After more then sixty years of study, the subject of metal-insulator
transitions has become a very wide and complicated field of research.
In many complex materials, particularly in transition metal oxides
and other strongly correlated systems, the emergence of the metal-insulator
transition is often accompanied by changes in magnetic or structural
symmetry. In many such cases the transition is dominated by material-specific
details, and it is often of first order, not a critical point. Somewhat
surprisingly, the situation is in fact simpler in presence of sufficient
amounts of disorder, where the metal and the insulator have the same
symmetries and the transition is not associated with any uniform ordering.
In such cases, strong evidence exists indicating that the transition
is a genuine quantum critical point of a fundamentally new variety,
with properties that most likely dominate the behavior of many materials.

At first glance disorder may be viewed simply as a nuisance and an
inessential complication. In recent years, though, fascinating new
evidence is emerging (Fig. \ref{complex}) revealing that genuinely
new phenomena \cite{RoP2005review} arise in presence of disorder
and impurities. Put quite simply, a given configuration of impurities
may locally favor one or another of several competing phases of matter.
In many instances this gives rise to strongly inhomogeneous states
that feature an enormous number of low-lying metastable states - giving
rise to quantitatively new excitation, slow relaxation, and glassy
fluctuations and response. These phenomena often dominates the observable
properties in many systems, ranging from colossal magneto-resistance
(CMR) manganites and cuprates,
to diluted magnetic semiconductors, and even Kondo alloys. Even more
surprisingly, recent work suggest that a plethora of intermediate
heterogenous phases may emerge between the metal and the insulator,
possibly even in absence of disorder. Such \textit{complexity} emerges
as a new paradigm \cite{dagotto-2005-309} of the metal-insulator
transition region, most likely requiring a description in terms of
probability distribution function functions (PDFs) rather then simple
minded order parameters.

\subsection{MIT in the strong correlation era}

The subject of MITs came to a renewed focus in the last two decades,
following the discovery of high temperature superconductivity, which
triggered much activity in the study of ``bad metals''
(for a recent perspective, see Ref. \cite{basov2011natphys}) . Many
of the materials in this family consist of transition metal or even
rare earth elements, corresponding to compounds which are essentially
on the brink of magnetism. Here, conventional approaches proved of
little help, but recent research has lead to a veritable avalanche
of new and exciting ideas and techniques both on the experimental
and the theoretical front. In many ways, these developments have changed
our perspective on the general problem of the metal-insulator transition,
emphasizing the deep significance of the physics of strong correlation.
A common theme for all these systems seems to be the transmutation
of conduction electrons into localized magnetic moments \cite{andersonlocrev},
a feature deeply connected to the modification of the fundamental
nature of elementary excitations as the metal-insulator transition
is crossed.

\subsection{The scope of this overview}

In this overview we will not discuss all the possible examples of
metal-insulator transitions, but will focus on the basic physical
mechanisms that can localize the electrons in absence of magnetic
or charge ordering, and produce well defined quantum critical behavior.
Our emphasis will be on results providing evidence that strong correlation
physics dominates such quantum critical points, where physical pictures
based on weak-coupling approaches prove insufficient or even misleading.

\section{Basic mechanisms of metal-insulator transitions}

What determines whether a material is a metal or an insulator? In
most cases the answer is provided by simply examining the electronic
band structure \protect\cite{Ashcroft} and the pattern
of chemical bonding of a given compound. At present, solid state physicists
and quantum chemists have an impressive toolbox of theoretical methods
to determine the band structure with sometimes surprising accuracy.
Quite generally, one calculates all the accessible electronic levels
\protect\cite{slater34rmp} for the valence electrons in a solid
and populates them according to the Pauli principle. If the highest
occupied electronic state - the Fermi energy - is within a band gap,
then the material is an insulator, since it takes a finite (generally
large) energy to excite the electron to the lowest accessible state
in order to carry electrical current. Otherwise, when electronic bands
are partially filled, then we expect metallic behavior.

\subsection{Band transitions}

Can one induce a metal-insulator transition within this band theory
picture? This is possible if the gap can be induced to open at the
Fermi surface by rearranging the charge or the spin density of the
electrons in the ground state, i.e. when the system undergoes an ordering
transition. Typically, this corresponds to some kind of Fermi surface
instability where a charge or spin density wave \protect\cite{Gruner00}
formation leads to unit cell doubling. An important early example
was the Slater theory \protect\cite{slater51} of itinerant antiferromagnets,
where the gap opens due to magnetic ordering. Such instabilities are
also common in low dimensional solids such as organic charge-transfer
salts \protect\cite{Gruner00}, leading to rich phase diagrams with
many exotic properties. Such ordering transitions typically take place
at finite temperature, and can often be successfully described using
conventional approaches based on the band theory picture. These situations
have attracted considerable attention in recent years, but we will
not them discuss them further in this overview.

\subsubsection{When does band theory work?}

The success of the band theory approach was so impressive that already
in 1930s Slater announced \protect\cite{slater34rmp} that the solid
state physics is a solved problem, and that we only need fast computers
to accurately predict physical properties of any material. In the
last few decades the computers did become amazingly fast - but in
many interesting cases the band structure approach proved insufficient.
When does that happen? To understand this important issue from a general
point of view we need to recall the implicit assumptions of the band
structure approach.

The band theory picture describes the dynamics of one electron moving
through a solid, while the effects of all the other electrons is approximated
by modifying the effective potential energy surface - the pseudopotential
- on which it moves. This approximation is generally expected to be
valid whenever the kinetic energy of the electrons is dominant over
the other energy scales in the problem. A most naive estimate would
involve simply calculating the so-called $r_{s}$-number: $r_{s}=E_{c}/E_{F}$,
where $E_{c}$ is the average Coulomb energy per particle and $E_{F}$
is the Fermi energy. The $r_{s}$-number ranges between 3 and 5 even
in good metals, and thus one would naively think that band theory
should never work. However, one has to keep in mind the following
important facts \protect\cite{Ashcroft} that minimize the role of
interactions:
\begin{itemize}
\item In metals screening reduces the magnitude of electron-electron and
electron-impurity interactions to a significant degree.
\item The largest part of the Coulomb energy (Hartree and exchange terms)
contributes to redefining the pseudopotential, and only the ``correlation''
energy gives rise to many-body effects.
\item The Pauli principle considerably restricts the phase space for electron-electron
scattering, as described by Fermi liquid renormalizations. 
\end{itemize}
As a result, the excitations in an electron gas can be viewed as a
dilute collection of quasi-particle excitations, as described by the
Landau's Fermi liquid theory \protect\cite{landauFL1,landauFL3}.
In a nutshell, a Fermi liquid is {}``protected''\ by a large kinetic
energy scale of the electrons in their ground state -- a direct result
of their Fermi statistics. In good metals, the Fermi energy is typically
in the electron-Volt range and the effects of electron correlations
and impurity scattering can be treated as small perturbations \protect\cite{agd}.

\subsubsection{And...when does it fail?}

In materials close to the metal-insulator transition the situations
is very different. Here, the Fermi energy is typically small, and
the {}``quantum protectorate'' of the Pauli principle starts to
weaken. This situation is found, for example, in
\begin{itemize}
\item Narrow band materials such as transition-metal oxide $V_{2}O_{3}$ .
\item Doped semiconductors such as Si:P or diluted two-dimensional electron
gases.
\item Doped magnetic (Mott) insulators such as the famous high-$T_{c}$
cuprate $La_{2-x}Sr_{x}CuO_{4}$. 
\end{itemize}
In all these cases the potential energy terms coming from either residual
electron-electron interactions or due to disorder (electron-impurity
interaction) become comparable to the Fermi energy, and the ground
state of the system can undergo a sudden and dramatic change - the
electrons become bound or {}``localized''. The material ceases to
conduct although band theory does not predict any gap at the Fermi
surface. In the following we briefly describe the early ideas on how
this can take place, and discuss some general features of such quantum
critical points.

\subsection{Interactions: the Mott transition}

Many insulating materials have an odd number of electrons per unit
cell, thus band theory would predict them to be metals - in contrast
to experiments. Such compounds (e.g. transition metal oxides) often
have antiferromagnetic ground states, leading Slater to propose that
spin density wave formation \protect\cite{slater51}is likely at
the origin of the insulating behavior. This mechanism does not require
any substantial modification of the band theory picture, since the
insulating state is viewed as a consequence of a band gap opening
at the Fermi surface.

According to Slater \protect\cite{slater51}, such insulating behavior
should disappear above the Neel temperature, which is typically in
the $10^{2}$ K range. Most remarkably, in most antiferromagnetic
oxides, clear signatures of insulating behavior persist at temperatures
well above any magnetic ordering, essentially ruling out Slater's
weak coupling picture.

What goes on is such cases was first clarified in early works by Mott
\cite{mott1949}and Hubbard \protect\cite{hubbard3}, tracing the
insulating behavior to strong Coulomb repulsion between electrons
occupying the same orbital. Within this picture, which is appropriate
for narrow-band systems \cite{mott-book90}, the electrons tunnel
between weakly hybridized atomic orbitals, as described by a Hubbard
Hamiltonian\begin{equation}
H_{HUB}  =  -\sum_{\left\langle ij\right\rangle \sigma}\left(tc_{i\sigma}^{\dagger}c_{j\sigma}^{\phantom{\dagger}}+\mathrm{h.}\,\mathrm{c.}\right)
 +\sum_{j\sigma}\epsilon_{j}c_{j\sigma}^{\dagger}c_{j\sigma}^{\phantom{\dagger}}+U\sum_{j}c_{j\uparrow}^{\dagger}c_{j\uparrow}^{\phantom{\dagger}}c_{j\downarrow}^{\dagger}c_{j\downarrow}^{\phantom{\dagger}}.\end{equation}
 Here, the operator $c_{i\sigma}^{\dagger}$ creates an electron of
spin $\sigma$ in the $i$-th orbital, $t$ is the tunneling element
describing the inter-orbital hybridization, $\epsilon_{j}$ represents
the corresponding site energy, and $U$ describes the on-site Coulomb
repulsion.

\begin{figure}[h]
\begin{center}
\includegraphics[width=4in]{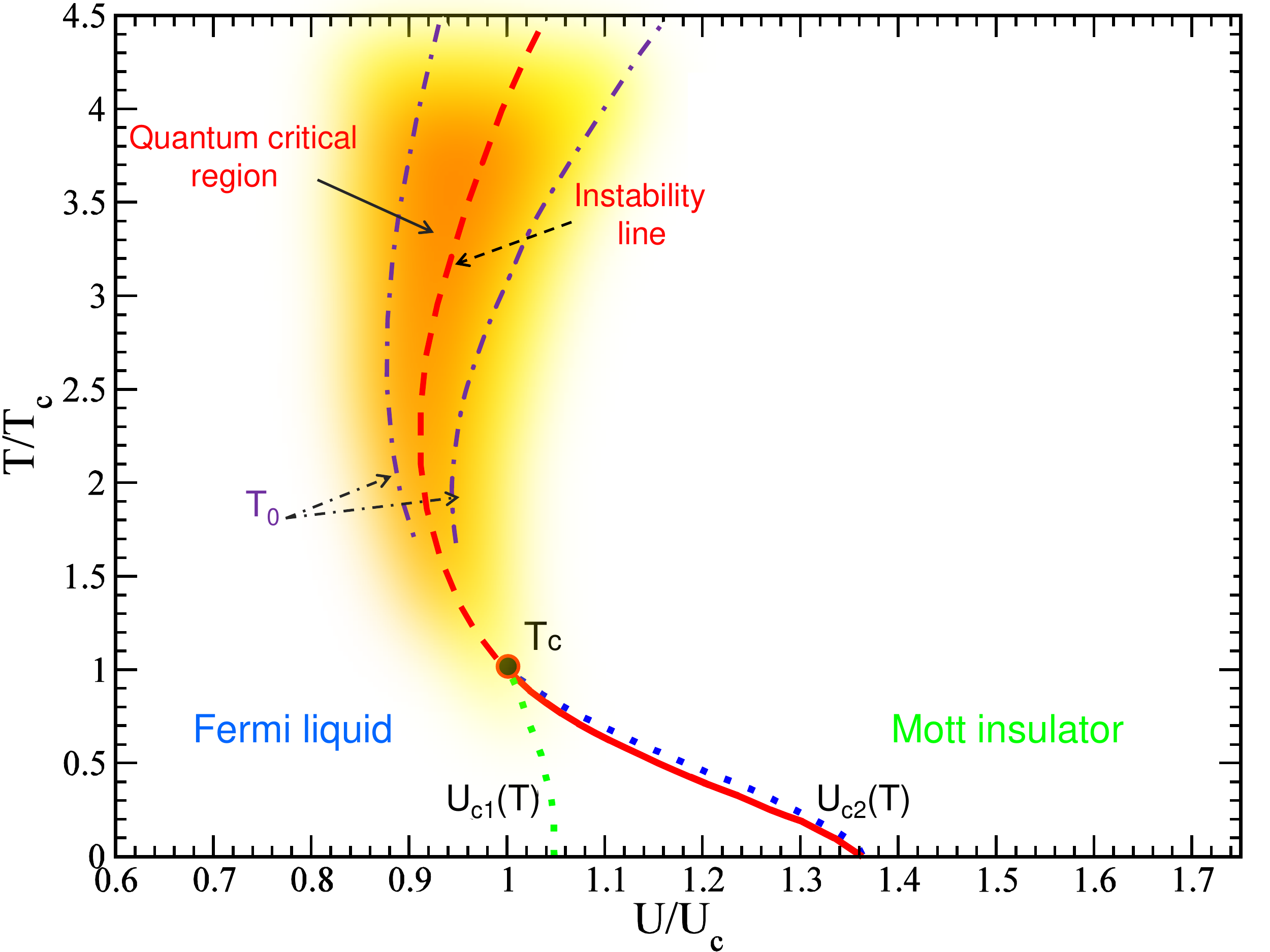}
\end{center}
\caption{Phase diagram for a fully frustrated half-filled Hubbard model calculated
from DMFT theory. At low temperatures the Fermi liquid and the Mott
insulating phases are separated by a first order transition line,
and the associated coexistence region. Very recent work \protect \cite{terletska-mott11prl} established
that at $T>T_{c}$ the intermediate metal-insulator crossover region
show all the features expected for the quantum critical regime, including
the characteristic scaling behavior for the family of resistivity
curves. }
\label{hubbard} 
\end{figure}

When the lattice has integer filling per unit cell, then electrons
can be mobile only if they have enough kinetic energy ($E_{K}\sim t$)
to overcome the Coulomb energy $U$. In the narrow band limit of $t\ll U$,
the electrons do not have enough kinetic energy, and a gap opens in
the single-particle excitation spectrum, leading to Mott insulating
behavior. This gap $E_{g}\approx U-B$ (here $B\approx$ $2zt$ is
the electronic bandwidth; $z$ being the lattice coordination number)
is the energy an electron has to pay to overcome the Coulomb repulsion
and leave the lattice site. In the ground state, each lattice site
is singly occupied, and the electron occupying it behaves as a spin
1/2 local magnetic moment. These local moments typically interact
through magnetic superexchange interactions \cite{anderson59superexchange}of
the order $J\sim U/t_{ij}^{2}$, leading to magnetic ordering at temperatures
of order $T_{J}\sim J$. The insulating behavior, however, is not
caused by magnetic ordering, and will persist all the way to temperatures
$T_{Mott}\sim E_{g}\gg T_{J}$. In oxides, $T_{Mott}\sim E_{g}\sim10^{3}-10^{4}K$
is typically on the atomic ($eV$) scale, while magnetic ordering
emerges at temperatures roughly an order of magnitude lower $T_{J}\sim100-300K$.

We note that Mott's simple argument for the stability of the interaction-driven
insulator does not directly rely on a periodicity of a lattice. Even
if the site energies $\varepsilon_{i}$ or the hopping elements $t_{ij}$
are random variables of moderate variance, the Mott gap will persist,
provided that the on-site repulsion $U$ is large enough as compared
to the typical kinetic energy. This is true, since the Mott gap essentially
measures the energy for an electron to hop to the nearest lattice
site - an inherently local process which does not depend much on long-range
periodicity of the lattice. Precisely this behavior is what takes
place in a doped semiconductor deep in the insulating phase, which
should be viewed as a strongly disordered Mott insulator. Clear evidence
for the correctness of this picture is provided by optical experiments
\cite{thomas81prb}, which provide unambiguous evidence of the coexistence
of the Mott gap at sufficiently low doping levels.

When the kinetic energy and the Coulomb interaction are comparable,
the system finds itself in the vicinity of the Mott transition (Fig.
\ref{hubbard}). Experimentally, the bandwidth can often be controlled
by modifying the orbital overlap $t$, thus the electronic bandwidth.
In several transition metal oxides, for example, this is possible
by applying external hydrostatic pressure. From the theoretical perspective,
describing the vicinity of the Mott transition proves quite difficult
due to the lack of a small parameter characterizing this nonperturbative
regime. Still, after more then thirty years of work on the problem,
several theoretical approaches have emerged, which provide the physical
picture of the transition region. Early arguments of Mott and Hubbard
make it clear that the gap will close for $U\lesssim B$, but the
precise form of the critical behavior remained elusive.

\subsubsection{Correlated metallic state close to the Mott transition}

An important step in elucidating the approach to the Mott transition
from the metallic side was provided by the pioneering work of Brinkmann
and Rice \protect\cite{brinkmann70prb}. This work, which was motivated
by experiments on the normal phase of $^{3}He$ (for a review , see:
\protect\cite{vollhardt84rmp}), predicted a strong effective mass
enhancement close to the Mott transition. In the original formulation,
as well as in its subsequent elaborations (slave boson mean-field
theory \protect\cite{kotliarruckenstein}, dynamical mean-field theory 
\protect\cite{georgesrmp}), the effective mass is predicted to continuously
diverge as the Mott transition is approached form the metallic side\begin{equation}
\frac{m^{\ast}}{m}\sim(U_{c}-U)^{-1}.\end{equation}
\begin{figure}[h]

\begin{centering}
\includegraphics[width=3.3399in,height=3.2872in]{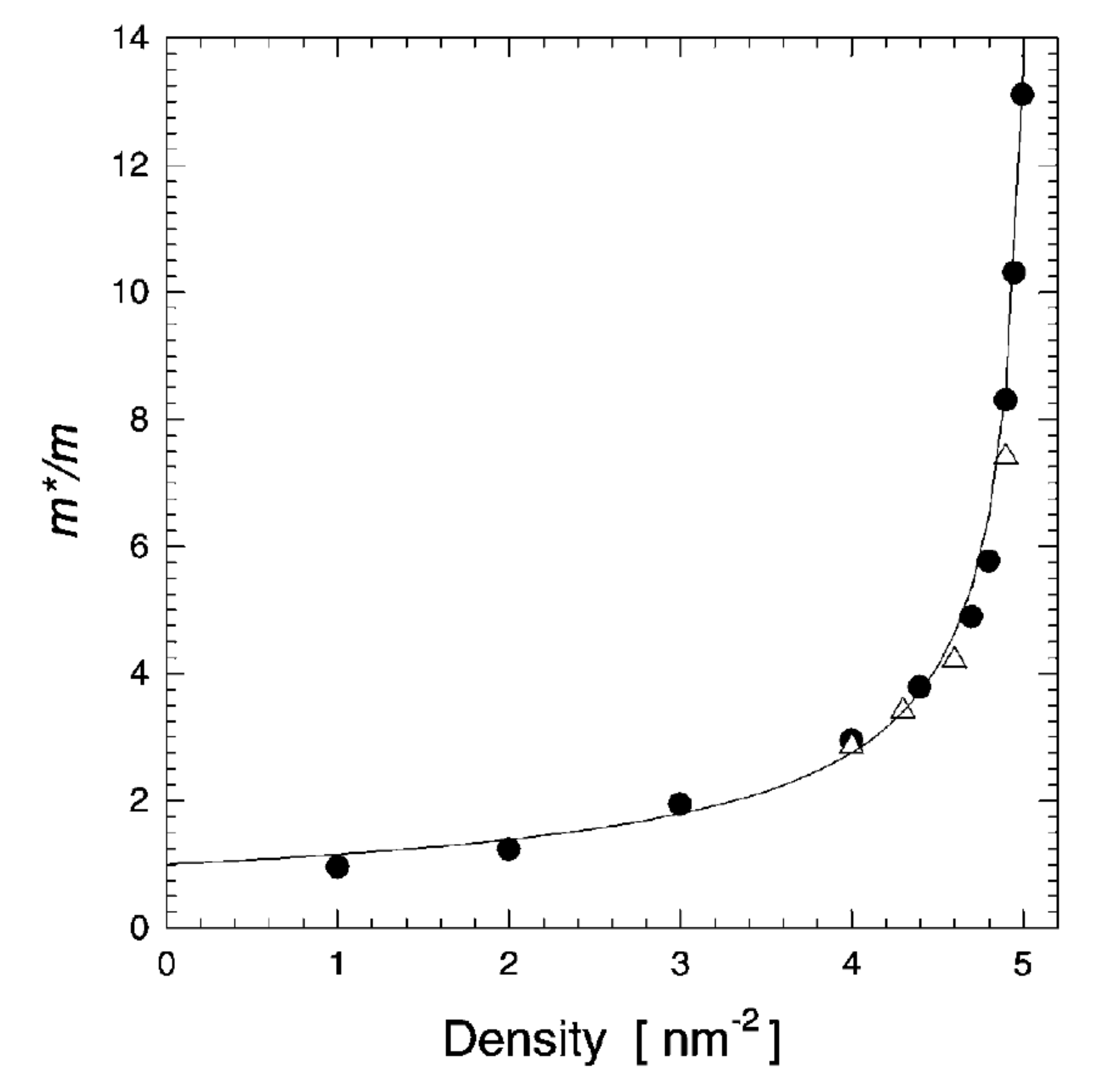}
\par\end{centering}

\caption{Clear evidence of strong mass enhancements can be seen in experiments
on mono-layer $He^{3}$ films on graphite \ \protect\cite{casey03prl}
. In this system, the solid phase (Mott insulator) can be approached
when the density is increased by the application of hidrostatic pressure.}

\centering{}\label{casey03prl}
\end{figure}

A corresponding coherence (effective Fermi) temperature \begin{equation}
T^{\ast}\sim T_{F}/m^{\ast}\end{equation}
is predicted above which the quasiparticles are destroyed by thermal
fluctuations. As a result, one expects a large resistivity increase
around the coherence temperature, and a crossover to insulating (activated)
behavior at higher temperatures. Because the low temperature Fermi
liquid is a spin singlet state, a modest magnetic field of the order\begin{equation}
B^{\ast}\sim T^{\ast}\sim(m^{\ast})^{-1}\end{equation}
 is expected to also destabilize such a Fermi liquid and lead to large
and positive magnetoresistance.

The physical picture of the \ Brinkmann-Rice seems to suggest that
the Mott transition should be viewed as a quantum critical point,
where powerlaw behavior of characteristic crossover scales is expected
as the transition is approached. On the other hand, the Mott transition
discussed here describes the opening of a correlation-induced spectral
gap in absence of magnetic ordering, i.e. within the paramagnetic
phase. Such a phase transition is, therefore, not associated with
spontaneous symmetry breaking associated with any static order parameter.
Why should the phase transition then have any second-order (continuous)
character at all?

The answer is that... in fact it does not! Later work \cite{moeller99prb,park08prl},
which presented better descriptions of inter-site correlations, suggested
that such a transition should generically have \ a (weakly) first
order character, in agreement with early ideas of Mott \cite{mott1949}.
The effective mass, even if it does not exactly diverge at the transition,
is still expected to be significantly enhanced in its close vicinity.
Such behavior is indeed seen in various Mott systems \cite{georgesrmp}
that have been studied, including bulk and two-dimensional $He^{3}$
liquids (Fig. \ref{casey03prl}), transition metal oxides, and organic
charge-transfer salts.

\subsubsection{Physical content of the effective mass enhancement}

How should we physically interpret the large effective mass enhancement
which is seen in all these systems? What determines its magnitude
if it does not actually diverge at the transition? An answer to this
important question can be given using a simple thermodynamic argument,
which does not rely on any particular microscopic theory or specific
model. In the following, we present this simple argument for the case
of a clean Fermi liquid, although its physical context is, of course,
much more general.

In any clean Fermi liquid \protect\cite{agd} the low temperature
specific heat assumes the leading form\begin{equation}
C(T)=\gamma T+\cdots,\end{equation}
 where the Sommerfeld coefficient \begin{equation}
\gamma\sim m^{\ast}.\end{equation}
 In the strongly correlated limit ($m^{\ast}/m\gg1$) this behavior
is expected only at $T\lesssim T^{\ast}\sim(m^{\ast})^{-1}$, while
the specific heat should drop to much smaller values at higher temperatures
where the quasiparticles are destroyed. Such behavior is indeed observed
in many systems showing appreciable mass enhancements.

On the other hand, from general thermodynamic principles, we can express
the entropy as \begin{equation}
S(T)=\int_{0}^{T}dT\frac{C(T)}{T}.\end{equation}
 Using the above expressions for the specific heat, we can estimate
the entropy around the coherence temperature\begin{equation}
S(T^{\ast})\approx\gamma T^{\ast}\sim O(1).\end{equation}
 The leading effective mass dependence of the Sommerfeld coefficient
$\gamma$ and that of the coherence temperature $T^{\ast}$ cancel
out!

Let us now explore the consequences of the assumed (or approximate)
effective mass divergence at the Mott transition. As $m^{\ast}\longrightarrow\infty$,
the coherence temperature $T^{\ast}\longrightarrow0+$, resulting
in large residual entropy\begin{equation}
S(T\longrightarrow0+)\sim O(1).\end{equation}
 We conclude that the effective mass divergence indicates the approach to
a phase with finite residual entropy!

Does not this result violate the Third Law of Thermodynamics?! And
how can it be related to the physical picture of the Mott transition?
The answer is, in fact, very simple. Within the Mott insulating phase
the Coulomb repulsion confines the electrons to individual lattice
sites, turning them into spin 1/2 localized magnetic moments. To the
extend that we can ignore the exchange interactions between these
spins, the Mott insulator can be viewed as a collection of free spins
with large residual entropy $S(0+)=R\ln2$. This is precisely what
happens within the Brinkmann-Rice picture; similar results are obtained
from DMFT, a result
that proves exact in the limit of large lattice coordination  \protect\cite{georgesrmp}. %
\begin{figure}[h]

\begin{centering}
\includegraphics[width=4in]{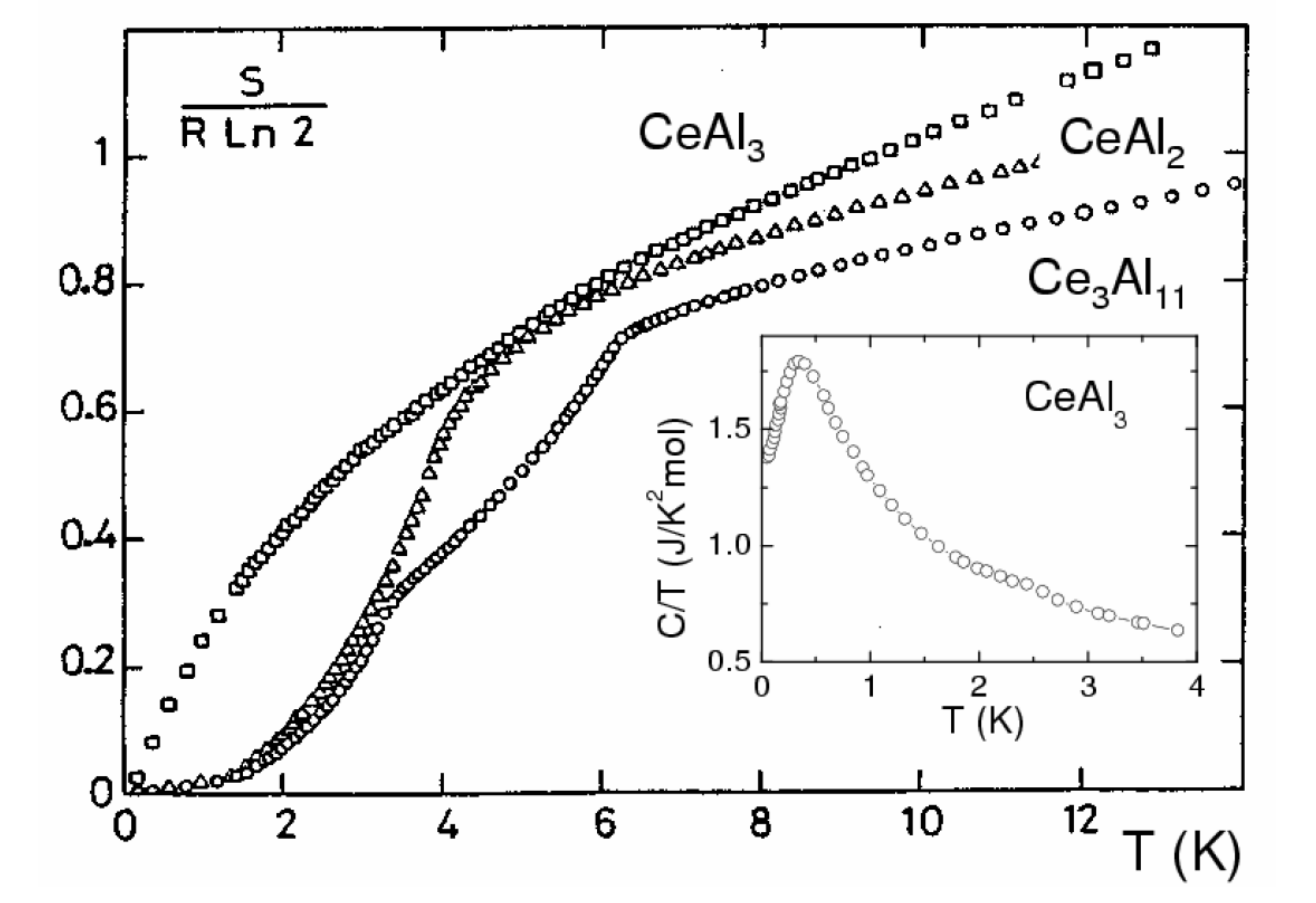}
\par\end{centering}

\caption{Temperature dependence of entropy extracted from specific heat (inset)
experiments \protect\cite{Flouquet2005} on several heavy-fermion
materials. Essentially the entire doublet entropy $S=R\ln2$ is recovered
by the time the temperature has reacued $T^{\ast}\approx10K$, consistent
with a large mass enhancement $m^{\ast}\sim1/T^{\ast}$. }

\centering{}\label{flouquet05lanl}
\end{figure}

In reality, the exchange interactions between localized spins always
exist, and they generally lift the ground state degeneracy, restoring
the Third Law. This happens below a low temperature scale $T_{J}$,
which measures the effective dispersion of inter-site magnetic correlations
\cite{moeller99prb,park08prl} emerging from such exchange interactions.
In practice, this correlation temperature $T_{J}$ can be very low,
either due to effects of geometric frustration, or additional ring-exchange
processes which lead to competing magnetic interactions.

We conclude that the effective mass enhancement, whenever observed
in experiment, indicates the approach to a phase where large amounts
of entropy persist down to very low temperatures. Such situations
very naturally occur in the vicinity of the Mott transition, since
the formation of local magnetic moments on the insulating side gives
rise to large amounts of spin entropy being released at very modest
temperatures. A similar situation is routinely found  \protect\cite{stewartrev1,hewson,Flouquet2005}
 in the so-called ``heavy fermion compounds(e.g. rare-earth intermetallics) 
featuring huge effective mass enhancements.
Here, local magnetic moments coexist with conduction electrons giving
rise to the Kondo effect, which sets the scale for the Fermi liquid
coherence temperature $T^{\ast}\sim1/m^{\ast}$, above which the entire
free spin entropy $S(T^{\ast})\sim R\ln2$ is recovered (Fig. \ref{flouquet05lanl}).
This entropic argument is, in turn, used to experimentally prove \ the
existence of localized magnetic moments within the metallic host.

We should mention that other mechanisms of effective mass enhancement
have also been considered. General arguments \cite{millis} indicate
that $m^{*}$ can diverge when approaching a quantum critical point
corresponding to some (magnetically or charge) long-range ordered
state. This effect is, however, expected only below an appropriate
upper critical dimension \cite{sachdevbook}, reflecting an anomalous
dimension of the incipient ordered state. In addition, this is mechanism
produced by long wavelength order-parameter fluctuations, and is thus
expected to contribute only a small amount of entropy per degree of
freedom, in contrast to local moment formation. It is interesting
to mention that weak-coupling approaches, such as the popular {}``on-shell''
interpretation of the Random-Phase Approximation (RPA)  \cite{quinn75prl}, often predict inaccurate
or even misleading predictions \cite{dassarma05prb} for the effective
mass enhancement behavior. As discussed in Chapter 6, more
accurate modern theories such as DMFT can be used to benchmark these
and other weak coupling theories, and reveal the origin of some of
the pathologies found when they are applied to strong coupling.

\subsubsection{Finite temperature metal-insulator coexistence region}

\begin{figure}[h]

\begin{centering}
\includegraphics[width=4in]{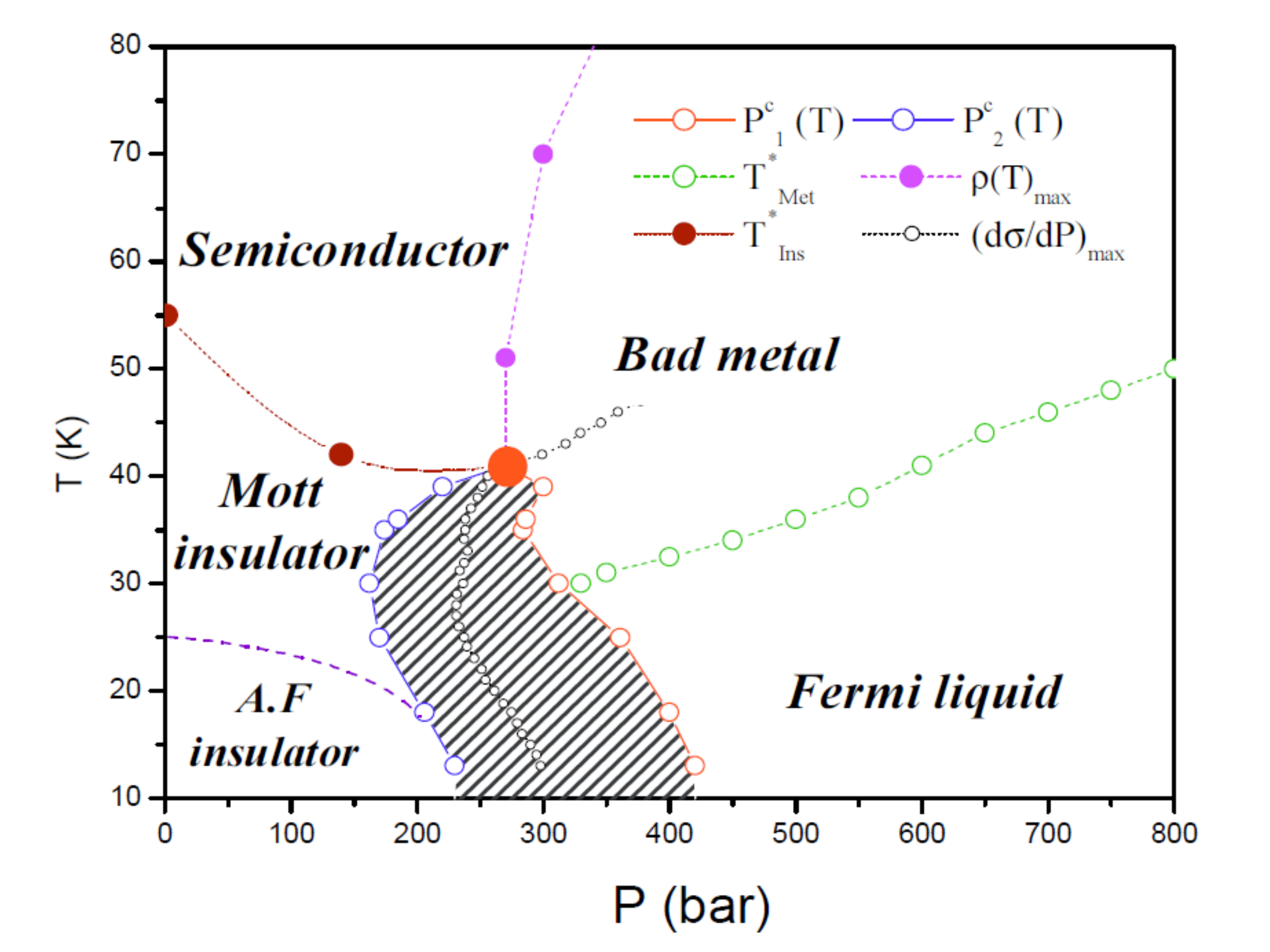}
\par\end{centering}

\caption{Phase diagram of the organic salt $\kappa-(BETD-TTF)_{2}Cu[N(CN)_{2}]Cl$
\protect\cite{limelette03prl}. In this material, increasing hydrostatic
pressure broadens the electronic bandwith, favoring the metallic state.
The first-order Mott transition extends at finite temperature up to
the critical end-point at $T_{c}\approx40K$. The corresponding coexistence
region (shaded) displays hysteresis in transport.}

\centering{}\label{limelette03prl}
\end{figure}

Since the Fermi liquid metal and the paramagnetic Mott insulator do
not differ on symmetry grounds, there is no reason why these two phases
cannot coexist in a finite range of parameter space. Indeed, recent
theories \protect\cite{georgesrmp} as well as several experimental
studies \protect\cite{limelette03prl,limelette03science} in clean
Mott system find such a metal-insulator coexistence region (Fig. \ref{limelette03prl})
leading to a finite temperature first-order phase transition line.
Experimentally, an appreciable drop of resistivity is seen as the
system is driven though such a finite temperature Mott transition,
which separates the Mott insulating state and the metallic (Fermi
liquid) state. Similarly as in standard liquid-gas systems, the coexistence
region, and the associated first-order line, terminate at the critical
end-point at $T=T_{c}$. The corresponding critical behavior has been
carefully studied in recent experiments \protect\cite{limelette03science}
on chromium-doped $V_{2}O_{3}$, and was found to belong to the standard
Ising universality class.

\subsubsection{Weak disorder near Mott transitions}

All quantities display a discontinuity across any first-order phase transition,
but this jump can be reduced in presence of impurities or disorder,
which tend to locally favor one or the other phase. Well-known droplet
arguments \protect\cite{re:Imry75}, then suggest that sufficiently
strong disorder can completely suppress such a first-order transition
\protect\cite{re:Berker91}, eliminating the finite temperature metal-insulator
coexistence region. To illustrate this argument, consider an uncompensated
doped semiconductor \cite{paalanen91}, where the bandwidth of the
impurity band can be tuned by varying the donor concentration $n$.
Assume that $n$ is chosen to lie just below its critical value $n_{c}(T)$,
so that even a small increase in $n$ would drive the system through
a Mott transition, leading to a large resistivity drop. Here, the
reduced donor density $\delta n(T)=(n-n_{c}(T))/n_{c}(T)$ plays a
role of the magnetic field in an ordinary ferromagnet, which can be
used for $T<T_{c}$ to drive the system through a first-order transition
where the magnetization jumps. A similar first order Mott metal-insulator
transition is precisely what one would expect in our case as well
- if only the donors were ordered with perfect periodicity. In fact,
the original work of Mott \protect\cite{mott1949} considered precisely
such a scenario, where one imagines varying the donor concentration,
while ignoring the local density fluctuations.

Unfortunately, in a real material, donor ions assume random positions
within the semiconducting host. A given region of size $L$ can have
a local concentration $n(L)\gtrsim n_{c}$, favoring the formation
of a metallic {}``droplet''. For $\delta n_{L}=n(L)-n_{c}>0$, the
energy of the metallic phase will be lower by the amount\begin{equation}
\Delta E_{M-I}(L)=\varepsilon\delta n_{L},\end{equation}
 where $\varepsilon$ is a constant measuring the density-dependent
(free) energy difference between the metal and the insulator. Creating
such a droplet will create a domain wall, which costs surface energy\begin{equation}
E_{s}=\sigma L^{d-1},\end{equation}
 where $\sigma$ is proportional to the surface tension of the droplet.
The droplet will be formed only if \begin{equation}
\Delta E_{M-I}(L)>\sigma L^{d-1}.\end{equation}
 Note, however, that $\Delta E_{M-I}(L)\sim\delta n_{L}$, which is
a random quantity. To calculate the probability that a droplet of
size $L$ will be formed, we need to calculate the probability of
a density fluctuation \begin{equation}
\delta n_{L}>\sigma L^{d-1}/\varepsilon.\end{equation}

Assuming that the donor density fluctuations are uncorrelated on large
enough scales, the probability distribution is given by the {}``central-limit
theorem''\begin{equation}
P(\delta n_{L})\sim\exp\left\{ -\frac{1}{2}\frac{\delta n_{L}^{2}}{L^{d}W^{2}}\right\} ,\end{equation}
 where $W$ is a constant measuring density fluctuations (disorder
strength) on the microscopic scale. The probability that the droplet
of size $L$ will for will, therefore, be of order\begin{equation}
P(L)\sim\exp\left\{ -\frac{1}{2}\frac{\sigma^{2}L^{2d-2}}{\varepsilon^{2}L^{d}W^{2}}\right\} =\exp\left\{ -\frac{1}{2}\frac{\sigma^{2}L^{d-2}}{\varepsilon^{2}W^{2}}\right\} .\end{equation}
 As we can see, for $d>2$, large droplets are exponentially suppressed,
i.e. their concentration is exponentially small, and for sufficiently
weak disorder ($W\ll\sigma/\varepsilon$), even the very small droplets
are exponentially rare. The first-order transition remains sharp,
at least at low enough temperatures.

But what happens when the temperature is increased and we approach
the critical end-point at $T=T_{c}$? Here, we expect the droplet
surface tension to decrease as a power of the correlation length $\xi\sim\left(T_{c}-T\right)^{-\nu}$
(in mean-field theory $\nu=1/2$), so even weak disorder starts to
have an appreciable effect. More precisely, here $\sigma(T)\sim\xi^{-3}$,
and $\varepsilon(T)\sim\xi^{-1}$ (see, for example Ref. \cite{goldenfeldbook}),
and even small droplets start to proliferate. The transition is then
smeared down to temperatures such that $\varepsilon(T)W/\sigma(T)\sim O(1)$,
i.e. $W\sim\xi^{-2}\sim\left(T_{c}-T\right)^{-2\nu}$. We conclude
that the in presence of weak disorder, the critical temperature is
depressed by\begin{equation}
\delta T_{c}(W)=T_{c}(0)-T_{c}(W)\sim W^{1/2\nu}.\end{equation}

When disorder is sufficiently strong ($W\sim\sigma(0)/\varepsilon(0)$),
the first order jump is completely eliminated at finite temperature,
and only a smooth metal-insulator crossover remains. Such behavior
is clearly seen in all standard doped semiconductors, where the positional
disorder in donor or acceptor ions is so strong that no evidence of
finite temperature Mott transition can be seen. A sharp distinction
between the metal and the insulator is then found only at $T=0$,
where the transition reduces to the conventional quantum critical
point. Of course, the resulting critical behavior is completely different
from that of a clean system, where all the conduction electrons simultaneously
turn into local magnetic moments at a well defined critical
concentration $n_{c}$.

In the random case the local density undergoes strong spatial fluctuations.
As a result, many regions form where the local density is much lower
then the average. Here, one expects the electrons to undergo local
Mott localization. In the remaining regions the local density is higher
the average, and the electrons remain itinerant. This simple physical
picture thus suggest a two-fluid behavior \cite{paalanenetal88} of
conduction electrons and local magnetic moments - a situation very
different then what one expects in a weakly disordered metal. Describing
such disorder-enhanced strong correlation effects proves theoretically
to be extremely difficult, since the theory must account for the effective interaction
between such disorder-induced local moments and the remaining itinerant
electrons. Developing such a theory should include an appropriate
\ description of the corresponding Kondo screening processes in a
disordered environment \cite{dkk}. This remains one of the most challenging
open problems in this field, although important advances have recently
been accomplished \cite{RoP2005review} based on dynamical mean-field
approaches.%
\begin{figure}[h]

\begin{centering}
f\includegraphics[width=4in]{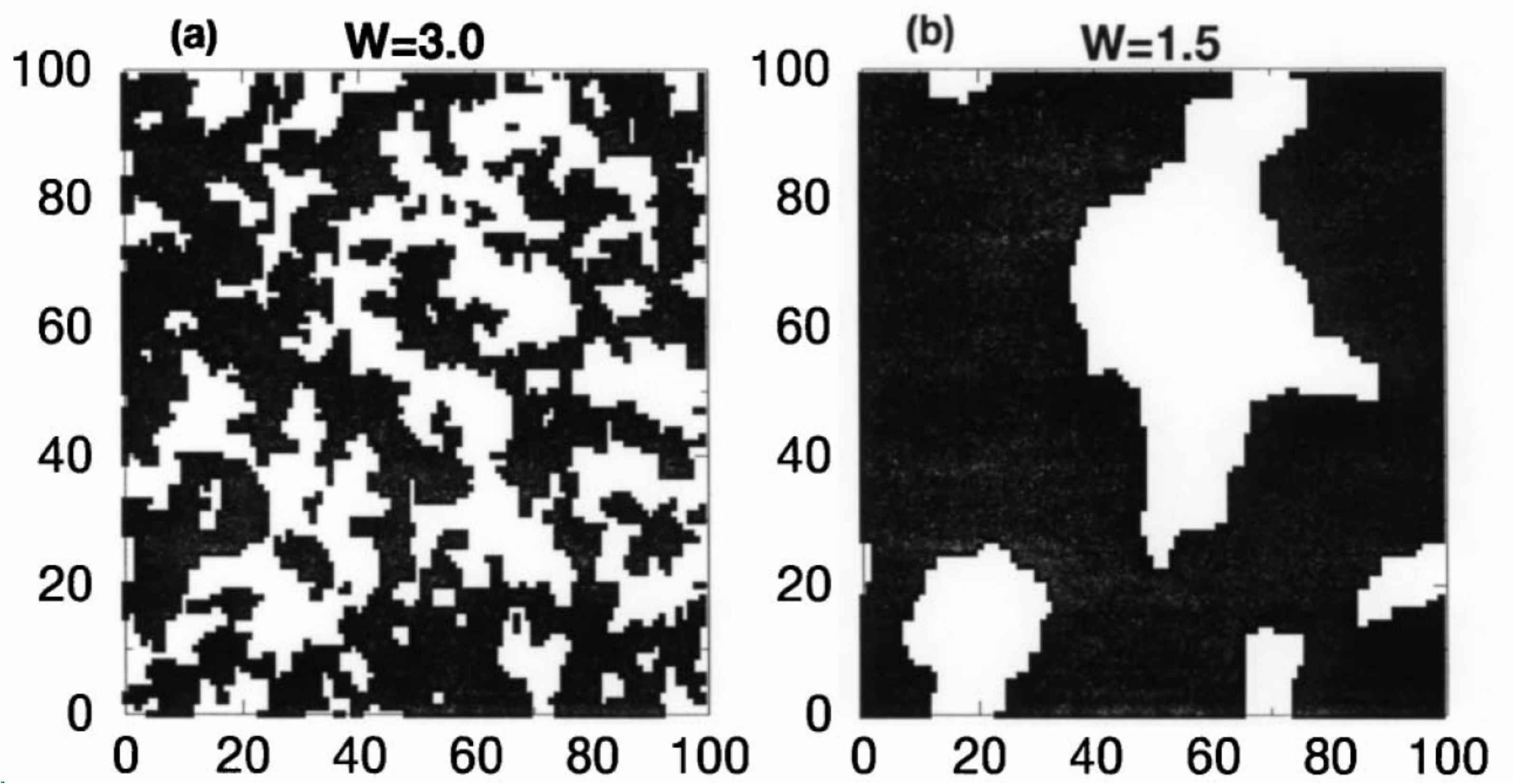}
\par\end{centering}

\caption{The simplest model for disorder-induced cluster states near first-order
phase transitions is provided by the random-field Ising model \protect\cite{re:Imry75}.
An illustration is provided by recent simulation results \protect\cite{moreo00prl},
which show how in d=2, stronger disorder (W=3 - pannel (a)) creates
many small size clusters, while only few large ones remain for sufficiently
weak disorder (W=1.5 - pannel (b)). Similar behavior is expected \protect\cite{re:Imry75}
near any disorder-smeared first order phase transition.}

\centering{}\label{more03prl}
\end{figure}

An especially interesting situation is found in $d=2$, where both
the bulk energy gain $\Delta E_{M-I}(L)=\varepsilon\delta n_{L}\sim LW$,
and the surface energy $E_{s}=\sigma L$ both scale linearly with
the droplet size $L$, allowing for \textit{arbitrarily large droplets}.
The typical droplet size thus diverges, and the first-order transition
is suppressed for arbitrarily weak disorder. This situation may be
relevant for high mobility (weak disorder) two-dimensional electron
gases \cite{abrahams-rmp01}. Here, behavior reminiscent of a Mott
transition ($m^{\ast}\sim(n-n_{c})^{-1}$) seems to emerge \cite{kravchenko-2004-67}
only at $T=0$, while only a smooth metal-insulator crossover persisting
at finite temperatures. We should note, however, that the above expressions
are valid only for sufficiently large droplets containing many impurities,
such that Gaussian statistics applies. At weak disorder, the average
distance between impurities $\ell$ is large, and the disorder can
produce only droplets larger then a certain minimum size \,$L_{\min}\gg\ell$.
We thus expect that reasonably large scale inhomogeneities should
emerge (Fig. \ref{more03prl}) when weak disorder is introduced near
first-order metal-insulator transitions in $d=2$.

\subsubsection{Is Wigner crystallization a Mott transition in disguise?}

The original ideas of Mott \cite{mott1949}, who thought about doped
semiconductors, envisioned electrons hopping between well localized
atomic orbitals corresponding to donor ions. In other Mott systems,
such as transition metal oxides, the electrons travels between the
atomic orbitals of the appropriate transition metal ions. In all these
cases, the Coulomb repulsion restricts the occupation of such localized
orbitals, leading to the Mott insulating state, but it does not provide
the essential mechanism for the formation such tightly bound electronic
states. The atomic orbitals in all these examples result from the
(partially screened) ionic potential within the crystal lattice.

The situation is more interesting if one considers an idealized situation
describing an interacting electron gas in absence of any periodic
(or random) lattice potential due to ions. Such a physical situation
is achieved, for example, when dilute carriers are injected in a semiconductor
quantum well \cite{AFS}, where all the effects of the crystal lattice
can be treated within the effective mass approximation \cite{Ashcroft}.
This picture is valid if the Fermi wavelength of the electron is much
longer then the lattice spacing, and the quantum mechanical dynamics
of the Bloch electron can be reduced to that of a free itinerant particle
with a band mass $m_{b}$. In such situations, the only potential
energy in the problem corresponds to the Coulomb repulsion $E_{C}$
between the electrons, which is the dominant energy scale in low carrier density 
systems. At the lowest densities, $E_{C}\gg E_{F}$, and the electrons
form a Wigner crystal lattice \cite{wigner34pr} to minimize the Coulomb
repulsion.

\begin{figure}[h]

\begin{centering}
\includegraphics[width=3.4411in,height=1.0585in]{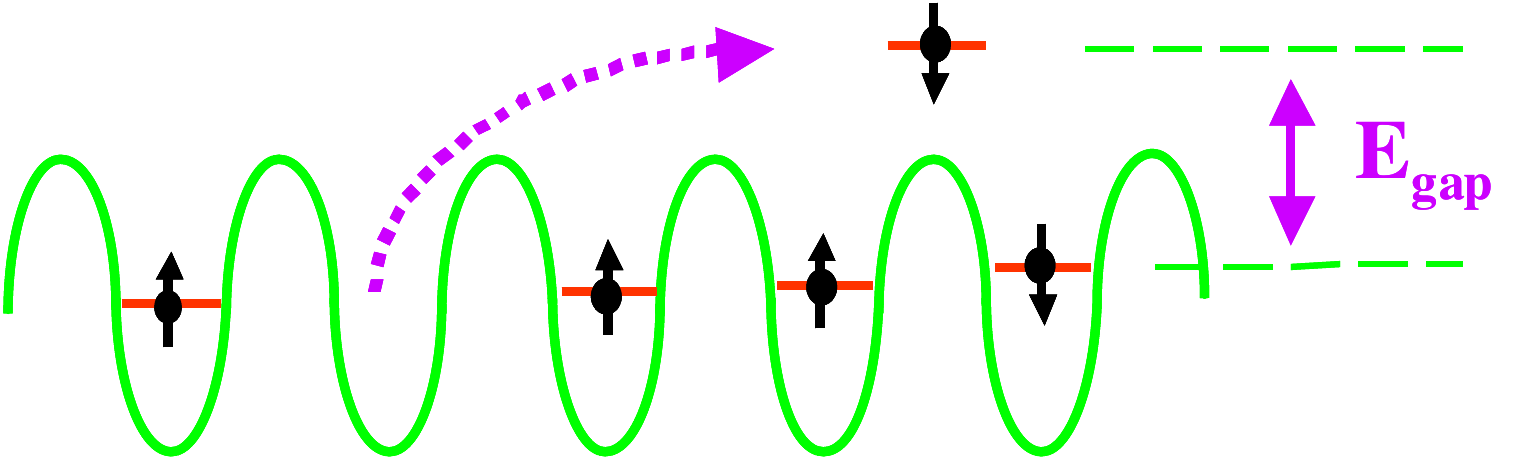}
\par\end{centering}

\caption{In a Wigner crystal, each electron is confined to a potential well
produced by Coulomb repulsion from neighboring electrons, forming
a spin 1/2 local moment. The lowest energy particle hole excitation
creates a vacancy-interstitial pair \protect\cite{candidi01prb},
which costs an energy $E_{gap}$ comparable to the Coulomb repulsion.}

\centering{}\label{wigner}
\end{figure}

Here, each electron is confined not by an ionic potential, but due
to the formation of a deep potential well produced by repulsion from
other electrons. The same mechanism prevents double occupation of
such localized orbitals, and each electron in the Wigner lattice reduces
to a localizes $S=1/2$ localized magnetic moment. A Wigner crystal
is therefore nothing but a magnetic insulator: a Mott insulator in
disguise. At higher densities, the Fermi energy becomes sufficiently
large to overcome the Coulomb repulsion, and the Wigner lattice melts
\cite{ceperley89prb}. The electrons then form a Fermi liquid. The
quantum melting of a Wigner crystal is therefore a metal-insulator
transition, perhaps in many ways similar to a conventional Mott transition.
What kind of phase transition is this? Despite years of effort, this
important question is still not fully resolved.

What degrees of freedom play the leading role in destabilizing the
Wigner crystal as it melts? Even in absence of an accepted and detailed
theoretical picture describing this transition, we may immediately
identify two possible classes of elementary excitations which potentially
contribute to melting, as follows.
\begin{itemize}
\item \textit{Collective charge excitations} ({}``elastic''\ deformations)
of the Wigner crystal. In the quantum limit, these excitations have
a bosonic character, but they persist and play an important role even
in the semi-classical ($k_{B}T\gg E_{F}$) limit, where they contribute
to the thermal melting of the Wigner lattice \cite{thouless78jphysc}.
These excitations clearly dominate in high magnetic field \cite{chen-2006-2},
where both the spin degrees of freedom and the kinetic energy are
suppressed due to Landau quantization. 
\item \textit{Particle-hole excitations} leading to vacancy-interstitial
pair formation (Fig. \ref{wigner}). These excitations have a fermionic
character, where the spin degrees of freedom play an important role.
Virtual excitations of this type give rise to superexchange processes
which produce magnetic correlations \cite{candido-2004-70} within
the Wigner crystal. These excitations would exist even if dynamic
deformations of the Wigner lattice are suppressed, for example by
impurity pinning. Recent quantum Monte-Carlo simulations indicate\cite{candidi01prb}
that the effective gap for vacancy-interstitial pair formation seems
to collapse precisely around the melting of the Wigner crystal. If
these excitation dominate, then the melting of the Wigner crystal
is a process very similar to the Mott metal-insulator transition,
and may be expected to produce a strongly correlated ($m^{\ast}/m\gg1$)
Fermi liquid on the metallic side. Behavior consistent with this possibility
has recently been documented \cite{kravchenko-2004-67} in several
two-dimensional electron systems. A microscopic theory for a simplified
model of such Wigner-Mott transitions has recently been solved \cite{camjayi-2008-4,camjayi10prb},
clarifying the mechanism for the effective mass enhancement on the
conducting side. 

\end{itemize}
We should mention, however, that the Wigner crystal melting in zero
magnetic field is believed \cite{ceperley89prb} to be a weakly first
order phase transition. Conventional (e.g. liquid-gas or liquid-crystal)
first order transitions are normally associated with a density discontinuity
and global phase separation within the coexistence dome. For charged
systems, however, global phase separation is precluded by charge neutrality 
\protect\cite{gorkov-JETP87}. In this case, one may expect the emergence of
various modulated intermediate phases, leading to bubble or stripe
\protect\cite{spivak-prl05}, or possibly even ``stripe
glass'' \protect\cite{schmalian-prl00} order. While
convincing evidence for the relevance of such {}``nano-scale phase
separation'' has been identified \cite{terletska11prl} in certain 
systems \cite{jan07prb}, recent work seems to indicate
\cite{waitnal06prb,ceperley09prl} that such effects may be negligibly
small for Wigner crystal melting.

\subsection{Localization by disorder}

A small concentration of impurities or defects simply produces random
scattering of mobile electrons. In ordinary metals, the kinetic energy
of electrons is generally so large, that the random potential due
to impurities can be treated as a small perturbation. In this case,
the Drude theory \protect\cite{Ashcroft} applies, where the conductivity
takes the form \begin{equation}
\sigma\approx\sigma_{o}=\frac{ne^{2}\tau_{tr}}{m},\label{eq:drudeconduc}\end{equation}
 where $n$ is the carrier concentration, $e$ the electron charge
and $m$ its band mass. According to Matthiessen's rule \protect\cite{Ashcroft}, the transport
scattering rate takes additive contributions from different scattering
channels, viz.\begin{equation}
\tau_{tr}^{-1}=\tau_{el}^{-1}+\tau_{ee}^{-1}(T)+\tau_{ep}^{-1}(T)+\cdots.\label{eq:mathiessen}\end{equation}
 Here, $\tau_{el}^{-1}$ is the elastic scattering rate (describing
impurity scattering), and $\tau_{ee}^{-1}(T)$, $\tau_{ep}^{-1}(T)$,...,
describe inelastic scattering processes from electrons, phonons, etc.
It is important to note that in this picture the resistivity $\rho=\sigma^{-1}$
is strictly a monotonically increasing function of temperature \begin{equation}
\rho(T)\approx\rho_{o}+AT^{n},\label{eq:lowtresist}\end{equation}
where $A>0$, and the exponent $n$ depends on the scattering process
($n=1$ for electron-phonon scattering; $n=2$ for electron-electron
scattering, etc.). The residual resistivity $\rho_{o}=\sigma^{-1}(T=0)$
is thus viewed as a measure of impurity (elastic) scattering.

In contrast, in low carrier density systems, the impurity potential
is comparable or larger then the Fermi energy, and the electrons can
get trapped, i.e. {}``localized'' by the impurities. Of course,
this process generally leads to a sharp metal-insulator transition
only at $T=0$, since at finite temperature the electrons can overcome
the impurity binding potential through thermal activation. In the
low temperature limit, a continuous metal -- insulator transition
is typically found \cite{rosenbaum80prl}, where the conductivity
decreases in a \ power -- law fashion\[
\sigma(T=0)\sim(n-n_{c})^{\mu},\]
where the conductivity exponent $\mu$ characterizes the critical
point. The precise value of the critical exponent $\mu$, as well
as other aspects of the transition depend strongly on the specific
form of disorder, and especially on the characteristic lengthscale
for disorder.

\subsubsection{Percolation transition}

The simplest situation where disorder can cause electron localization
is found in cases where the random potential is sufficiently {}``smooth''.
More precisely, consider the situation where the spatial correlation
length for the random potential is much larger then the phase coherence
length $L_{\phi}$ \protect\cite{leeramakrishnan}.

\begin{figure}[h]
\begin{centering}
\includegraphics[width=3.0113in,height=3.378in]{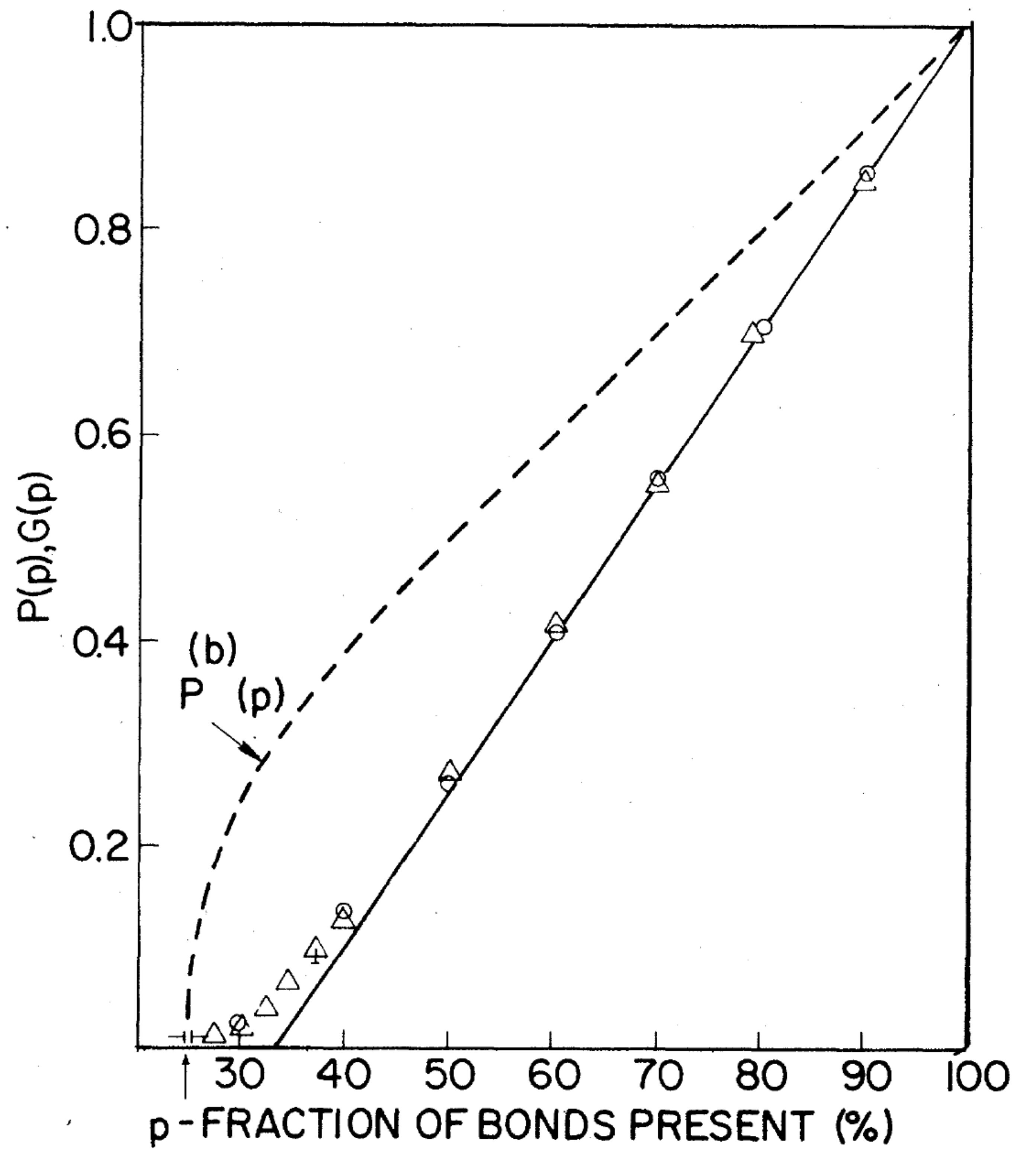}
\par\end{centering}

\caption{Critical behavior at the bond percolation for a $3D$ cubic lattice,
following \protect\cite{kirkpatrick73rmp}. Percolation probability(dashed
line) $P(p)$ and conductance (symbols) $G(p)$ are shown as a function
of the bond probability $p$. The solid line is the prediction of
an (approximate) effective medium theory.}

\centering{}\label{percolation}
\end{figure}

In this regime, the influence of the random potential can be described
in a semi-classical picture, where quantum interference processes
can largely be ignored, and the calculation of the resistivity reduces
to solving a percolation problem \protect\cite{stauffer-book} describing
a random resistor network.

As the electrons are added to the system, the electron liquid will
first fill the deepest potential wells, thus forming small metallic
{}``puddles''. At low temperature, regions where the random potential
$V(\mathbf{r})>E_{F}$ is essentially free of electrons and therefore
represent insulating areas. When the electron density (and thus $E_{F}$)
increases, the metallic puddles grow and eventually connect at the
percolation threshold; the system becomes metallic.

The critical behavior of the conductivity within such a percolation
approach has been studied in detail \protect\cite{stauffer-book},
giving\begin{equation}
\mu\approx\left\{ \begin{array}{c}
2.0\; for\; d=3,\\
1.3\; for\; d=2.\end{array}\right.\end{equation}
 Note (see Fig. \ref{percolation}) that such dimensionality-dependent
critical behavior is seen only within a narrow critical region. In
$d=3$, for example, the conductance behavior is rougly linear, in
agreement with the prediction of the effective-medium theory \protect\cite{kirkpatrick73rmp}
which does not capture such dimensionality dependence. This theory,
as any mean-field description, works equally well in any dimension
- everywhere except within a narrow critical region, where long-wavelength
fluctuations produce dimensinality-dependent behavior.

The percolation theory predictions can be directly compared experiments,
and agreement is found in a number of systems such as granular metals
\cite{granular07rmp}, where the characteristicv inhomogeneities scale
is sufficiently large. The percolation scenario seems also to apply
\cite{shahar97prl} to certain experiments showing apparent violation
of the expected quantum critical scaling in quantum Hall plateau transitions.
Similar behavior is found, for example, in manganese oxide materials
showing colossal magneto-resistance, where disorder induces nano-scale
phase separation \protect\cite{dagotto-book}, but the transport
behavior can be well described using an effective random resistor
model and the underlying percolation processes. Here the droplets
sizes are not necessarily very large, but the temperatures are sufficiently
elevated to produces sufficiently short $L_{\phi}$. Such percolative phase 
coexistence has very recently been observed by nano-scale x-ray imaging 
on a thermally-driven Mott transition in $VO_2$ \cite{basov11prb}.

In other systems, most notable uncompensated doped semiconductors,
low temperature studies find $\mu\approx0.5$, indicating drastic
departure from the percolation picture. Can a better description of
quantum effects provide the answer? Or does one have to include strong
correlation effects as well? To address this question, it is useful
to first discuss the prediction of a full quantum theory of localization 
for noninteracting electrons in presence of disorder.

\subsubsection{Tunneling vs. localization}

The semiclassical percolation picture assumes that in the insulating
phase the electrons are confined to potential wells. It ignores the
possibility of tunneling through barriers separating the wells - a
process that is allowed by quantum mechanics. \ Although the tunneling
probability is exponentially small with the barrier size, it always
remains finite. \ Naively, one would expect that the tunneling processes
would lead to an exponentially small but finite conductivity even
in the regime where classical percolation would predict insulating
behavior. \ According to this argument, quantum mechanical tunneling
would smear the metal insulator transition, and no true insulating
behavior would be possible, even at $T=0$. \ This argument is, of
course, incorrect because is based on an incomplete description of
quantum mechanics. \ It does take into account the tunneling effect,
but it ignores the crucial interference processes without which electronic
bound states could not be formed. To fully appreciate this, we should
recall that interference processes are what gives rise to the formation
of quantized electronic orbits even for simple bound states within
atoms or molecules.

\subsubsection{Anderson localization}

The possibility that true electronic bound states can be formed in
presence of a random potential was first discussed by Anderson in
1958 \protect\cite{anderson58}. This pioneering work argued
that sufficiently strong randomness will localize all the electronic
states within a given band, leading to a sharp metal-insulator the
transition at $T=0$. Anderson's original argument takes the local
point of view, which provides a very transparent physical picture
of how localization can occur, as follows.

Suppose that an experiment or a computer simulation can examine only
local quantities associated with a particular lattice site. Can such
a study determine whether the material is a metal or an Anderson insulator?
The answer \protect\cite{anderson58} is - somewhat surprisingly
- yes! One simply has to determine the escape rate $\hbar/\tau_{esc}$
from the local orbital. The recipe how to do this has a long history,
and is provided by Fermi's Golden Rule: \begin{equation}
\frac{\hbar}{\tau_{esc}}=t^{2}\rho_{loc}(\varepsilon).\end{equation}
 where $t$ is the appropriate matrix element between the local orbital
and its environment. The local density of available states $\rho_{loc}(\varepsilon),$
to which an electron of energy $\varepsilon$ can escape, is proportional
to the wavefunction amplitude on this site \begin{equation}
\rho_{loc}(\varepsilon)\sim|\psi_{i}(\varepsilon)|^{2}.\end{equation}
 If electronic states in the relevant energy range are all localized
(i.e. bound), then only a small number of such states have appreciable
overlap with the given site. Therefore, in the case of the Anderson
insulator, the local density of states (LDOS)will consist of only
a few discrete $\delta$-function peaks (Fig. 1.11) with appreciable weight -
which are very improbable to reside at the Fermi energy.%
\begin{figure}[h]
\includegraphics[width=5in]{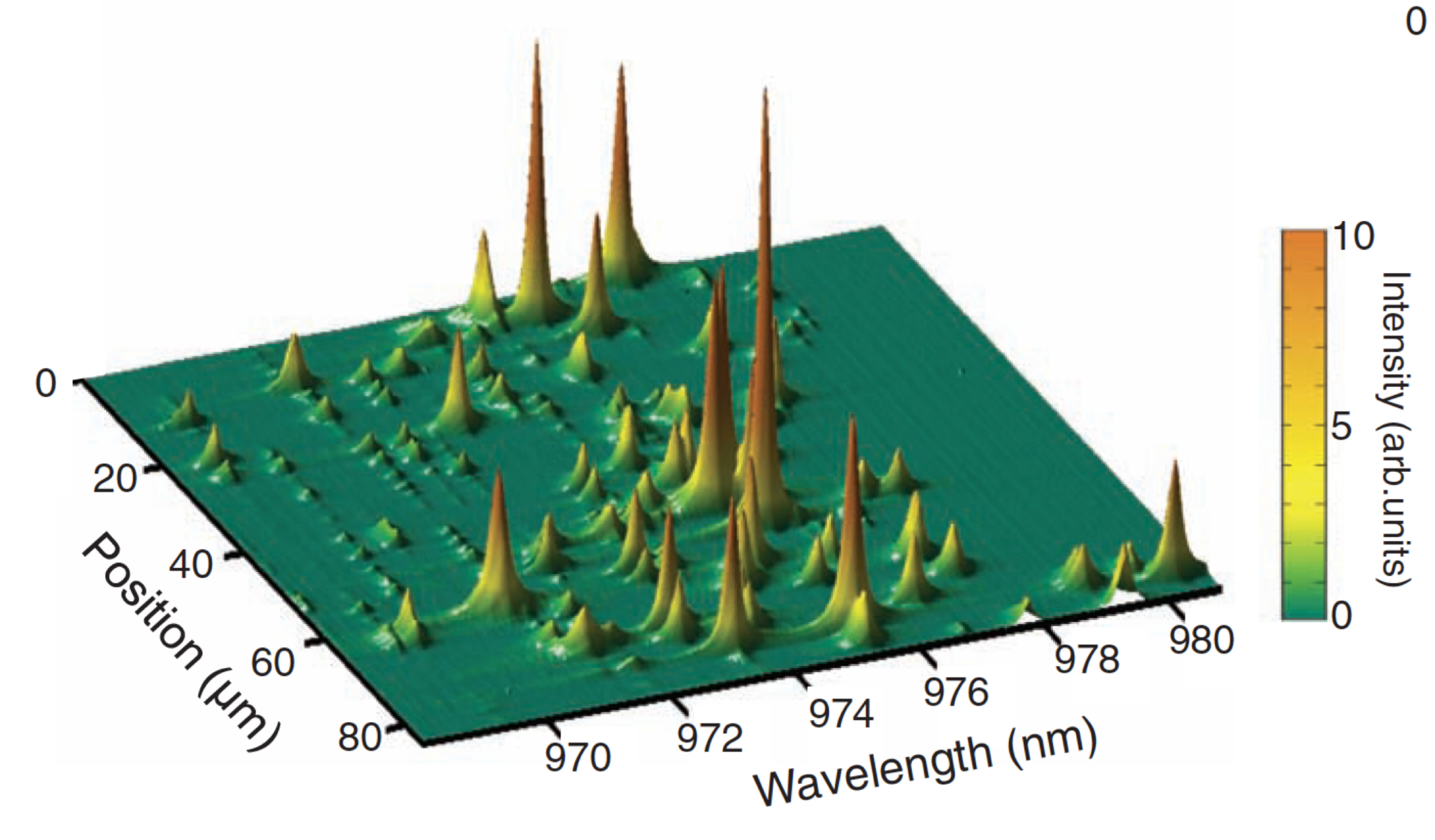}

\caption{Wavefunction amplitude $|\psi_{i}|^{2}$ of strongly localized states.
Since the concept of Anderson localization applies to any wave in
random media, recent efforts have documented localization of light
in disordered optical media. Shown here are spectra experimentally
observed in photonic crystals \protect \cite{Sapienza2010science}. }

\end{figure}

At strong disorder $W$ $\gg B$, the typical density of available
states and thus the \textit{typical escape rate} from a given site
can be shown to vanish \protect\cite{anderson58}, and the electron
remains localized. In the Anderson insulator, the spectrum of the
local environment of any given site has a gap with a certain typical
size measuring the localization strength. According to this point
of view, the ultimate localization mechanism in both the Mott and
the Anderson insulator is somewhat similar: the electron cannot find
levels to which it can escape.

Anderson's original arguments demonstrated that sufficiently
strong disorder is able to localize all electronic states within a narrow
band of electronic states. At weaker disorder, only the states near the band edge are expected
to localize, but other states at energies $E>E_{c}$ - the so-called
``mobility edge'' - remain extended. As
the Fermi energy is increased (for example by carrier doping), the
system undergoes an Anderson metal-insulator transition. Following
the development of the scaling theories of localization \protect\cite{gang4},
and especially due to recent progress in numerical studies of the
problem, the corresponding critical behavior is now well understood
for noninteracting electrons. These studies established (see also Chap. 3)
 that for noninteracting
electrons at $T=0,$ all the electronic states remain localized for
dimensions $d\leq2$, while a continuous metal-insulator transition
is found in higher dimensions. According to most recent estimates,
the corresponding critical exponent \begin{equation}
\mu\approx1.58\end{equation}
 in $d=3$, and becomes even larger in higher dimensions (see below).

It is interesting to note that the conductivity exponent at $d=3$
Anderson transition is not too different from that of percolation
theory ($\mu_{perc}\approx2$). In most experimental systems where
the metal-insulator transition has been studied, the observed exponent
is much smaller ($\mu\lesssim1$). This indicates that disorder-driven
mechanisms which ignore electron-electron interactions cannot hope
to explain the behavior at the metal-insulator transition.

\subsection{Basic interaction effects in disordered systems}

In most realistic systems both the disordered strength and the electron-electron
interactions have comparable magnitudes. In such cases, the localization
mechanisms of Mott and Aderson cannot be considered separately, since
each will influence and affect the other. Despite recent advances
(see Chap.6), a complete theory
of such a Mott-Anderson transition remains incomplete. Nevertheless,
early physical arguments of Mott \protect\cite{mott-book90} already
indicate how new and more complicated behavior must emerge when both
mechanisms are at play.

\subsubsection{The Mott-Anderson transition}

To see this most simply, consider a disordered Hubbard model in the
strongly localized (atomic) limit. In the absence of disorder, each
site has two energy levels, $\varepsilon_{0}=0$ and $\varepsilon_{1}=U$,
where $U$ is the on-site interaction potential. If the system is
half-filled, then each site is singly occupied; the levels $\varepsilon_{1}$
remain empty. We have one local magnetic moment at each site, and
a gap equal to $U$ to charge excitations.

When disorder is added, each of these energy levels is shifted by
a randomly fluctuating site energy $-W/2<\varepsilon_{i}<W/2$. The
situation remains unchanged for $W<U$, as all the levels $\varepsilon_{1}^{\prime}(i)=U+\varepsilon_{i}$
remain empty (for half-filling the chemical potential is $\mu=E_{F}=U/2$).
For larger disorder, those sites with $\varepsilon_{i}>U/2$ have
the level $\varepsilon_{0}^{\prime}(i)=0+\varepsilon_{i}>\mu$ and
are empty. Similarly, those sites with $\varepsilon_{i}<-U/2$ have
the excited level $\varepsilon_{1}^{\prime}(i)=U+\varepsilon_{i}<\mu$
and are doubly occupied. Thus for $W>U$ a fraction of the sites are
either doubly occupied or empty. The Mott gap is now closed, although
a fraction of the sites still remain as localized magnetic moments.
We can describe this state as an inhomogeneous mixture of a Mott and
an Anderson insulator (as shown in the center of Fig. 12). However,
the empty and doubly occupied sites have succeeded to completely fill
the gap in the average single particle density-of-states (DOS).%
\begin{figure}[h]
\includegraphics[width=5.0194in,height=1.9666in,keepaspectratio]{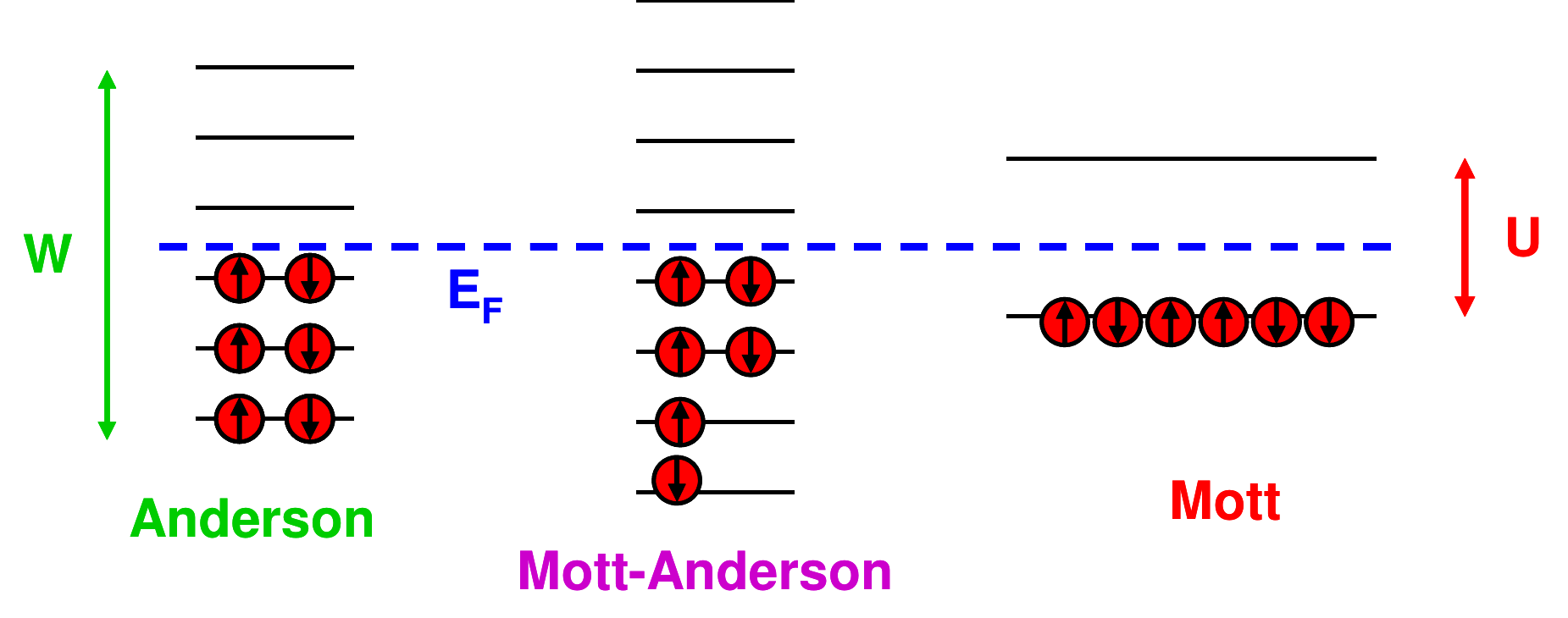}

\caption{Energy level occupation in the strongly localized (atomic) limit for
an Anderson (left), a Mott (right), and a Mott-Anderson (center) insulator.
In a Mott-Anderson insulator, the disorder strength W is comparable
to the Coulomb repulsion U, and a two-fluid behavior emerges. Here,
a fraction of localized states are doubly occupied or empty as in
an Anderson insulator. Coexisting with those, other states remain
singly occupied forming local magnetic moments, as in a Mott insulator.
Note that the spins of the local moments may be randomly oriented
indicating the absence of magnetic ordering. The chemical potential
is represented by the dashed line.}

\label{mott-anderson}
\end{figure}

This physical picture of Mott, which is schematically represented
in Fig.~\ref{mott-anderson}, is very transparent and intuitive.
The nontrivial question is how the strongly localized (atomic) limit
is approached as one crosses the metal-insulator transition from the
metallic side. To address this question one needs a more detailed
theory for the metal-insulator transition region, which was not available
when the questions posed by Mott and Anderson were put forward.

\subsubsection{Coulomb gap}

Local moment formation leading to the Mott or the Mott-Anderson insulating
state is most important for narrow bands where the on-site Coulomb
repulsion ({}``Hubbard U'') dominates. This mechanism is most effective
close to half-filling, since local moment formation requires exactly
one electron per orbital. In this regime the long-range (inter-site)
component of the Coulomb interaction plays a secondary role, because
on-site repulsion opposes charge rearrangement. Such a situation is
found in narrow impurity bands (deeply insulating regime) of uncompensated
doped semiconductors such as \emph{Si:P}. Deep in the insulating regime,
each electron forms a hydrogenic bound state with exactly one Phosphorus
ion, forming a spin $S=1/2$ local magnetic moment, and charge rearrangements
are suppressed.

\begin{figure}[h]
\begin{center}
\includegraphics[width=4.5in]{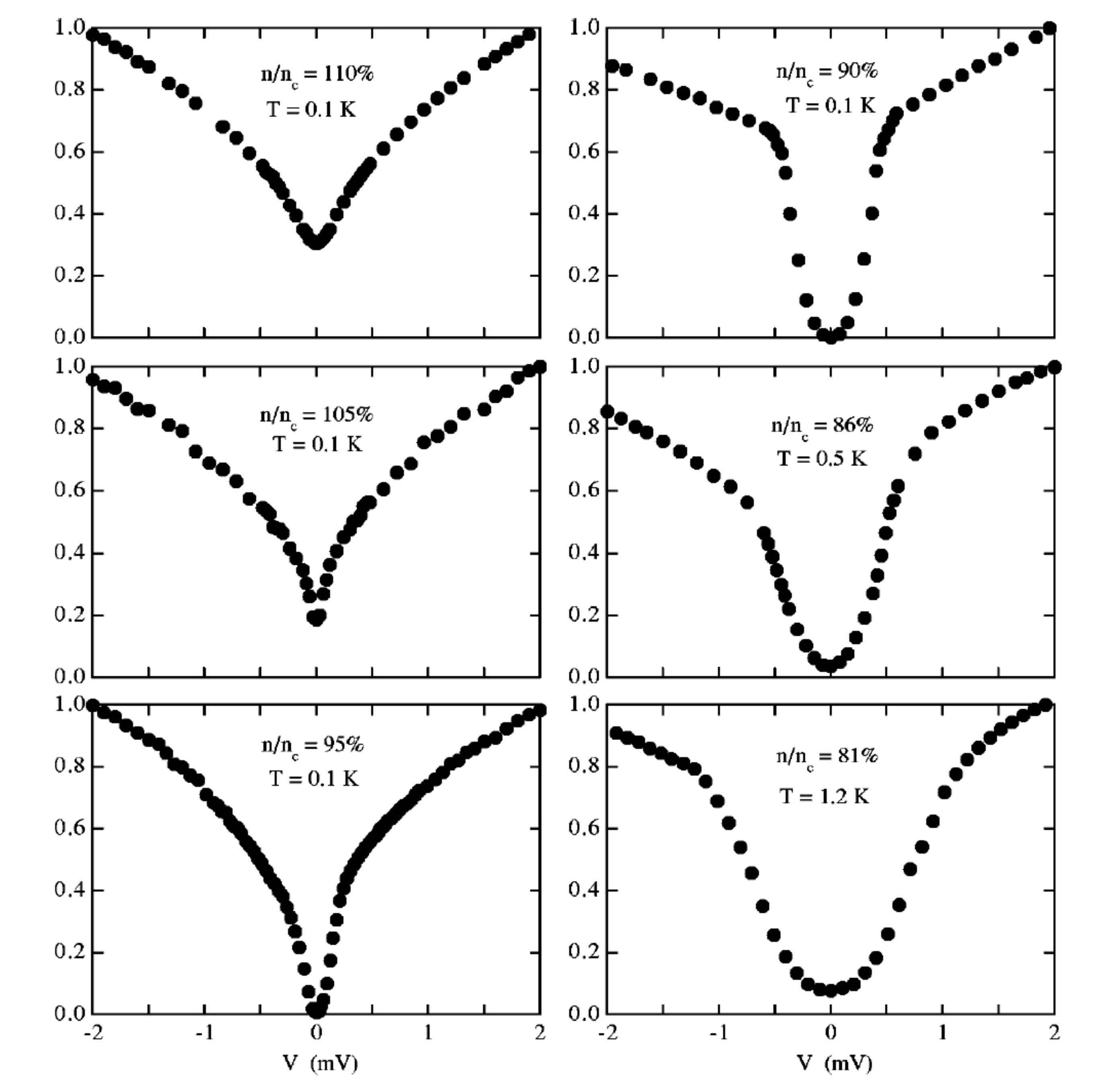}
\end{center}
\caption{Tunneling density of states spectra observed \protect \cite{PhysRevB.60.1582}
accross the metal-insulator transition in \emph{Si:B}. The Coulomb gap
is seen to gradually close and change shape as the transition is crossed.
This crossover behavior was interpreted to reflect the emergence of screening as one approaches the metallic
phase. }

\end{figure}

A more complicated situation is found \cite{doped-book} away from
half filling, which can be realized, for example, in partially compensated
\emph{Si:P,B}. Here the electrons can occupy different localized states,
and many charge rearrangements are possible. This is the regime considered
by the well-known theory of Efros and Shklovskii \cite{re:Efros75,doped-book},
which focuses on a classical model of spinless electrons distributed
among stongly localized states, as givn by the Hamiltonian\begin{equation}
H=\sum_{j\ne i}\frac{e^{2}}{\mathbf{\kappa|r_{i}-r_{j}|}}(n_{j}-\overline{n})(n_{j}-\overline{n})+\sum_{j\ne i}\varepsilon_{i}n_{i}.\end{equation}

 Here $n_{j}=0,1$ is the occupation number of the remaining localized
states with (bare) energy $\varepsilon_{i}$ at position $\mathbf{r}_{i}$,
$\overline{n}$ is the average occupation per site, and $\kappa$ is the dielectric
constant of the insulator. For localized electrons, the single-particle
(tunneling) density of electronic states (DOS) is then simply the
probability distribution of the local energy levels \begin{equation}
N(\varepsilon)=<\delta(\varepsilon-\varepsilon_{i})>\end{equation}
of an electron occupying a localized state is shifted (renormalized)
by the electrostaic potential produced by the electrons on all remaining
sites.

\begin{equation}
\varepsilon_{i}\rightarrow\varepsilon_{i}^{R}=\varepsilon_{i}+\sum_{j\ne i}\frac{e^{2}}{\mathbf{\kappa|r_{i}-r_{j}|}}(n_{j}-n).\end{equation}

This electrostatic shift depends, of course, on the precise electronic
configuration, favoring those charge configurations which lower the
energies of the occupied states. In absence of disorder, this effect
leads to charge ordering - Wigner crystallization - opening a hard
gap at the Fermi energy. 

When sufficiently strong disorder is present, the Wigner gap is smeared,
leading to the soft ``Coulomb gap'' (Fig. 1.13). The original argument of
Efros and Shklovskii \cite{re:Efros75} rested on a stability argument
for the ground state with respect to any single-particle displacement
from a given occupied site i to an empty site j. Stability requires
every such excitation to cost positive energy, giving\begin{equation}
\varepsilon_{j}-\varepsilon_{i}-\frac{e^{2}}{\mathbf{\kappa|r_{i}-r_{j}|}}>0.\end{equation}

If the states i and j are within energy $\varepsilon$ from the Fermi
level, i.e. $|\varepsilon_{i}-\varepsilon_{F}|<\varepsilon$, then
their typical separation in space is $r(\varepsilon)=e^{2}/\kappa\varepsilon$
is large if $\varepsilon$ is small. The DOS $N(\varepsilon)\sim\frac{d}{d\varepsilon}[r^{-3}(\varepsilon)]$
then has a soft gap around the Fermi energy. In three dimensions the
result is

\begin{equation}
N(\varepsilon)=\frac{3\kappa^{3}}{\pi e^{6}}(\varepsilon-\varepsilon_{F})^{2}.\end{equation}

More generally, for interactions of the form $V(r)\sim r^{-\alpha},$
with $\alpha<d$ (in $d$ spatial dimensions), a generalization of this argument
gives \cite{pankov05prl} $N(\varepsilon)\sim(\varepsilon-\varepsilon_{F})^{\beta},$
with $\beta=(d-\alpha)/\alpha$. Arguing that incomplete (scale-dependent)
screening modifies the form Coulomb interaction within the quantum
critical region of the metal-insulator transition, recent work suggested 
\cite{PhysRevB.60.1582} that in this case $\alpha=2$, leading to 
$\beta=1/2$ in $d=3.$ While such critical scaling seems consistent
with the experimental results on $Si:B$,
a more precise treatment of quantum effects will be needed for a more
convincing theory of the Coulomb gap in the quantum critical region. 

It is very important to reiterate that the Anderson localization mechanism
by itself does not lead to opening of any kind of gap at the transition.
This scenario, which provided a popular and attractive scenario for
interpreting many transport experiments, rests on the concept of a
``mobility edge'' - the energy separating the extended from the
localized states. The critical concentration then obtains when all
the localized states are filled up and the Fermi energy reaches the
mobility edge. Precisely at the critical point one expects \protect\cite{mott-book90}  the electronic
states ``above'' the Fermi energy to be extended, while those
``below'' the mobility edge to remain localized. A test of these
ideas has very recently become possible through high resolution scanning
tunneling microscopy (STM) experiments \cite{yazdani10science} which
can directly determine the degree of wavefunction localization in
an energy-resolved fashion. In contrast to the conventional mobility
edge scenario, these experiments provided striking evidence that,
close to the critical concentration, the electronic states precisely
at the Fermi energy are the ones most strongly localized. This experiment,
which will be discussed in more detail in Chapter 7, provided
direct evidence that interaction effects cannot be neglected 
near the metal-insulator transition, where pseudogap opening plays
a key role in controlling the localization of electronic states. Is
the relevant interaction mechanism directly related to Coulomb glass
phenomena? Only time will tell. Still, all these experiments make
it clear that the fundamental -- but yet unresolved -- physics questions
posed by the early work of Mott cannot be ignored.

\subsubsection{Coulomb glass}

The long-range Coulomb interactions produce, however, another important
effect. Because Coulomb repulsion favors a uniform charge configuration
while disorder opposes it, these competing interactions give rise
to frustration and the emergence of many meta-stable electronic states.
As described in Chapter 8, this typically leads gradual glassy freezing
of electrons, to slow relaxation and ``aging'', in a fashion surprisingly similar to glassy phenomena
in spin glasses or super-cooled liquids. 

The precise relation of the Coulomb gap formation and the glassy freezing
has, however, long remained controversial and ill-understood \cite{clareyu93prl}. On the
one hand, even the early arguments of Efros and Shklovskii indicate
that the long-range nature of the Coulomb interaction is a key feature
for the formation of the Coulomb gap. On the other, thermal and/or
quantum fluctuations can allow charge rearrangements, generically
leading to phenomenon of screening, practically eliminating the long-range
part of the interaction. In contrast, if the electrons are (partially
or fully) frozen in a glassy state, then screening may remain incomplete,
allowing the long-range nature of the Coulomb interaction to manifest
itself. This physically plausible idea \cite{leeramakrishnan} has
found support in very recent theoretical work \cite{pastor-prl99,horbach02prb,pankov05prl},
which argues that the two phenomena typically go hand-in-hand. 

Both the formation of the Coulomb gap and the emergence of glassy
features are, at this time, well established features in strongly
disordered insulators. But how should this influence the approach
to the metal-insulator transition? Recent theoretical \cite{pastor-prl99,mitglass-prl03} and experimental
\cite{bogdanovich-prl02,mag-glass} works have suggested that in some
cases it may dramatically affect the critical region, perhaps even
leading to an intermediate metallic-glass phase with unusual transport
properties \cite{Denis}. This important question remains far from
settled. Still, its fundamental importance has very early been recognized
by Mott \cite{mott-book90}, who noted that the phenomenon of screening
must be dramatically modified as one crosses from a metal to an insulator.
In short, the localization of electrons immediately produces the demise
of screening, so the two phenomena must be profoundly linked. Mott's
dream was to understand how this ``unscreening'' occurs at the
metal-insulator transition, a physical question of basic importance,
but one that is typically not addressed by most conventional theories,
thus remaining a major challenge for future work.

\section{Current theories of the metal-insulator transition}

\subsection{MIT as a critical point}

\subsubsection{Absence of minimum metallic conductivity}

The early ideas of Mott and Anderson identified the basic mechanism
for the metal-insulator transition, but did not provide specific and
detailed prediction for the critical behavior, or even the precise
nature of this phase transition. In fact, up to the late 1970s, Mott's
arguments \cite{mott-book90} suggested that the transition is discontinuous,
where a \textit{minimum metallic conductivity} should exist on the
metallic side even at $T=0$. Mott's early argument examined the transport
behavior based on Drude's picture, where increasing disorder simply
reduces the elastic mean-free path $\ell$. Since the scattering rate
$\tau_{tr}^{-1}$ from any impurity assumes a maximum possible value
which is finite (the so-called {}``unitarity limit'') , the corresponding
mean-free path $\ell=v_{F}\tau_{tr}$ cannot be shorter then a microscopic
lower cutoff $a$ of the order of the lattice spacing. Therefore,
it was argued, the conductivity of any metal is bounded from below
by the {}``Mott limit''\[
\sigma\geq\sigma_{\min}=\frac{ne^{2}a}{mv_{F}}.\]

Early low temperature experiments on many materials seemed to confirm
these predictions by only reporting metallic conductivities in excess
of $\sigma_{\min}$. The metal-insulator transition in disordered
systems was thus assumed to have a first-order character, similarly
as the Mott transition in clean systems. Similarly, high temperature
behavior in a number of metallic systems was found to display {}``resistivity
saturation'', where the Mott limit is approached due to incoherent
(inelastic) scattering.

With the development of more advance cryogenic techniques, lower temperatures
and more precise measurements became available. Our perspective on
the fundamental nature of the transition has been deeply influenced
by the ground-breaking experiments on \emph{Si:P} in early 1980s \protect\cite{rosenbaum80prl,paalanen82}.
These experiments provided evidence (Fig. 1.14) of metallic conductivities as
much as two orders of magnitude smaller then $\sigma_{min}$. They also
made it clear that the metal-insulator transition in doped semiconductors
is a continuous (second order) phase transition \protect\cite{paalanen82},
which bears many similarities to conventional critical phenomena.
This important observation has sparked a veritable avalanche of experimental
\protect\cite{paalanen91,sarachik95} and theoretical \protect\cite{wegner76z.phys.b,wegner79,gang4,wegner80}
works, most of which have borrowed ideas from studies of second order
phase transitions. Indeed, many experimental results were interpreted
using scaling concepts \protect\cite{leeramakrishnan}, culminating
with the famed scaling theory of localization \protect\cite{gang4},
and the subsequent extensions to incorporate the interaction effects
\protect\cite{fink-jetp83,fink-jetp84,cclm-prb84,kirkpatrick-rmp94}.

\begin{figure}[h]
\begin{center}
\includegraphics[width=3.4in]{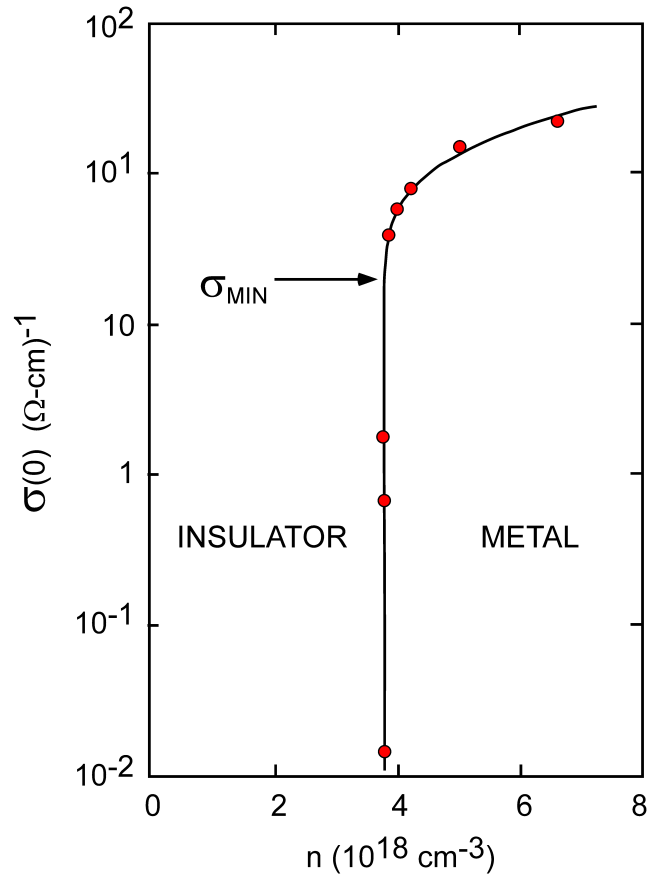}
\end{center}
\caption{Critical behavior of the conductivity extrapolated to $T\rightarrow0$
for uncompensated \emph{Si:P} \protect\cite{rosenbaum80prl}. Sharp power-law
behavior with critical exponent $\mu\approx1/2$ is extending over
a surprisingly large concentration range. Finite values of the conductivity
much smaller then $\sigma_{M}$ (shown by arrow) are observed close
to the transition.}

\centering{}\label{SiP}
\end{figure}

\subsubsection{Phenomenological scaling formulation}

Microscopic theories describing the MIT remain controversial and somewhat
incomplete. We should stress, however, that scaling behavior near
second order phase transitions is a much more robust and general property
then any particular approximation scheme or microscopic model. Historically,
the scaling hypothesis of Widom \protect\cite{widom65} has been
put forward for conventional (classical) critical phenomena much before
the microscopic theory of Wilson \protect\cite{Wilson1} became available.
It has provided crucial guidance for experimentalist to systematically
analyze the experimental data, and has provided a framework and direction
for the development of microscopic theories.

A phenomenological scaling hypothesis can be formulated for quantum
criticality \ \protect\cite{sachdevbook} as well, in direct analogy
to conventional critical phenomena. In particular, if the scaling
description is valid, then a single correlation length $\xi\sim\delta n^{-\nu}$
exists characterizing the system, and the corresponding time scale
$\tau_{\xi}\sim$ $\xi^{z}$, both of which diverge in a powerlaw
fashion at the critical point. Here, $\delta n=(n-n_{c})/n_{c}$ is
the dimensionless distance from the transition, and we have introduced
the correlation length exponent $\nu$ and the {}``dynamical exponent''
$z$. Because of the Heisenberg Uncertainty Principle, the corresponding
energy (temperature) scale \[
T^{\ast}\sim\frac{\hbar}{\tau_{\xi}}\sim\delta n^{\nu z}\]
vanishes as the critical point is approached.

For the MIT the conductivity plays a role similar to an order parameter,
as it vanishes at $T\longrightarrow0$ in the insulating phase. Its
sharp critical behavior at $T=0$ is rounded at finite temperature,
suggesting a scaling behavior similar to that of a ferromagnet in
an external (symmetry breaking) field. The conductivity can therefore
be written in a scale invariant form as \[
\sigma(\delta n,T)=b^{-\mu/\nu}\; f_{\sigma}\;(b^{1/\nu}\delta n,b^{z}T).\]
 Here, $T$ is the temperature, $b$ is the length rescaling factor,
and $\mu$ is the conductivity exponent.

The $T=0$ behavior $\sigma\left(T=0\right)\sim\delta n^{\mu}$ can
be obtained by working at low temperatures and choosing $b=\delta n^{-\nu}\sim\xi$.
We obtain the scaling form \begin{equation}
\sigma(T)=\delta n^{\mu}\widetilde{\phi}_{\sigma}(T/\delta n^{\nu z}),\label{eq:scalingcond2}\end{equation}
 where $\widetilde{\phi}_{\sigma}(y)=f_{\sigma}(1,y)$. Finite temperature
corrections in the metallic phase are obtained by expanding\begin{equation}
\widetilde{\phi}_{\sigma}(y)\approx1+ay^{\alpha},\label{eq:condscalingfunction}\end{equation}
 giving the low temperature conductivity of the form \begin{equation}
\sigma(\delta n,T)\approx\sigma_{o}(\delta n)+m_{\sigma}(\delta n)T^{\alpha}.\label{eq:lowtcond}\end{equation}
 Here, $\sigma_{o}(\delta n)\sim\delta n^{\mu}$, and $m_{\sigma}(\delta n)\sim\delta n^{\mu-\alpha\nu z}$.
Since the \textit{form} of the scaling function $\phi_{\sigma}(y)$
is independent of the distance to the transition $\delta n$, the
exponent $\alpha$ must take an universal value in the entire metallic
phase, and therefore can be calculated by perturbation theory at weak
disorder. For example, the interaction corrections in $d=3$ \protect\cite{leeramakrishnan}
lead to $\alpha=1/2$.

This scaling argument provides a formal justification for using the
predictions from perturbative quantum corrections as giving the leading
low temperature dependence in the entire metallic phase. Note, however,
that the prefactor $m_{\sigma}(\delta n)$ (i.e. its dependence on
$\delta n$) is not correctly predicted by perturbative calculations,
since it undergoes Fermi liquid renormalizations which can acquire
a singular form in the critical region near the metal-insulator transition.

The temperature dependence at the critical point (in the critical
region) can be obtained if we put $\delta n=0$, and choose $b=T^{-1/z}$,
giving \begin{equation}
\sigma_{c}(T)=\sigma(\delta n=0,T)\sim T^{\mu/\nu z}.\label{eq:criticalcond}\end{equation}
 We can also write \begin{equation}
\sigma(n,T)=T^{\mu/\nu z}\phi_{\sigma}(T/T_{o}(n)),\end{equation}
 where $\phi_{\sigma}(x)=x^{\mu/\nu z}\widetilde{\phi}_{\sigma}(x)$,
and the crossover temperature $T_{o}(\delta n)\sim\delta n^{\nu z}$.

\subsubsection{How to experimentally find quantum criticality?}

To demonstrate quantum critical scaling around a MIT one must adopts
the following systematic procedure in analyzing experimental data:
\begin{enumerate}
\item Plot $\sigma(\delta n,T)$ vs. $T$ for several carrier concentrations
$n$. Simple powerlaw behavior $\sigma_{c}\sim T^{x}$ is expected
only at the critical point, and we can identify the critical concentration
($n=n_{c}$) as the only curve that looks like a straight line when
the date are plotted on a log-log scale.
\item From the slope of the $\sigma_{c}(T)$ we find the critical exponent
$x=\mu/\nu z$.
\item Having determined $\sigma_{c}(T)=\sigma(0,T)$ we can now plot $\phi_{\sigma}(T/T_{o}(\delta n))=\sigma(\delta n,T)/$
$\sigma_{c}(T)$ as a function of $T/T_{o}(\delta n)$. The crossover
temperature is determined for each concentration $n$ in order to
collapse all the curves on two branches (metallic and insulating)
of the scaling function $\phi_{\sigma}(y)$. This procedure does not
assume any particular functional form (density dependence) for the
crossover scale $T_{o}(\delta n)$.
\item Next, we plot $T_{o}(\delta n)$ as a function of $\delta n$ on \ log-log
scale to determine the corresponding exponent $\nu z$. If scaling
works, then $T_{o}(\delta n)$ should vanish as the transition is
approached, and we expect to find the same exponent from both sides.
The conductivity exponent is then obtained from $\mu=x\nu z$.
\item A crosscheck can be obtained from extrapolating the metallic curves
$\sigma(\delta n,T)\longrightarrow\sigma_{o}(\delta n)$ to $T\longrightarrow0$,
and by determining the exponent $\mu$ from the relation
$\sigma_{o}(n)\sim\delta n^{\mu}$. 
\end{enumerate}
All the above expressions are quite general, and can be considered
to be a phenomenological description of the MIT. This is of particular
importance in instances where a scaling approach is utilized to systematically
analyze the experimental data in instances where an accepted microscopic
theory is not available. Such a situation is found in several two-dimensional
systems, where beautiful and convincing scaling behavior is observed,
providing evidence that the MIT is a well defined quantum critical
point.

\subsubsection{How to theoretically approach quantum criticality?}

To better understand the \textit{physical content} of the scaling
approach to the \ MIT, we should contrast it to standard approaches
to critical phenomena, which are by now perfectly well understood.
In most cases, the theory is built based on the emergence of spontaneous
symmetry breaking within the low temperature (ordered) phase. Based
on identifying the appropriate order parameter $\varphi$ describing
such symmetry breaking, one typically proceeds in the following steps:
\begin{enumerate}
\item Formulate an appropriate Landau theory, which defines how the free
energy $F[\varphi]$ depends on the (spatially fluctuating) order
parameter $\varphi(\mathbf{x})$.
\item Mean-field theory (MFT) is obtained by minimizing $F[\varphi]$ and
ignoring the spatial fluctuations of $\varphi(\mathbf{x}).$
\item Examine the effects of long wavelength spatial fluctuations of the
order parameter beyond MFT.
\item The most singular effects of spatial fluctuations are found close
to critical points, and are re-summed using renormalization group
(RG) methods. 
\end{enumerate}
When this program is implemented in practice, one finds that the mean-field
description (steps 1 and 2) suffices everywhere except in a very narrow
interval around the critical point. In fact, it is precisely by examining
the leading corrections to MFT one is able to theoretically estimate
the size of the so-called {}``Ginzburg'' region where non-MFT behavior
can be observed. Accounting for non-MFT behavior within such a critical
region is much more difficult, and requires the powerful arsenal of
renormalization group (RG) methods \cite{goldenfeldbook}. The practical
calculations simplify considerably near the upper critical dimension
$d_{uc}$, where the fluctuations corrections are logaritmically weak,
and can be effectively re-summed by a perturbative RG methods, using
$\varepsilon=d_{uc}-d$ as a small parameter in the theory \cite{wilsonfisher72prl}. For standard
magnetic critical phenomena $d_{uc}=4$, this RG program has been
effectively implemented, and all the appropriate critical exponents
calculated to leading order in $\varepsilon=4- d$. When
the results are extrapolated to $d=3$, very impressive agreement
with both experiments and numerical simulation results has obtained.
The theory of conventional (thermal) critical phenomena can therefore
be considered a closed book.

\subsubsection{What to do if an order-parameter description is not available?}

We emphasize that the above {}``standard''\ approach to criticality
relies on being able to identify an appropriate symmetry breaking
scheme. For phenomena such as magnetic and charge ordering or superconductivity,
this program can be straightforwardly extended to the quantum domain,
as first discussed by John Hertz \protect\cite{hertz}. In such cases,
especially for insulating magnets, the familiar approach of Landau
theory and weak-coupling RG methods has been studied in detail, and 
met some success in describing quantum criticality \protect\cite{sachdevbook}.

When it comes to the metal-insulator transition, the situation is
more complicated. Here, despite convincing experimental evidence for
criticality, the conventional approach cannot be applied directly.
The fundamental difficulty lies in the absence of an obvious symmetry
breaking scheme needed to build a Landau theory. For this reason,
a simplistic mean-field description of the MIT is not readily available,
and one is forced to look for alternative approaches.

A clue on how this may be possible is again found by analogy to conventional
critical phenomena, where one generally expects the fluctuations to
increase in importance in low dimensions. Their effects are particularly
strong near the lower critical dimension (LCD) $d_{lc}$, where they
are able to completely suppress the ordering. These phenomena allow
for a particularly elegant approach in systems with continuous broken
symmetry where $d_{lc}=2$. Here, an alternative perturbative RG treatment
based on an $\varepsilon=d-2$ expansion has been developed following
early ideas of Polyakov \cite{Polyakov75physlettb}.

The simplest example is the behavior of a Heisenberg magnet near $d=2$,
where one examines the low temperature spin-wave corrections to the
spin stiffness. Finite corrections are found for $d>2$, while logarithmic
singularities of the form $T\ln L$ ($T$ is the temperature and $L$
is the system size) arise precisely in $d=2$. This result indicates
the instability of the ordered phase due to infinitesimal thermal
fluctuations indicating that $d_{lc}=2$, and allows for a perturbative
RG treatment based on expanding around two dimensions \cite{pelcovits77prb}.

In the case of disorder-driven MITs, one should examine the effects
of weak disorder on the stability of the metallic phase. If similar
singular corrections arise in low dimensions, then one can not only
hope to identify the LCD, but should also be able to develop a perturbative
RG scheme. This elegant approach does not require developing an appropriate
symmetry breaking scheme, thus bypassing the essential stumbling block
in the theory for the MIT. However, it focuses on those physical processes
that describe how weak disorder modifies a clean Fermi liquid.
In those instances where the system is close to instabilities of the
clean system, this approach may be insufficient because new types
of low energy excitations may become important. Such a situation may
be found if the clean system is sufficiently close to Mott or Wigner
transitions, in which case strong correlation effects may require
a different theoretical framework and approach. Independent of these
issues, the physical content of theories describing perturbative disorder
effects within conventional Fermi liquids is sufficiently rich and
nontrivial, and in the following we discuss its basic ideas and results.

\subsection{Scaling theories of disorder-driven transitions}

\subsubsection{Phenomenological $\beta$-function}

An elegant and compact description of the scaling behavior around
a critical point is provided by the $\beta$-function formulation.
What we want to emphasize here is that the $\beta$-function description
is simply a alternative language one can use, rather then a microscopic
theory. However, its straightforward application to the MIT is based
on several implicit assumptions, that allow for a simplified phenomenology,
and which we discuss as follows.

In its original formulation as presented by the {}``gang of four''
\cite{gang4}, the $\beta$-function describes how the conductance
of the system changes with the (effective) system size. In principle,
one could imagine taking a finite size chunk of disordered metal,
and attaching it to contacts. If the experiment is repeated for different
sample sizes, then one could experimentally determine how the conductance
depends on the system size.

Why use the conductance and what do we expect to find? The essential
physics is easy to see by thinking about what happens far from the
MIT. In a good metal disorder is weak, and the conductivity $\sigma$
is large and finite even at $T=0$, and the standard Drude theory
applies. From Ohm's law, the conductance $g$ then scales with the
system size $L$ as\[
g_{met}(L)=\sigma L^{d-2}.\]
 For increasing system sizes (in $d>2$), the conductance grows as
$L^{d-2}$. In the opposite limit of very strong disorder, we expect
all the electrons to form bound (localized) states with impurities.
If $\xi$ \ is the characteristic (localization) length of these
bound states, then the conductance is expected to decrease exponentially\[
g_{ins}(L)\sim\exp\{-L/\xi\}.\]

More generally, we may expect that $g(L)$ do increase with $L$ in
the metal, and decrease in the insulator. What is not a priory obvious
is how $g(L)$ behaves around the transition. The seminal work on
the scaling theory for Anderson localization concentrated on the logarithmic
rate of change of the conducatnce with lengthscale, by defining the
{}``$\beta$-function''\begin{equation}
\beta(g)=\frac{d(\ln g)}{d\ln L}.\end{equation}
 This quantity is expected to be positive in a metal and negative
in the insulator. Its precise form, or a possible dependence on the
sample size $L,$ is not a priori clear. 

To make more specific predictions on the critical regime, the {}``gang
of four'' made two \textit{key assumptions}, as follows:
\begin{enumerate}
\item $\beta(g)$ is a function on $g$ only, but does not depend explicitly
on $L$.
\item $\beta(g)$ is a smooth (analytic) function near the transition. 
\end{enumerate}
In particular, in an Ohmic metal, we find \begin{equation}
\beta_{met}=d-2,\end{equation}
 while in the localized insulator \begin{equation}
\beta_{ins}=\ln g.\end{equation}
 Since $\beta_{met}>0$, and $\beta_{ins}<0$, assumptions (1) and
(2) then suggest that $\beta(g)$ has to change sign at some \textit{finite}
value of the conductance $g=g_{c}$, and we can write\begin{equation}
\beta(g)\approx s\ln(g/g_{c}),\end{equation}
 where $s=\beta^{\prime}(g_{c})$ is the critical slope of the $\beta$-function.
Defining th logarithmic variable $t=\ln(g/g_{c}),we$ can now integrate
the $\beta$-function equation\[
\frac{d(\ln t)}{d\ln L}\approx s,\]
 from the microscopic cutoff $\ell$ to the sample size $L$, and
write\[
t(L)=t_{o}(L/\ell)^{s}.\]
 This integration has to be carried out up to the lengthscale (i.e.
the correlation length) $L=\xi$ such that the renormalized distance
to the transition $\delta g(L)=(g(L)-g_{c})/g_{c}\approx t(L)\sim O(1)$.
Since at short scales $t_{o}=t(\ell)\sim\delta g_{o}\sim\delta n$,
we find\[
\xi\sim t_{o}^{1/s}\sim\delta n^{1/s}.\]
 Thus, the correlation length exponent \[
\nu=1/s.\]
 At this scale the conductivity is expected to saturate to a (size-independent)
macroscopic value\[
\sigma\approx g_{c}\xi^{d-2}\sim\delta n^{(d-2)\nu}.\]
 We conclude that the assumptions implied by the $\beta$-function
formultion, i.e. the scaling theory of localization predict the validity
of {}``Wegner scaling\textquotedblright\ for the conductivity exponent\[
\mu=(d-2)\nu.\]

It is worth emphasizing that postulating a particular form for the
$\beta$-function is completely equivalent to formulating a scaling
hypothesis for the critical behavior, \textit{provided that the conductance
is assumed to be finite at the transition}. Under these conditions,
the scaling hypothesis states that the $T=0$ conductance of a finite
size system takes a scaling form\[
g(\delta n,L)=\; f_{\sigma}\;(b^{1/\nu}\delta n,b/L).\]
 Choosing $b=\delta n^{-\nu}\sim\xi$, we can write $g(\delta n,L)=\psi(\xi/L)$,
where $\psi(x)=f_{\sigma}(1,x)$. The condition that the conductance
is finite at the transition is equivalent to require that $\psi(0)=g_{c}$
is a finite constant. Using the definition of the $\beta$-functon,
we then find\[
\beta(g)=x^{-1}\psi^{\prime}(x)/\psi(x),\]
 where $x=\psi^{-1}(g)$. The form of the $\beta$-functon can be
directly extracted from the experimental data, provided that an appropriate
scaling behavior is found.

\subsubsection{Is there a $\beta$-function for percolation?}

The $\beta$-function formulation of the {}``gang of four'' \protect\cite{gang4}
may be viewed as a convenient phenomenological description of how
the conductance depends on lengthscale. While its implicit assumptions
prove correct for the problem of Anderson localization of noninteracting
electrons, one may ask the same question for other models of the metal-insulator
transition. In particular, the percolation problem represents a consistent
description of the transition in the semiclassical limit. Since percolation
\protect\cite{stauffer-book} is a well characterized (classical)
critical phenomenon, powerlaw scaling of all quantities is still valid,
and many of the same questions raised by the {}``gang of four''\ can
be again posed.

Since $d=2$ remain above the lower critical dimension for percolation,
Wegner scaling cannot be valid in this case. How is thecaling behavior
modified in this case, and what would the $\beta$-function look like?
Can it even be defined? How is the finite-temperature scaling modified,
and what form does it take?

These questions are important, since the precise answer allows us
to distinguish experimental systems where percolation behavior dominates
from those where genuine quantum critical behavior is at play. In
the case of the percolation transition, we can give precise and rigorous
answers to all these questions, as we discuss in the following.

To discuss transport behavior near the percolation transition, consider
a resistor network corresponding to a random mixture of a metallic
and and insulating component. Let the conductivities of the respective
components be $\sigma_{M}(T)$ and $\sigma_{I}(T)$, both of which
remain finite at $T\neq0$, but with $\sigma_{M}(0)=\sigma_{o}\neq o$,
while $\sigma_{I}(0)=0$. As the relative fraction $x$ of the insulating
component increases past the percolation concentration $x_{c}$, the
overall conductivity of the network behaves as\begin{equation}
\sigma(x,T=0)\sim(x_{c}-x)^{\mu},\end{equation}
 but such a sharp critical behavior emerges only at $T=0$. At any
finite temperature the sharp percolation transition is smeared, similarly
as when a symmetry breaking field is turned on in conventional critical
phenomena. The family of conductivity curves generated by varying
the percolation concetration $x$ and temperature $T$ at first glance
looks very similar to those expected at any quantum localization transition,
where sharp critical behavior also emerges only at $T=0$.

How can we distinguish the two phenomena? The simples way to do so
is by focusing on the behavior at the the critical concetration, where
the finite temperature conductivity is determined \protect\cite{straley77prb}
by the conductivity of the insulating components \begin{equation}
\sigma(x_{c,}T)\sim(\sigma_{I}(T))^{u}.\end{equation}
 The corresponding critical exponent $u=1/2$ in $d=2$, and $u\approx0.7$
in $d=3$ \protect\cite{straley77prb}. Therefore, one may first
determine the location of the critical point by extrapolating the
conductivity to $T=0$ from the metallic side. One should then plot
the temperature dependence at the critical point. At any quantum critical
point, we generally expect a powelaw temperature dependence, $\sigma_{c}(T)\sim T^{(d-2)/x}$.
In contrast, the conductivity of any insulator typically takes a exponential
form\begin{equation}
\sigma_{I}(T)\sim\exp\{-(T_{o}/T)^{\alpha}\}\,\end{equation}
 where $\alpha=1$ for simple activated behavior, $\alpha=1/(d+1)$
for Mott variable-range hopping \cite{mott-book90}, or $\alpha=1/2$
for Efros-Shklovskii hopping \cite{doped-book}. Since the measured
conductance is expected to be a power of $\sigma_{I}$, we conclude
that in presence of percolation, the conductivity will assume an insulating-like
(exponential) temperature dependence on only in the insulating phase,
but even at the critical point. Thus, when the conductivity is plotted
as a function of temperature on a log-log scale, one will find a single
straight line (indicated powerlaw behavior at the critical concentration)
in the case of quantum criticality, but no such curve will be found
in case of percolation.

One can be even more precise, and specify the precise scaling behavior
around such a percolation transition, which can be used to perform
an alternative scaling analysis and collapse the experimental curves
in presence of percolation. In this case, we expect \protect\cite{stauffer-book}
that the conductivity should assume the following scaling form\begin{equation}
\sigma_{perc}(\delta x,h,L)=b^{-(d-2+\zeta)}f_{perc}(b^{1/\nu}\delta x,hb^{z^{\prime}},b/L).\end{equation}
 Here, $\delta x=(x_{c}-x)$ is the distance to the critical point,
$h=\sigma_{I}/\sigma_{M}$ plays the role of the symmetry breaking
field. Choosing $b^{1/\nu}\delta x=1$ and taking $L\rightarrow\infty$,
we can write \begin{equation}
\sigma_{perc}(\delta x,h)=\delta x^{\mu}\phi_{perc}(h/\delta x^{\nu z^{\prime}}).\end{equation}
 The conductivity exponent \begin{equation}
\mu=(d-2+\zeta)\nu,\end{equation}
 which replaces Wegner scaling of the quantum case. Thus, to collapse
all the experimental curves, the argument of the scaling function
should contain $h\sim\sigma_{I}(T)$, and not the temperature $T$.
Working at the critical point ($\delta x=0$), we similarly find the
exponential relation\begin{equation}
u=\frac{d-2+\zeta}{z^{\prime}}.\end{equation}

Finally, let us examine the finite size scaling behavior at $h=0$.
Working again at the critical point, we find\begin{equation}
\sigma_{c}^{perc}(L)\sim L^{-(d-2+\zeta)},\end{equation}
 so that the critical conductance \begin{equation}
g_{c}^{perc}(L)\sim L^{-\zeta},\end{equation}
revealing the physical meaning of the anomalous dimension $\zeta$,
which describes transport on the critical percolation cluster. As
we can see, the critical conductance vanishes in the case of percolation,
in contrast to the scaling theory of localization. We conclude that
although percolation does display conventional powerlaw finite size
scaling at the critical point, a $\beta$-function description is
not possible. This behavior reflects the fact that $d=2$ is not the
lower critical dimension for percolation. The anomalous dimension
$\zeta$ is very generally expected to vanish at ordinary quantum
critical points describing conductor-insulator transitions. This result
can be shown \cite{wen92prb} to very generally follow from charge
conservation for any quantum criticality displaying simple single-parameter
scaling behavior. Although more complicated quantum scenarios are
in principle possible, no microscopic quantum model has been identified
to date showing $\zeta\neq0$, or equivalently a lower critical dimension
$d_{lc}<2$.

\subsubsection{Perturbative quantum corrections in disordered metals}

What is the physical mechanism that invalidate's Mott's bound on impurity
scattering? To understand this, recall that Mott used Drude's picture,
where scattering processes from each impurity or defect are assumed
to independent and uncorrelated. This assumption is indeed justified
at weak enough disorder, where the Drude prediction is recovered
as a leading order contribution. For stronger disorder, multiple-scattering
processes cannot be ignored, and they provide the so-called {}``quantum
corrections'' to Drude theory. In good metals the magnitude of the
quantum corrections is generally small, modifying the conductivity
by typically only a fraction of a percent. In this regime, the quantum
corrections can be systematically obtained as next-to-leading corrections
within weak-disorder perturbation theory for impurity scattering.

Detailed calculations and classification of all such perturbative
quantum corrections has been carried out in the late 1970s and early 1980s, and
are by now well understood \protect\cite{leeramakrishnan}.
They consist of several additive terms, \[
\sigma=\sigma_{o}+\delta\sigma_{wl}+\delta\sigma_{int},\]
 corresponding to the so-called {}``weak localization''\ and {}``interaction''
corrections. These {}``hydrodynamic'' corrections are dominated
by infrared singularities, i.e., they acquire non-analytic contributions
from small momenta or equivalently large distances. Specifically,
the weak localization corrections take the form\begin{equation}
\delta\sigma_{wl}=\frac{e^{2}}{\pi^{d}}\left[l^{-(d-2)}-L_{Th}^{-(d-2)}\right],\label{eq:weakloccond}\end{equation}
 where $l=v_{F}\tau$ is the mean free path, $d$ is the dimension
of the system, and $L_{Th}$ is the length scale over which the wave
functions are coherent. This effective system size is generally assumed
to be a function of temperature of the form $L_{Th}\sim T^{p/2}$,
where the exponent $p$ depends on the dominant source of decoherence
through inelastic scattering.

The situation is simpler in the presence of a weak magnetic field
or magnetic impurities. Here the weak localization corrections are
suppressed and the leading dependence comes from the interaction corrections
first discovered by Altshuler and Aronov \protect\cite{altshuler-79b}\begin{equation}
\delta\sigma_{int}=\frac{e^{2}}{\hslash}(c_{1}-c_{2}\widetilde{F}_{\sigma})(T\tau)^{(d-2)/2}.\label{eq:intercorrcond}\end{equation}
 Here, $c_{1}$ and $c_{2}$ are constants, and $\widetilde{F}_{\sigma}$
is an interaction amplitude.

In $d=3$, this leads to a square-root singularity $\delta\sigma_{int}\sim\sqrt{T}$,
and to a more singular logarithmic divergence $\delta\sigma_{int}\sim\ln(T\tau)$
in $d=2$. These corrections are generally more singular than the
temperature dependence of the Drude term, and thus they are easily
identified experimentally at the lowest temperatures. Indeed, the
$T^{1/2}$ law is commonly observed \protect\cite{leeramakrishnan}
in transport experiments in many disordered metals at the lowest temperatures,
typically below 500 mK.

Similar corrections have been predicted for other physical quantities,
such as the tunneling density of states and, more importantly, for
thermodynamic response functions. As in Drude theory, these quantities
are not expected to be appreciably affected by noninteracting localization
processes, but singular contributions are predicted from interaction
corrections. In particular, corrections to both the spin susceptibility
$\chi$, and the specific heat coefficient $\gamma=C_{V}/T$ were
expected to take the general forms \[
\delta\chi\sim\delta\gamma\sim T^{(d-2)/2},\]
 again leading to logarithmic corrections in $d=2$.

As in conventional Fermi liquid theories, these corrections emerged
already when the interactions were treated at the lowest, Hartree-Fock
level, as done in the approach of Altshuler and Aronov \protect\cite{altshuler-79b}.
Higher order corrections in the interaction amplitude were first incorporated
by Finkelshtein \protect\cite{fink-jetp83,fink-jetp84}, demonstrating
that the predictions remained essentially unaltered, at least within
the regime of weak disorder. In this sense, Fermi liquid theory has
been generalized to weakly disordered metals, where its predictions
have been confirmed in numerous materials \protect\cite{leeramakrishnan}.

\subsubsection{Anderson transition in $2+\varepsilon$ dimensions}

The essential idea of these approaches focuses on the fact that a
weak, logarithmic instability of the clean Fermi liquid arises in
two dimensions, suggesting that $d=2$ corresponds to the lower critical
dimension of the problem. In conventional critical phenomena, such
logarithmic corrections at the lower critical dimension typically
emerge due to long wavelength fluctuations associated with spontaneously
broken continuous symmetry. Indeed, early work of Wegner emphasized
\protect\cite{wegner79} the analogy between the localization transition
and the critical behavior of Heisenberg magnets. It mapped the problem
onto a field theoretical nonlinear $\sigma$-model and identified
the hydrodynamic modes leading to singular corrections in $d=2$.
Since the ordered (metallic) phase is only marginally unstable in
two dimensions, the critical behavior in $d>2$ can be investigated
by expanding around two dimensions. Technically, this is facilitated
by the fact that in dimension $d=2+\varepsilon$ the critical value
of disorder $W$ for the metal-insulator transition is very small
($W_{c}\sim\varepsilon$), and thus can be accessed using perturbative
renormalization group (RG) approaches in direct analogy to the procedures
developed for Heisenberg magnets. In this approach \protect\cite{wegner80,gang4},
conductance is identified as the fundamental scaling variable associated
with the critical point, which is an unstable fixed point of the RG
flows. This RG calculation provides a scaling description predicting
how the conductance depends on the system size, and thus produces
the desired $\beta$-function in $d=2+\varepsilon$ dimensions. To
leading order in $\varepsilon$ the resulting critical exponent $\nu$
is predicted \protect\cite{wegner80,gang4} to be \begin{equation}
\nu^{-1}=\varepsilon+O(\varepsilon^{2}).\end{equation}
 When this result is extrapolated to three dimensions ($\varepsilon=1$),
the conductivity exponent \begin{equation}
\mu=1+O(\varepsilon).\end{equation}

This early prediction was initially widely acclaimed as a plausible
theoretical explanation for the critical behavoior commonly observed
in several systems (e.g. compensated doped semiconductors, see below),
where $\mu\approx1$. From the theoretical side, this result was believed
to be exact from more then ten years, since very tedious subsequent
work established that higher order corrections to $\nu$, when evaluated
to second and even third order in $\varepsilon$ all vanished! Very
surprisingly, a tour-de-force calculation by Hikami (to $O(\varepsilon^{5})$)
succeeded in finding nonvanishing correction to this exponent, giving
an estimate \begin{equation}
\mu\approx0.67\end{equation}
in three dimensions. This calculation demonstrated that the exact
value ($\mu\approx1.58$ for the ``orthogonal ensemble''
describing ordinary potential scattering) lies hopelessly far from
the estimates based on the $\varepsilon$-expansion.

This imporant result indicates a potentially serious shortcoming of
any such perturbative RG schemes. As in ordrinary (magnetic) critical
phenomena, the $\varepsilon$-expansion around $d=2$ seems to have
extremely bad convergence properties \cite{PhysRevLett.71.384} when
extrapolated to $d=3$, making it a virtually useless theoretical
tool for making quantitative estimates for the critical exponents.
The reason for generally poor convengence of $d=2+\varepsilon$ expansions
is at present not fully understood, although related work \cite{PhysRevLett.71.1911}
suggested that it reflects an inability to incorporate topological
excitations ({}``hedgehog'' configurations for the $d=3$ Heisenberg
model they studied) within any perturbative scheme. This behavior
should be contrasted to that familar from the $4-\varepsilon$ RG
approaches in standard critical phenomena. Here, a consistent description
of the critical point is found even at the mean-field level, which
becomes exact for $d>d_{uc}=4$. The fluctuation effects described
by RG flows only provide the corrections to mean-field values for
the critical exponents, and in general proved to have much superior
convenrgence properties, often allowing surprisingly accurate estimates
when extrapolated to $d=3$. Interestingly, the complete lack of spatial
correlations within mean-field theory allows for \emph{all spin configurations}
to be considered on equal footing within the $4-\varepsilon$ scheme
, including both the smooth configurations described by perturbative
methods, \emph{and} the topological defects which they ignore. 

Returning to metal-insulator transitions, one may speculate that the
difficulties with peturbative approaches reflect that here the very
existence of the phase transition emerges only due to fluctuation
corrections; no transition is found at the mean-field (saddle-point)
level describing Drude-Boltzmann theory. Developing a more appriate
mean-field description of the MIT thus appears as one of the most
promising directions for future work.

\subsubsection{Generalized Fermi liquid theory for disordered electrons}

In the mid 1980s, these ideas were extended with a great deal
of effort in the formulation of a scaling theory of interacting disordered
electrons by Finkelshtein \protect\cite{fink-jetp83} and many followers
\protect\cite{cclm-prb84,kirkpatrick-rmp94}. While initially shrouded
by a veil of quantum field theory, these theories were later
given a simple physical interpretation in terms of Fermi liquid ideas
\protect\cite{ckl,kotliar-fl87} for disordered electrons. Technical
details of these theories are of considerable complexity, and the
interested reader is referred to the original literature \protect\cite{fink-jetp83,fink-jetp84,cclm-prb84,kirkpatrick-rmp94}.
Here we just summarize the principal results, in order to clarify
the constraints imposed by these Fermi liquid approaches.

Within the Fermi liquid theory for disordered systems \protect\cite{ckl,kotliar-fl87},
the low energy (low temperature) behavior of the system is characterized
by a small number of effective parameters, which include the diffusion
constant $D$, the frequency renormalization factor $Z$, and the
singlet and triplet interaction amplitudes $\gamma_{s}$ and $\gamma_{t}$.
These quantities can also be related to the corresponding quasi-particle
parameters which include the quasi-particle density of states \begin{equation}
\rho_{Q}=Z\rho_{o},\label{eq:quasipartdos}\end{equation}
 and the quasi-particle diffusion constant \begin{equation}
D_{Q}=D/Z\sim D/\rho_{Q}.\label{eq:quasipartdiff}\end{equation}
 Here, $\rho_{o}$ is the {}``bare'' density of states which describes
the noninteracting electrons. In the absence of interactions, the
single-particle density of states is only weakly modified by disorder
and remains noncritical (finite) at the transition \protect\cite{wegner80}.

Using these parameters, we can now express the thermodynamic response
functions as follows. We can write the compressibility \begin{equation}
\chi_{c}=\frac{dn}{d\mu}=\rho_{Q}[1-2\gamma_{s}],\label{eq:compress}\end{equation}
 the spin susceptibility \begin{equation}
\chi_{s}=\mu_{B}^{2}\rho_{Q}[1-2\gamma_{t}],\label{eq:susceptibility}\end{equation}
 and the specific heat \begin{equation}
C_{V}=2\pi^{2}\rho_{Q}T/3.\label{eq:specificheat}\end{equation}
 In addition, we can use the same parameters to express transport
properties such as the conductivity \begin{equation}
\sigma=\frac{dn}{d\mu}D_{c}=\rho_{Q}D_{Q},\label{eq:conductivity}\end{equation}
 as well as the density-density and spin-spin correlation functions
\begin{equation}
\pi(q,\omega)=\frac{dn}{d\mu}\frac{D_{c}q^{2}}{D_{c}q^{2}-i\omega}\;\;\;\;\;\;\;\;\chi_{s}(q,\omega)=\chi_{s}\frac{D_{s}q^{2}}{D_{s}q^{2}-i\omega}.\label{eq:correlations}\end{equation}
 Here, we have expressed these properties in terms of the spin and
charge diffusion constants, which are defined as \begin{equation}
D_{c}=\frac{D}{Z(1-2\gamma_{s})};\;\;\;\; D_{s}=\frac{D}{Z(1-2\gamma_{t})}.\label{eq:diffusionconstants}\end{equation}

Note that the quantity $D$ is \emph{not} the charge diffusion constant
$D_{c}$ that enters the Einstein relation, Eq. (\ref{eq:conductivity}).
As we can write $\sigma=\rho_{o}D$, and since $\rho_{o}$ is not
critical at any type of transition, the quantity $D$ (also called
the {}``renormalized diffusion constant'') has a critical behavior
identical to that of the conductivity $\sigma$. We also mention that
the quasi-particle diffusion constant $D_{Q}=D/Z$ has been physically
interpreted as the heat diffusion constant.

\subsubsection{Weak-coupling renormalization group approach of Finkelshtein}

The Fermi liquid relations provide constraints relating different
Landau parameters, but they cannot predict their specific values,
or how should they behave in the vicinity of the metal-insulator transition.
In practical applications of these approaches to disordered system
one assumes that the Landau parameters characterizing the clean Fermi
liquid are known (say from experiments), and then examines how they
are modified when impurities are added to the system. Precise prediction
for these impurity effects have been obtained in the limit of weak
disorder, where perturbative {}``quantum corrections'' to all quantities
have been calculated \cite{leeramakrishnan}. As in the noninteracting
limit, singular logarithmic corrections are found in $d=2$, signaling
that a systematic weak coupling approach can be developed, to investigate
the instability of the metallic phase under impurity scattering. While
conceptually simple and transparent, practical calculations to implement
this program proved of considerable complexity, even within the framework
of weak coupling approaches.

Perturbative renormalization group calculations \protect\cite{fink-jetp83,fink-jetp84,cclm-prb84,kirkpatrick-rmp94}
based on the $2+\varepsilon$ expansion have been used to make explicit
predictions for the values of the critical exponents for different
universality classes. The details of this formulation will not be
elaborated here, as it has been discussed in great detail in several
excellent reviews. Instead, we comment on the physical content of
these theories, and indicate what aspects of the problem may and which
may not be addressed using this framework.

\subsubsection{Physical content of Finkelstein's theory}

The mathematical complexity of of Finkelstein's formalism seems, at
first sight, to shroud its physical content with a veil of mystery,
and render it difficult to comprehent to all byt the very few specialists
working in the field. Later works \cite{cclm-prb84,ckl,Zala}, however,
succeeded to rederive most of Finkelstein's results using standard
(albeit much less elegant) diagrammatic approaches, allowing for a
simple and transparent physical interpretation. The main points include:
\begin{itemize}
\item This theory describes a disordered Fermi liquid, specifically the scale-dependence
of the disorder-modified Landau parameters. 
\item As in the original scaling theory of localization, the transition
is identified as an unstable fixed point for the conductance. If the
interaction amplitudes retain finite (albeit renormalized) values
at the fixed point, then the scaling scenario is essentially of the
same form as for noninteracting localization. This behavior is found
\cite{kirkpatrick-rmp94} for all the universality classes with broken
time reversal invariance (BTRI): in presence of an external magnetic
field, magnetic impurity scattering, or presence of spin-orbit scattering. 
\item The metallic phase, as in any Fermi liquid, retains the same qualitative
behavior at $T=0$ as a weakly disordered systems of noninetracting
electrons. In particular, this means that all thermodynamic quantities
(e.g. the spin susceptibility $\chi$ or the Sommerfeld coefficient
$\gamma=C/T$) remain finite away from the transition. 
\item The scaling hypothesis, which is built into this formalism, guarantees
that the low temperature behavior of all quantities is qualitatively
identical within the each (metallic or insulating) phase. For example,
the leading low temperature $\sqrt{T}$ corrections, first predicted
by using weak-disorder perturbation theory of Altshuler and Aronov,
should retain their form throughout the metallic phase. 
\end{itemize}
These results may lead - at least in principle - to a mathematically
consistent description of the metal-insulator transition in $d=3$.
In two dimensions, where Anderson localization is expected to distroy
any metallic phase, the situation is more subtle. An especially interesting
and still controversial situation is found for the {}``generic''
model, i.e. in absence of BTRI perturbations. Here, a qualitatively
new behavior is found due to interaction effects. The triplet interaction
amplitude $\gamma_{t}$ is found to grow withour limits under rescaling
(i.e. as $T$ is lowered), even at infinitesimal disorder \cite{fink-jetp83,fink-jetp84}.
As a result, the {}``interaction'' correction for the conductivity
(which to leading order is proportional to $\gamma_{t}$) is found
to produce an {}``antilocalization'' effect, which is predicted
to {}``overcome'' the usual weak localization term responisble for
the instability of the $d=2$ metallic phase. From this point of view,
Finkelstein's results open a consistent possibility that interactions
may be able to stabilize the metallic phase at sufficiently weak disorder,
thus leading to a sharp metal-insulator transition in $d=2$. 

Most remarkaby, recent experiments on a variety two-dimensional electron
gases have revealed behavior quite suggestive of the existence of
precisely such a transition \cite{abrahams-rmp01}. This contriversial
phenomenon, first discovered by Kravchenko in 1995 \cite{re:Kravchenko95},
has reignited interest in this fundamental question. Its many facets
continue to facinate theorists and experimentalists alike, but much
still remains to be understood, as described in more detail in Chap.
2. Even from the purel theoretical point of view, however,
the physical interpretation of what precisely happens in a $2D$ weakly
disordered Fermi liquid still remains puzzling and controversial.
The main issue is how to understand the singular behavior of the triplet
interaction amplitude $\gamma_{t}$, which is found to grow without
bounds at low temperature. In disordered Fermi liquid it indicates
enhanced spin susceptibility $\chi\sim\gamma_{_{t}}$, so this result
may signal some kind of magnetic instability of the metallic phase.
Various interpretations have been proposed ranging from disorder-induced
ferromagnetism \cite{kirkpatrick-rmp94}, to disorder-induced local
moment formation \cite{cclm-prb84,milovanovicetal89,paalanenetal88,vladtedgabi}.
While some progress has recently been made by Punnoose and Finkelstein
(see Chapter 4), using large-N methods , the
issue remains controversial and chalenging.

\subsubsection{Difficulties with runaway RG flows}

From a more general perspective, any solution with such {}``runaway''
coupling constant immediately brings into question the credibility
of perturbative RG approaches. It is interesting to recall that similar
situations have been encoutered in several other problems, where disorder
effects in interacting systems have proved difficult to describe using
perturbative RG methods. The first such example is the famed random-field
Ising model (RFIM) \cite{re:Imry75}, where a perturbative arguments
of Parisi and Sourlas (PS) suggested \cite{re:Parisi79} that the
critical behavior of the disorder problem can be mapped to the corresponding
clean problem in $d-2$ dimensions. Because the clean Ising model
does not order in $d=1,$ this work suggested that infinitesimal disorder
would be sufficient to destroy the ordered phase even in $d=3.$ Later
rigorous by Imbrie \cite{re:Imbrie84} rigorously showed that a finite
temperature ordering survives sufficiently weak disorder in $d=3$,
thus invalidating the PS mechanism, which was claimed to hold to all
orders in perturbation theory. The puzzle was soon resolved by more
careful functional renormalization group methods of Daniel Fisher
\cite{d.fisher85prb}, who showed that the perturbative fixed point
of PS proves unstable, leading to {}``runaway'' RG flows. Subsequent
work \cite{re:Mezard92,re:Mezard94,re:Dominicis96} found evidence
that this reflects the proliferation of metastable states in the RFIM,
a phenomenon which proves {}``invisible'' in perturbative treatment.
Interestingly, much later work  \cite{re:Kirkpatrick94,re:Belitz95}
attempted a non-pertubative (albeit uncontrolled) solution of Finkelstein's
nonlinear $\sigma$-model by proposing a new saddle point solution,
and a RG approach based on a $6-\varepsilon$ expansion, finding RG
flows of a structure very similar to the PS theory for the RFIM. The
authors interpreted these results as evidence that metastability and
non-perturbative glassy effects cannot be ignored within the Finkelstein's
model any more then they could for the RFIM. 

\begin{figure}[h]
\begin{center}
\includegraphics[width=0.8\columnwidth]{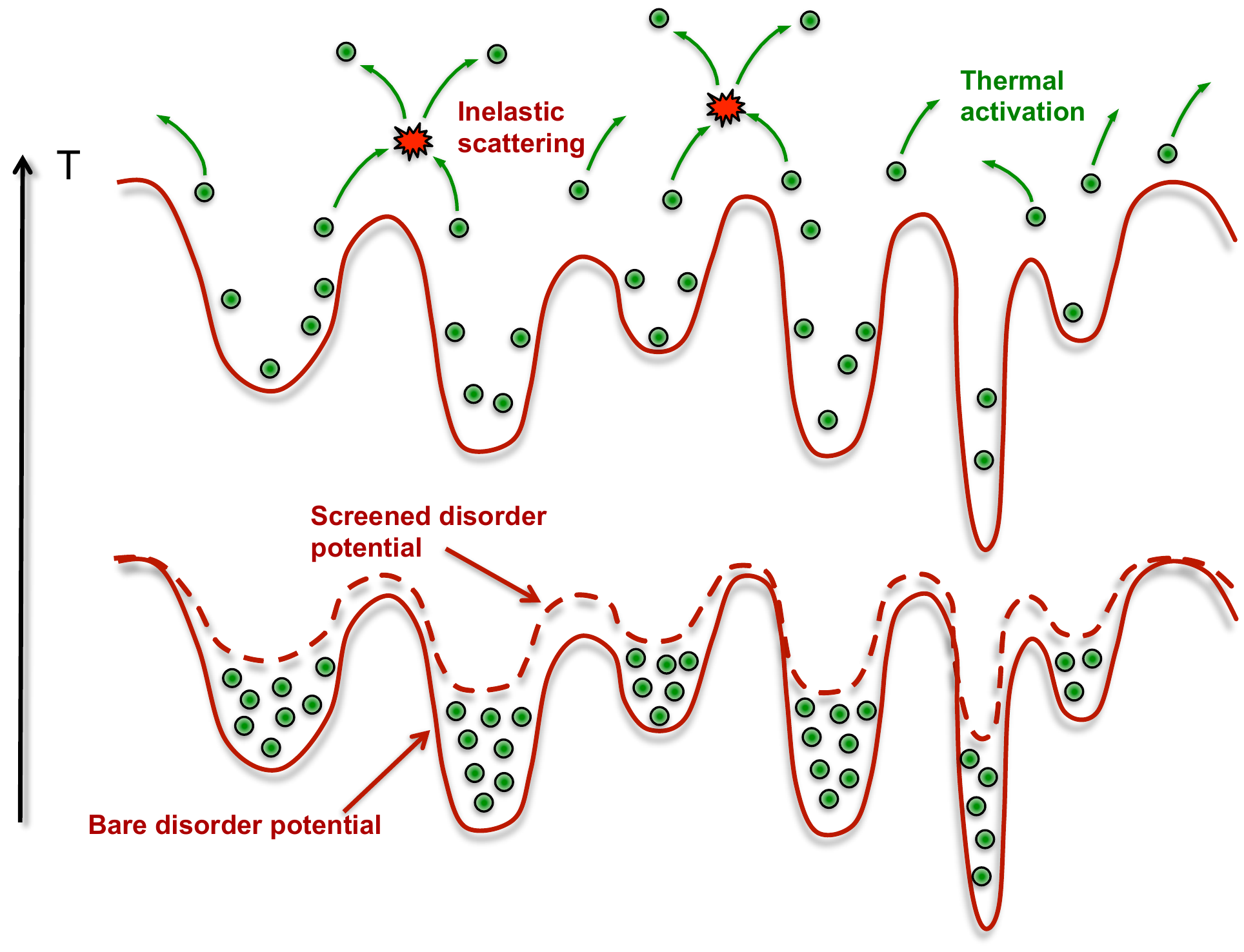}
\end{center}

\caption{In the disordered Fermi liquid picture, the leading low-temperaure
dependence of transport reflect elastic scattering off a renormalized,
but temperature-dependent random potential (dashed line). At low temperatures
(bottom), the potential wells {}``fill-up'' with electrons; in presence
of repulsive (Coulomb) interactions, the screened (renormalized) potential
has reduced amplitude (dashed line), leading to effectively weaker
disorder. As the temperature increases (top), electrons thermally
activate (shown by arrows) out of the potential wells, reducing the
screening effect. This physical mechanism, which operates both
in the ballistic and in the diffusive regime  \protect\cite{Zala}, is at the origin of
all {}``quantum corrections'' found within the Fermi liquid picture.
It is dominant, provided that inelastic electron-electron scattering
can be ignored. While this approximation is well justified in good
metals, inelastic scattering (star symbol) is considerably enhanced
in presence of strong correlation effects, often leading to disorder-driven
non-Fermi liquid behavior \protect\cite{RoP2005review} and electronic
Griffithe phases \protect\cite{mirandavlad1,tanaskovic-2005-95,andrade09prl}.}

\label{unscreening}
\end{figure}

Yet another example of non-perturbative disorder effects has been
found in recent studies of Hertz-Millis \cite{hertz,millis,sachdevbook}
models for itinerant quantum criticality. Here early works attempted
a perturbative RG treatment for weak disorder, only to find
runaway flows. Later works by Vojta and Hoyos \cite{hoyos07prl},
which used a complementary {}``strong-disorder renormalization group''
(SDRG) methods, deminstrated that for these models disorder completely
changes the physical nature of quantum criticality, leading to so-called
{}``infinite-randomness fixed point'' (IRFP) behavior \cite{RoP2005review,vojta-review06}.
In such cases, the behavior of the system is deminated by rare disorder
realizations, leading to the formation of Quantum Griffiths Phases
(QGP) surrounding the critical point. 

We should stress that these non-perturbative disorder effects do not
bring into question the general ideas of scaling, but they they do
seem to raise concerns about the application of weak coupling RG methods.
In cases such as the IRFP universality class, the renormalized effective
action describing the critical point assumes a qualitatively different
form then that of the clean system. This simple fact makes it abundantly
clear how and why perturbative methods may fail in a most dramatic
fashion. Whether a similar mechanism resolves the puzzles identified
by the perturbative solution of the Finkelstein model remains to be
established by future work.

\subsubsection{Conceptual limitations of the disordered Fermi liquid picture}

Despite a great deal of effort invested in such calculations, the
predictions of the perturbative RG approaches have met only limited
success in explaining the experimental data in the critical region
of the metal-insulator transition. We should emphasize, though, that
limitations associated with these weak-coupling theories do not invalidate
the potential applicability of Fermi liquid ideas \textit{per se}.
Having this in mind, it is also important to understand the fundamental
limitations of the disordered Fermi liquid picture which is the underpinning
of Finkelstein's field theory. The important physical conditions implicitly
assumed by this formulation include:
\begin{itemize}
\item It describes leading low-temperature excitations, which are adiabatically
connected to a noninteracting (but disordered) electronic system.
These excitations, therefore, assume fermionic character. Other (spin
or charge) collective excitations are assumed to play a
subleading role in this temperature regime. 

\item The fermionic excitations are assumed to be sufficiently dilute, such
that inelastic electron-electron scattering processes can be neglected.
The leading low temperature corrections then reflect elastic scattering
processes on a static, but temperature-dependent renormalized disorder
potential. Upon disorder averaging, this can produce different temperature
dependence in the diffusive and ballistic regimes. As explained in
recently by Zala \emph{et al.} \cite{Zala}, however, the basic physical
mechanism for all these corrections is one and the same (Fig. \ref{unscreening}).
We should emphasize that these processes dominate only at sufficiently
low temperatures, typically being only a small fraction of the Fermi
temperature. In presence of strong electronic correlations, the {}``coherence
temperature'' $T^{*}$ below which the Fermi liquid picture applies,
may be very low; a good example are so-called {}``heavy-fermion''
systems, where $T^{*}\ll T_{F}$ and a very broad incoherent metallic
regime is found. In presence of disorder, the coherence temperature
$T^{*}$ may be further reduced or even driven to zero \cite{RoP2005review}.
This situation is illustrated in certain exactly solvable models with
strong correlation and disorder \cite{dk-prl93,tanaskovicetal04},
which can be exatly solved in the limit of large coordination. Here
strong disorder fluctuations stabilize {}``non-Fermi liquid'' metallic
behavior at any non-zero temperature. 

\item The standard formulation of the disordered Fermi liquid theory is
derived at weak disorder, where the electrons are well delocalized
accross the entire sample. The Landau interaction amplitude are assumed
to be self-averaging, and are thus replaced by their averaged values.
At stronger disorder, very strong spatial flauctuations may emerge
\cite{RoP2005review}, so that the interaction effects may be much
more pronounced in certain regions of the sample. These effects are
most important in strongly correlated electronic systems, where the
assumption of self-averaging may completely break down, leading to
the formation of {}``electronic Griffiths phases'' \cite{mirandavlad1,tanaskovic-2005-95,andrade09prl}.
If this happens, then the simiplified version of disordered Fermi
liquid theory may prove insufficient or even misleading. 

\item The Fermi liquid pictures assumes a unique ground state, implicitly
ignoring the possibility metastable states resulting from the competition
of disorder and the long-range Coulomb interactions. The associated
electron glass behavior \cite{pastor-prl99} is a well establised
feature of disordered insulators, but its possible role in the critical
regime \cite{mitglass-prl03} has been ignored by the weak coupling
approaches. 
\end{itemize}

\subsection{Order-parameter approaches to interaction-localization}

\subsubsection{Need for an order-parameter theory: experimental clues}

In conventional critical phenomena, simple mean-field approaches such
as the Bragg-Williams theory of magnetism, or the Van der Waals theory
for liquids and gases work remarkably well - everywhere except in
a very narrow critical region  \cite{goldenfeldbook}. Here, effects of long wavelength fluctuations
emerge that modify the critical behavior, and its description requires
more sophisticated theoretical tools, based on renormalization group
(RG) methods. A basic question then emerges when looking at experiments:
is a given phenomenon a manifestation of some underlying mean-field
(local) physics, or is it dominated by long-distance correlations,
thus requiring an RG description? The answer for conventional criticality
is well know, but how about metal-insulator transitions? Here the
experimental evidence is much more limited, but we would like to emphasize
a few well-documented examples which stand out. 

\begin{itemize}
\item \emph{Doped semiconductors such as Si:P.} These are the most carefully
studied \cite{doped-book} examples of the MIT critical behavior.
Here the density-dependent conductivity extrapolated to $T=0$ shows
sharp critical behavior \cite{paalanen91} of the form $\sigma\sim(n-n_{c})^{\mu}$,
where the critical exponent $\mu\approx1/2$ for uncompensated samples
(half-filled impurity band), while dramatically different $\mu\approx1$
is found for heavily compensated samples of Si:P,B, or in presence
of strong magnetic fields. Most remarkably, the dramatically differences
between these cases is seen over an extremely broad concentration
range, roughly up to several times the critical density. Such robust
behavior, together with simple (apparent values for the critical exponents,
seems reminiscent of standard mean-field behavior in ordinary criticality.

\begin{figure}[h]
\begin{center}
\includegraphics[width=0.65\textwidth]{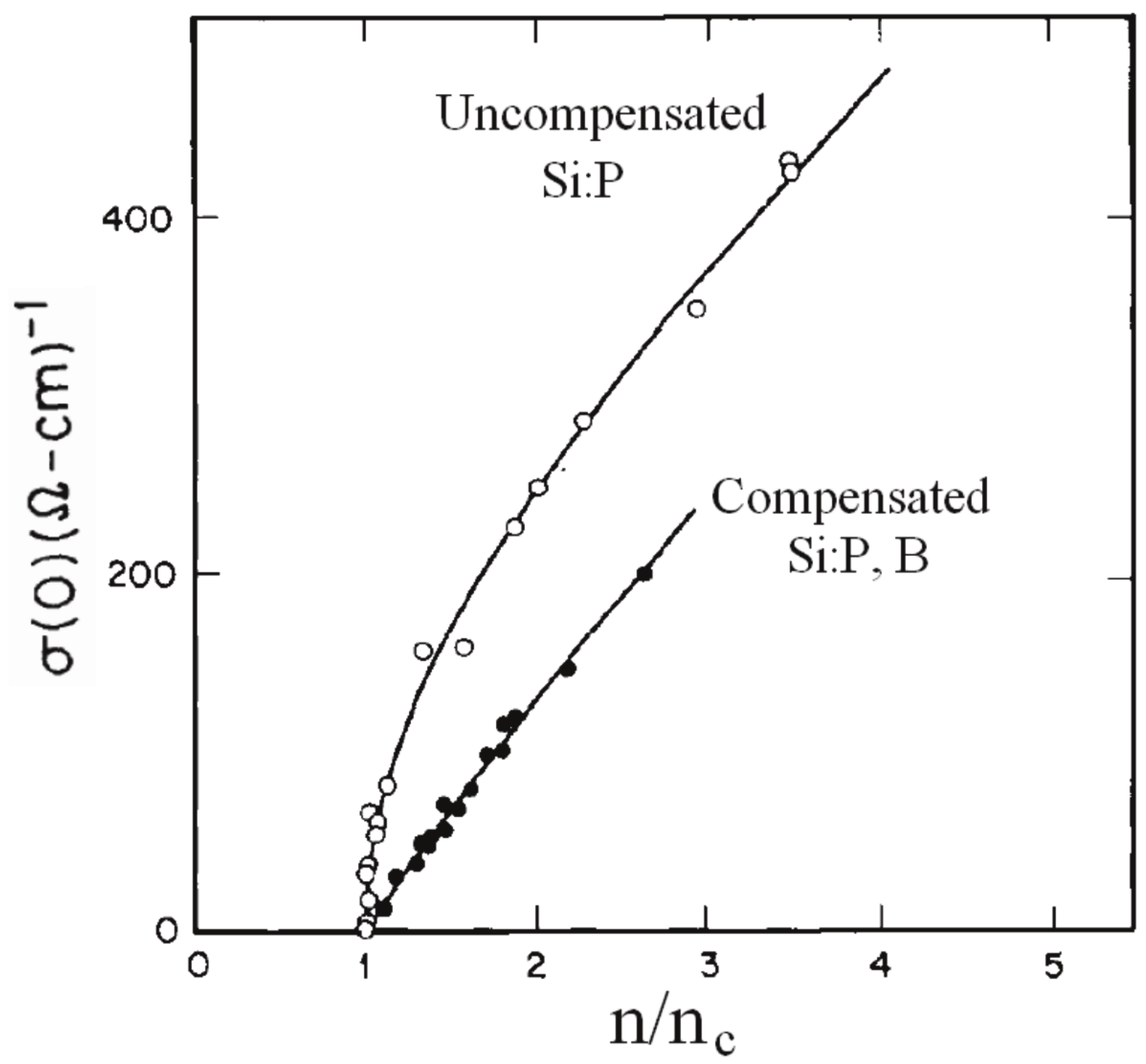}
\end{center}
\caption{Critical behavior of the conductivity for uncompensated $Si:P$ and compensated
Si:P,B \protect\cite{paalanen91}. The conductivity exponent $\mu\approx1/2$
in absence of compensation, while $\mu\approx1$ in its presence.
Clearly distinct behavior is observed in a surprisingly broad range
of densities, suggesting mean-field scaling. Since compensation essentially
corresponds to carrier doping away from a half-filled impurity band
\protect\cite{doped-book}, it has been suggested \protect\cite{leeramakrishnan}
that the difference between the two cases may reflect the role of
strong correlations.}
\end{figure}

\item \emph{Two-dimensional metal-insulator transitions.} Signatures of a remarkably sharp metal-insulator transition
has also been observed \cite{re:Kravchenko95,re:Popovic97,abrahams-rmp01}
in several examples of two-dimensional electron gases (2DEG) such
as silicon MOSFETs. While some controversy regarding the nature or
even the driving force for this transition remains a subject of intense
debate, several experimental features seem robust properties common
to most studied samples and materials. In particular, various experimental
groups have demonstrated \cite{re:Kravchenko95,re:Popovic97} striking
scaling of the resistivity curves in the critical region, which seems
to display \cite{simonian97prb} remarkable mirror symmetry ({}``duality'')
\cite{gang4me} over a surprisingly broad interval of parameters.
In addition, the characteristic behavior extends to remarkably high
temperatures, which are typically \emph{comparable the Fermi temperature}
\cite{abrahams-rmp01}. One generally does not expect a Fermi liquid
picture of diluted quasiparticles to apply at such {}``high energies'',
or any correlation length associated with quantum criticality to remain
long.
\end{itemize}

\begin{figure}[h]
\vspace{12pt}
\begin{center}
\includegraphics[width=2.5in]{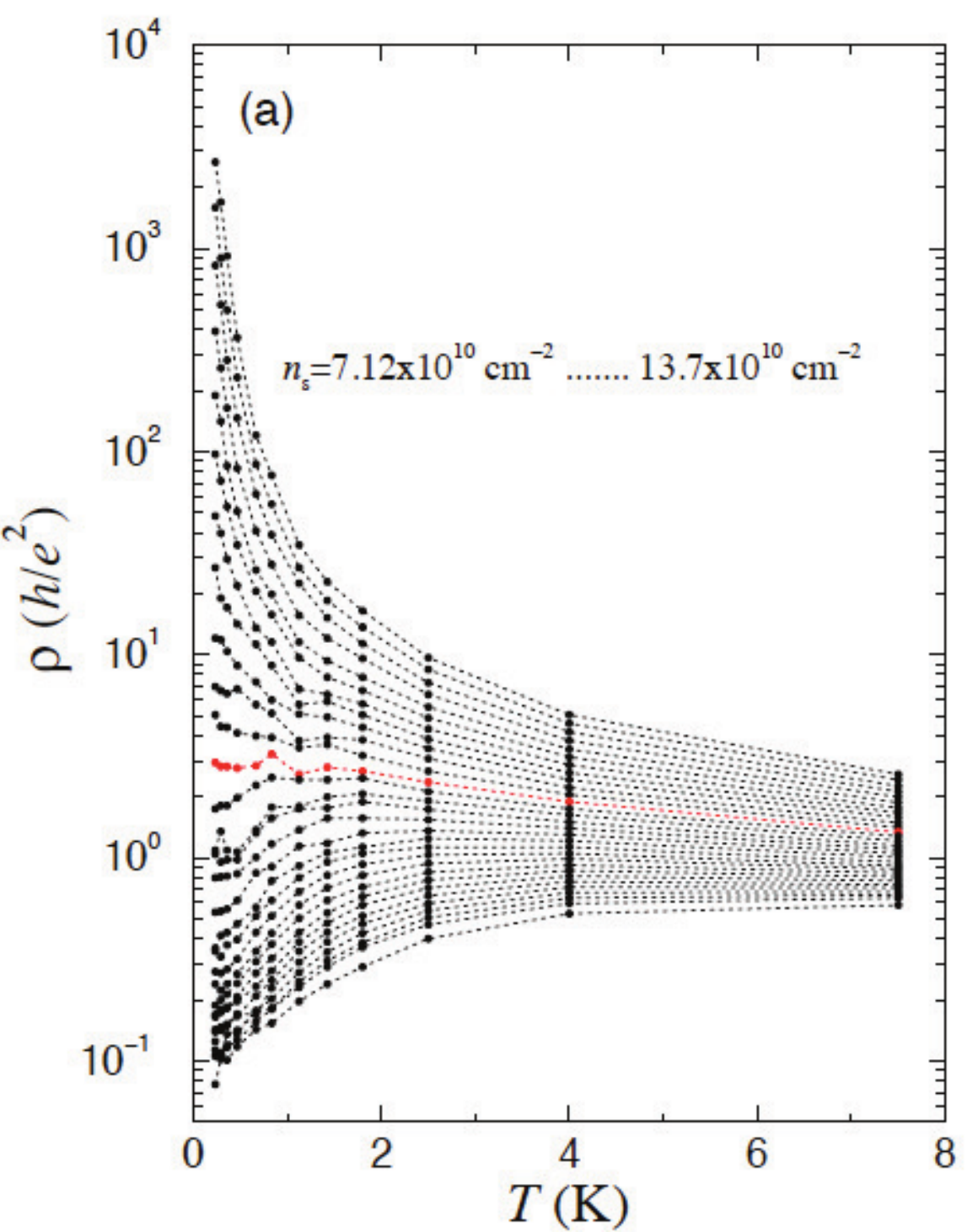}\includegraphics[width=2.7in]{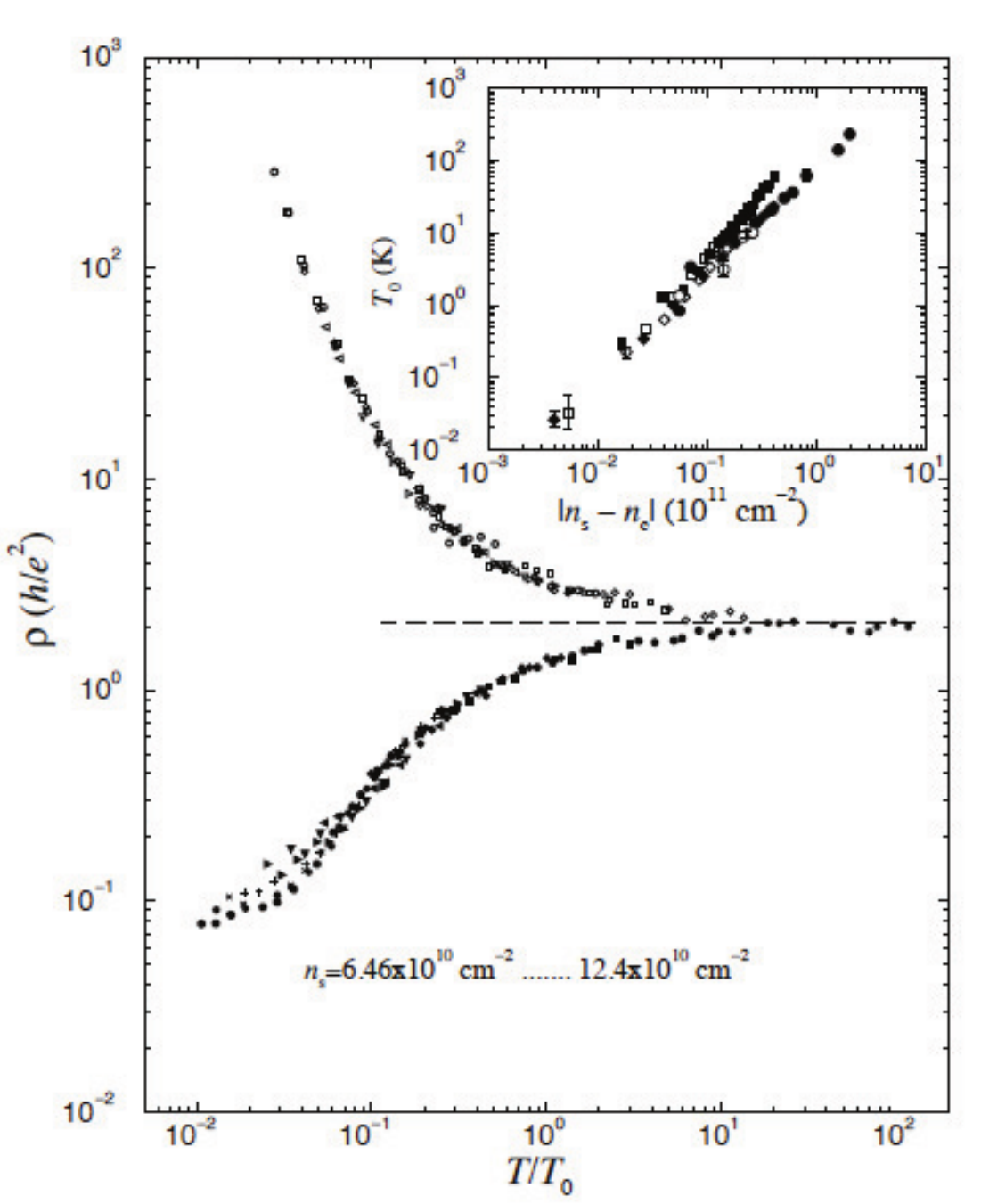}
\end{center}\vspace{24pt}
\caption{The resistivity curves (left panel) for a two-dimensional electron
system in silicon \protect\cite{re:Kravchenko95} show a dramatic
metal-insulator crossover as the density is reduced below $n_{c}\sim10^{11}cm^{-2}$.
Note that the system has {}``made up its mind'' whether to be a
metal or an insulator even at surprisingly high temperatures $T\sim T_{F}\approx10K$.
The right panel displays the scaling behavior which seems to hold
over a comparable temperature range. The remarkable {}``mirror symmetry''
\protect\cite{simonian97prb} of the scaling curves seems to hold
over more then an order of magnitude for the resistivity ratio. This
surprising behavior has been interpreted \protect\cite{gang4me}
as evidence that the transition region is dominated by strong coupling
effects characterizing the insulating phase. }

\end{figure}
\begin{figure}[h]
\begin{center}
\includegraphics[width=4in]{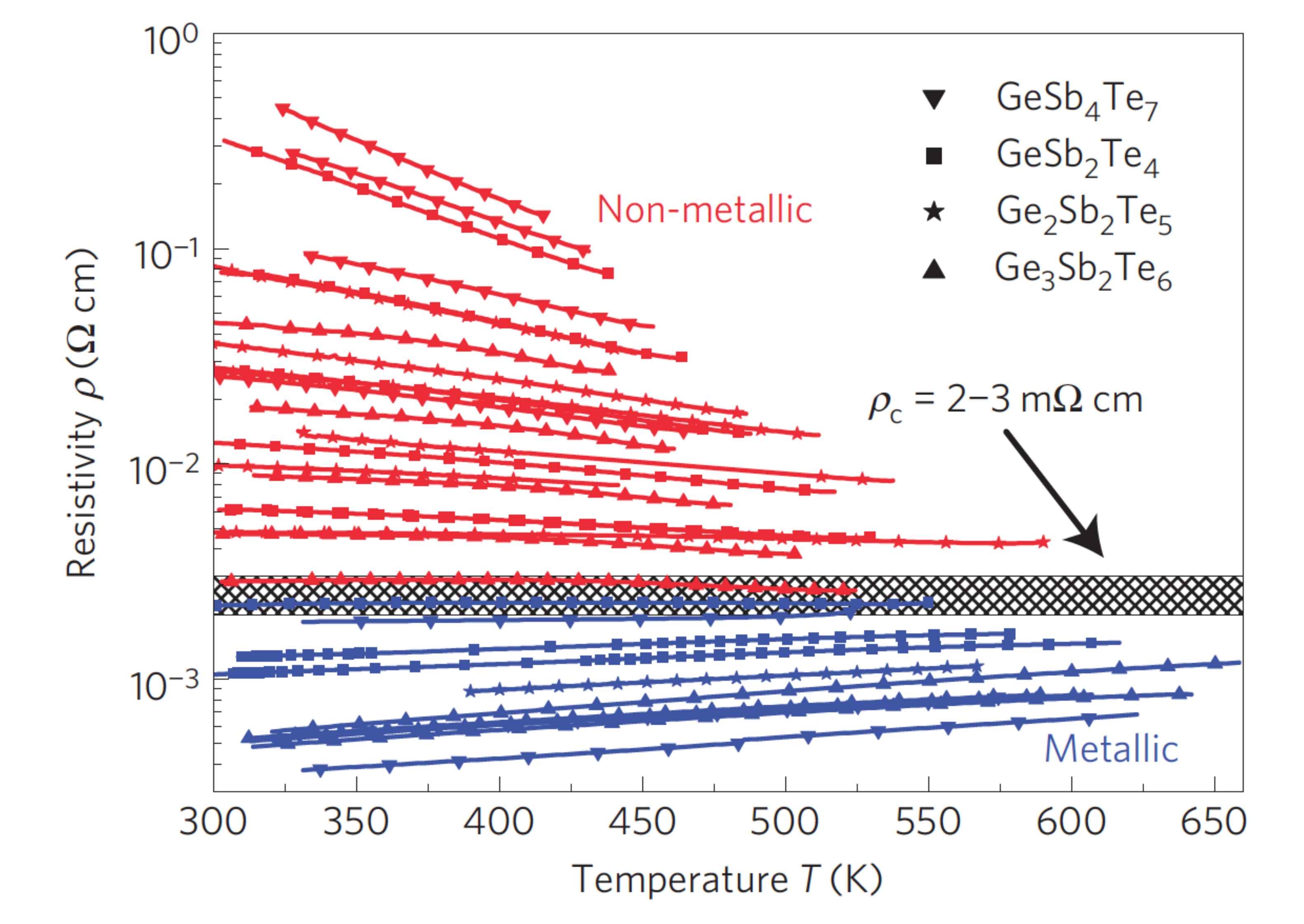}
\end{center}
\caption{High-temperature transport in the disorder-driven metal-insulator
transition region in recently discovered {}``phase change'' materials
\protect\cite{siwgriest11natphys} . Here disorder is tuned at fixed
carrier concentration by partially annealing the parent amorphous
insulator. The change of the sign of TCR is found to occur precisely
at the {}``separatrix'' coinciding with the estimated Mott limit. }

\label{phase change}
\end{figure}

\pagebreak
\begin{itemize}
\item \emph{High temperature anomalies - Mooij correlation.} Surprisingly
similar metal-insulator crossover is seen at elevated temperatures
in several systems \cite{leeramakrishnan} where the MIT is driven
by increasing the level of disorder at fixed electron density. This
behavior, first identified by Mooij \cite{mooij}, has recently been
studied in phase change materials \protect\cite{siwgriest11natphys}. Here,
the temperature coefficient of the resistivity (TCR) $ $is found
to change sign, indicating the crossover from metallic ($d\rho/dT>0$)
to insulating(-like) ($d\rho/dT>0$) transport, around the values
of the resistivity close to the {}``Mott limit'' $\rho_{c}=1/\sigma_{min}$
(Fig. \ref{phase change}). This behavior is consistent with the early
ideas of Mott, suggesting that in metals, as temperature is increased,
the resistivity should display\emph{ resistivity saturation} \cite{fisk76prl}
as soon as the mean-free path approaches the atomic scale. Although
reminiscent to what is seen in ($d=2$) silicon MOSFETs, this behavior
observed in bulk ($d=3$) systems is inconsistent with the expectations
based on the scaling theories of localization (e.g. a divergent critical
resistivity $\rho_{c}(T)\sim1/T^{\frac{d-2}{z}}$). Since any concievable
coherence length must be short at such elevated temperatures, local
(incoherent) scattering processes are likely to be at the origin of
this puzzling behavior. This should be contrasted to the well-known
weak localization and interaction corrections in a disordered Fermi
liquid, processes dominated by coherent miltiple-scattering processes
at long lengthscales.
\end{itemize}

\begin{itemize}
\item \emph{High temperature violations of the Mott limit.} Examples of
metallic ($d\rho/dT>0$) transport with resistivities dramatically
exceeding the Mott limit have been reported only in sufficiently clean
systems not too far from the Mott insulating state. This curious behavior
was first noted shortly after the discovery of cuprate superconductors
\cite{hussey04philmag}, and was quickly interpreted as a {}``smoking
gun'' of non-Fermi liquid physics and strong correlation effects.
Further examples have been documented in organic Mott systems \cite{limelette03prl}
and transition metal oxides such as $V_{2}O_{3}$ \cite{limelette03science}.
In all these cases, the violation of the Mott limit was found at temperatures
exceeding the Fermi-liquid coherence scale $T^{*}$, and was observed
to coincide with the suppression of the corresponding Drude peak in
the optical conductivity. Similarly to resistivity saturation, such
{}``bad metal'' behavior seems to emerge only in the high temperature
incoherent regime, where local scattering processes dominate.
\end{itemize}
\begin{figure}[h]
\begin{center}
\includegraphics[width=4in]{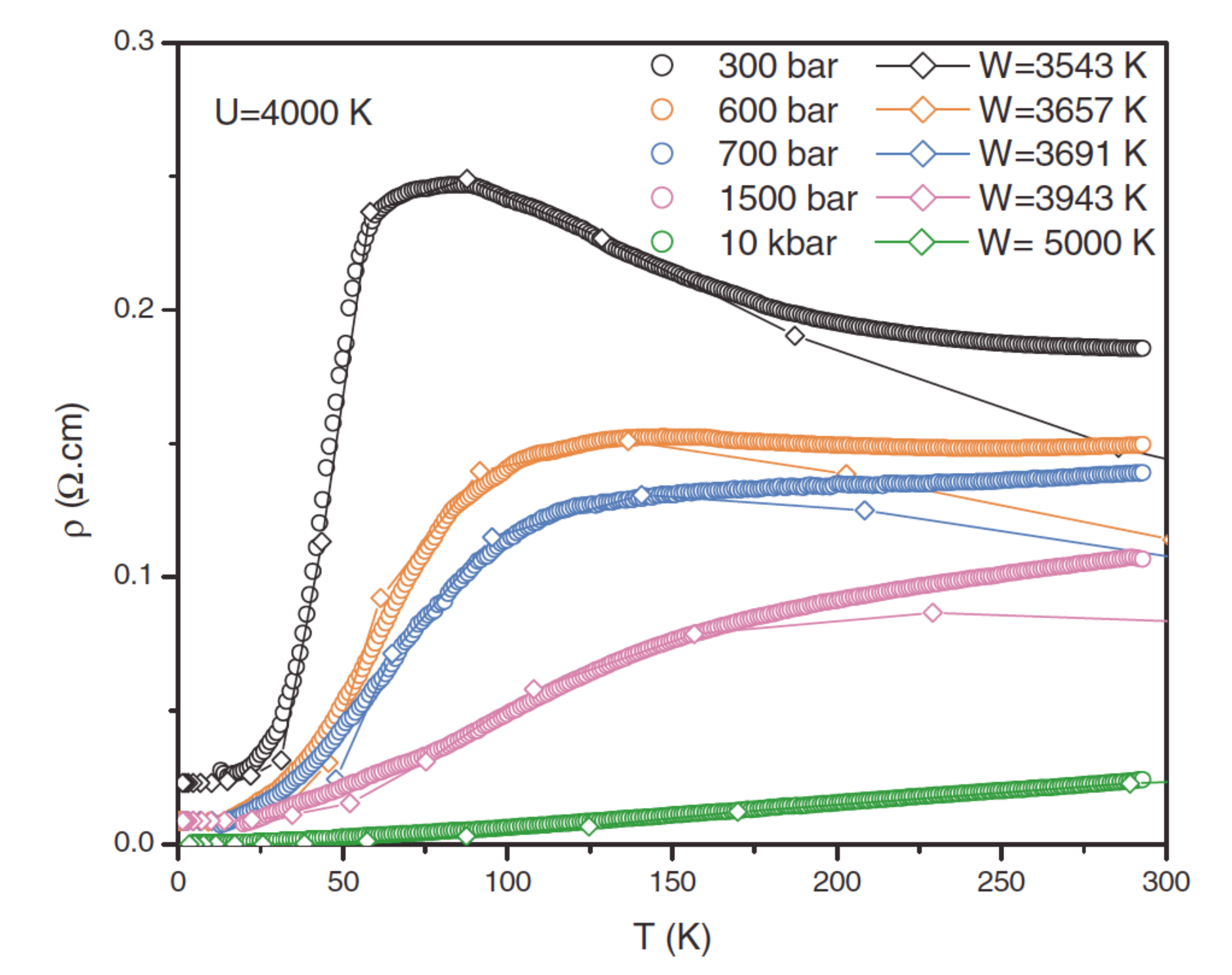}
\end{center}
\caption{Transport behavior in a strongly correlated metal $\kappa-(BETD-TTF)_{2}Cu[N(CN)_{2}]Cl$,
approching the pressure-driven Mott transition. Here, Fermi liquid
transport ($\rho\sim T^{2})$ is found only at the lowest temperatures.
The unusual transport behavior at higher temperatures, displaying
pronounced resistivity maxima well exceeding the Mott limit, has been
quantitatively explained using DMFT theories \protect\cite{limelette03prl,radonjic10prb}.}

\end{figure}

All these experiments taken together provide strong hints that in
many systems of current interest an appropriate mean-field description
is what is needed. It should provide the equivalent of a Van der Waals
equation of state, for the metal-insulator transition problem of disordered
interacting electrons. Unfortunately, such a theory has long been
elusive, primarily due to a lack of a simple order-parameter formulation
for this problem. Very recently, an alternative approach to the problem
of disordered interacting electrons has been developed, based on dynamical
mean-field (DMFT) methods \cite{georgesrmp}. This formulation is
largely complementary to the scaling approach, and has already resulting
in several striking predictions. In the following, we briefly describe
this method, and summarize the main results that have been obtained
so far.

\subsubsection{The DMFT physical picture}

The main idea of the DMFT approach is in principle very close to the
original Bragg-Williams (BW) mean-field theories of magnetism \cite{goldenfeldbook}.
It focuses on a single lattice site, but replaces \cite{georgesrmp}
its environment by a self-consistently determined {}``effective medium'',
as shown in Fig. 1.3.

\begin{figure}[h]
\centerline{\includegraphics[width=4.5in]{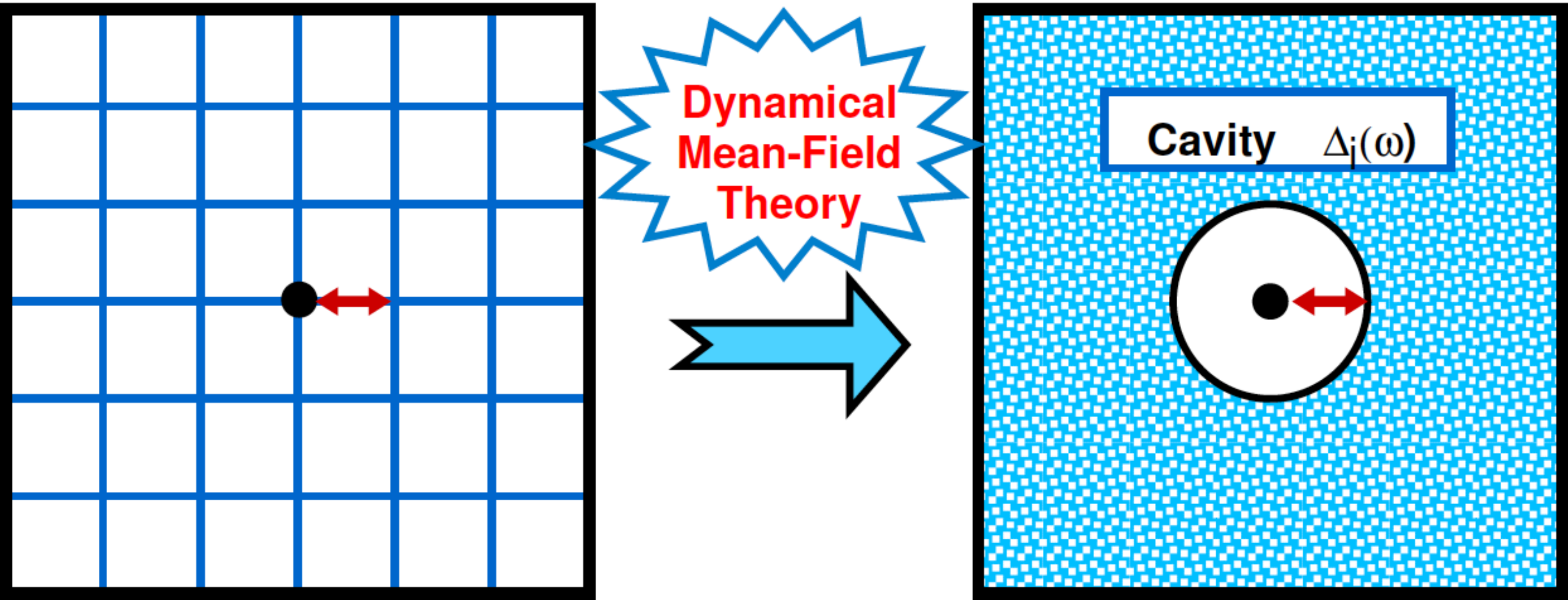}} 

\caption{In dynamical mean-field theory, the environment of a given site is
represented by an effective medium, represented by its ``cavity
spectral function'' $\Delta_{i}(\omega)$. In a \emph{disordered}
system, $\Delta_{i}(\omega)$ for different sites can be very different,
reflecting Anderson localization effects \protect\cite{anderson58}.}

\end{figure}

In contrast to the BW theory, the environment cannot be represented
by a static external field, but instead must contain the information
about the dynamics of an electron moving in or out of the given site.
Such a description can be made precise by formally integrating out
all the degrees of freedom on other lattice sites.
In presence of electron-electron interactions, the resulting local
effective action has an arbitrarily complicated form. Within DMFT \cite{georgesrmp},
the situation simplifies, and all the information about the environment
is contained in the local single particle spectral function $\Delta_{i}(\omega)$.
The calculation then reduces to solving an appropriate quantum impurity
problem supplemented by an additional self-consistency condition that
determines this {}``cavity function'' $\Delta_{i}(\omega)$. 

The precise form of the DMFT equations depends on the particular model
of interacting electrons and/or the form of disorder, but most applications
\cite{georgesrmp} to this date have focused on Hubbard and Anderson
lattice models. The approach has been very successful in examining
the vicinity of the Mott transition in clean systems, and it
has met spectacular successes in elucidating
various properties of several transition metal oxides, heavy fermion
systems and Kondo insulators \cite{LDA-DMFT06rmp} .

The central quantity in the DMFT approach is the local ``cavity''
spectral function $\Delta_{i}(\omega)$. From the physical point of
view, this object essentially represents the {\em available electronic
states} to which an electron can {}``jump'' on its way out of a
given lattice site. As such, it provides a natural order parameter
description for the MIT \cite{london}. Of course, its form can be
substantially modified by either the electron-electron interactions
or disorder, reflecting the corresponding modifications of the electron
dynamics. According to Fermi's Golden Rule, the transition rate to
a neighboring site is proportional to the density of final states
- leading to insulating behavior whenever $\Delta_{i}(\omega)$ has
a gap at the Fermi energy. In the case of a Mott transition in the
absence of disorder, such a gap is a direct consequence of the strong
on-site Coulomb repulsion, and is the same for every lattice site.%
\begin{figure}[h]
\centerline{\includegraphics[width=5in]{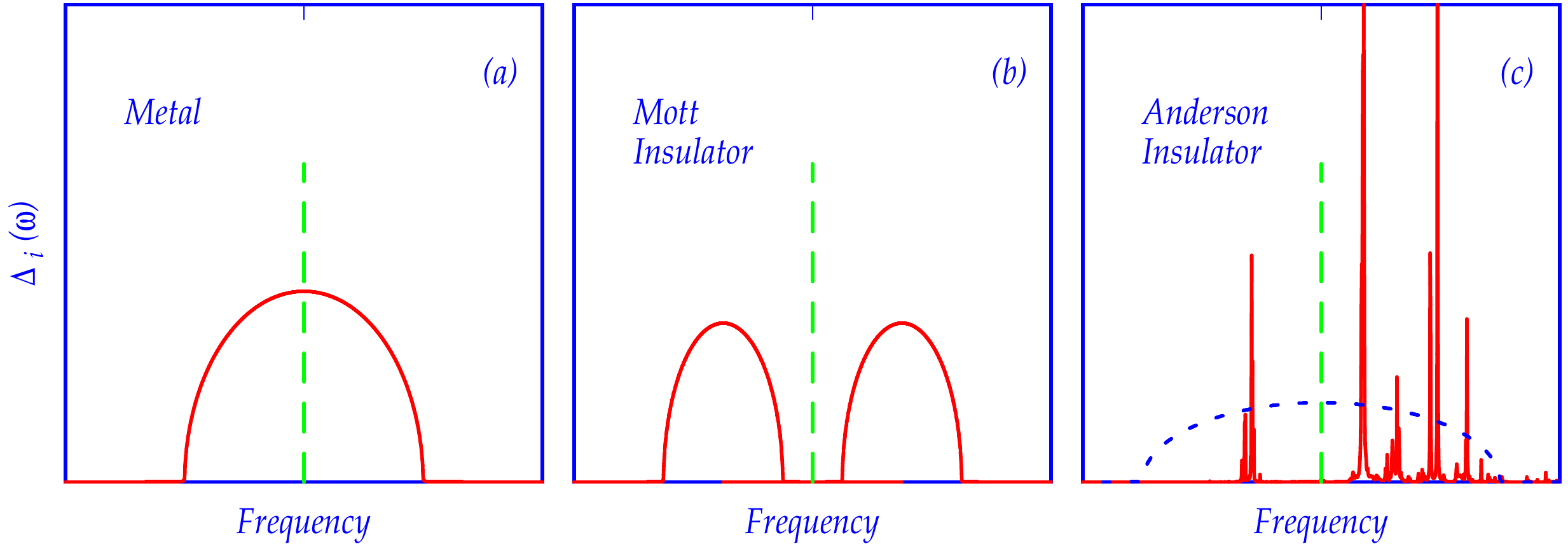}} 

\caption{The local cavity spectral function $\Delta_{i}(\omega)$ as the order
parameter for the MIT. In a metal (a) there are available electronic
states near the Fermi level (dashed line) to which an electron from
a given site can delocalize. Both for a Mott insulator (b) and the
Anderson insulator (c) the Fermi level is in the gap, and the electron
cannot leave the site. Note that the {\em averaged} spectral function
(dotted line in (c)) has no gap for the Anderson insulator, and thus
cannot serve as an order parameter.}

\label{cavity}
\end{figure}

The situation is more subtle in the case of disorder-induced localization,
as first noted in the pioneering work of Anderson \cite{anderson58}.
Here, the \emph{average} value of $\Delta_{i}(\omega)$ has no gap
and thus cannot serve as an order parameter. However, as Anderson
noted a long time ago, {}``...no real atom is an average atom...''
\cite{andersonlocrev}. Indeed, in an Anderson insulator, the environment
{}``seen'' by an electron on a given site can be very different
from its average value. In this case, the\emph{ typical} {}``cavity''
spectral function $\Delta_{i}(\omega)$ consists of several delta-function
(sharp) peaks, reflecting the existence of localized (bound) electronic
states (Fig. \ref{cavity}). Thus a \emph{typical }site is embedded
in an environment that has a \emph{gap }at the Fermi energy - resulting
in insulating behavior. We emphasize that the location and width of
these gaps strongly vary from site to site. These strong fluctuations
of the local spectral functions persist on the metallic side of the
transition, where the typical spectral density $\Delta_{typ}=\exp<\ln(\Delta_{i})>$
can be much smaller than its average value \cite{tmt}. Clearly, a
full \emph{distribution function} is needed \cite{andersonlocrev}
to characterize the system. The situation is similar as in other disordered
systems, such as spin glasses \cite{re:Mezard86}. Instead of simple
averages, here the entire distribution function plays a role of an
order parameter, and undergoes a qualitative change at the phase transition.
The DMFT formulation thus naturally introduces self-consistently defined\emph{
dynamical }order parameters, that can be utilized to characterize
the qualitative differences between various phases. In presence of
disorder, these order parameters have a character of distribution
functions, which change their qualitative form as we go from the normal
metal to the non-Fermi liquid metal, to the insulator.

\subsubsection{Physical content of DMFT approaches}

DMFT and its various generalizations are designed to capture strong
but local correlation effects. Here we will not discuss the technical
details on these approaches, which will be discussed, in some detail,
in Chapter 6. Instead, we give a brief summary of the advantages
and the limitations of the existing DMFT theories, in applications
to strongly correlated electronic systems with and without disorder. 
\begin{itemize}


\item The approach is formally exact in the limit of large coordination,
but in general it represents a ``conserving'' approximation scheme \cite{LDA-DMFT06rmp}, in the sense
of Baym and Kadanoff. Although it reproduces
Fermi liquid behavior of metals at low temperature, it is not restricted
to this regime. In fact, DMFT is \emph{more accurate (essentially
exact) in the high-temperature incoherent regime}, because this is
where any spatiall correlations ignored by DMFT become negligible. 

\item The local correlation effects are described through the local self-energies
$\Sigma_{i}(\omega)$, defining both the quasiparticle effective mass
$m^{*}=1-(\partial\Sigma_{i}/\partial\omega)_{\omega=0},$ and the
inelastic scattering rate $ $$\hbar\tau_{in}^{-1}(\omega)=-Im\Sigma_{i}(\omega=0).$
In presence of disorder these quantities display spatial fluctuations
\cite{RoP2005review}, and need to be characterized with appropriate
distribution functions. 

\item DMFT cannot be used to properly describe those phenomena which are
dominated by long-wavelength spatial fluctuations. It cannot provide
a description of anomalous dimensions of various physical quantities
within the narrow ``Ginzburg'' region around a critical point,
at least for ordinary criticality. On the other hand, recent work
suggested the possibility of ``Local Quantum Criticality'' \cite{qimiao01nature},
where DMFT-like approaches may represent the proper description. 

\end{itemize}

We should stress that the local dynamical description description
provided by DMFT is a feature built-in from the start. How broad is
the range of parameters and models where this approximation is valid
is not obvious or precisely known at this time, despite the many successful
applications \cite{LDA-DMFT06rmp} of DMFT to various experimental
systems. Interestingly, very recent work based on ``holographic
duality'' \cite{sachdev10prl,mcgreevy10physics} ideas borrowed from
super-string theory suggests that certain strong-coupling classes
of QCPs may exist, having strictly local character. If these arguments
prove correct, they may provide fundamental insight into why and where
the local ideas implied by DMFT can be expected to apply, and in which
cases the do not.

\subsubsection{Applications and extensions of DMFT}

The first practical application of DMFT has focused at the Mott transition
in a single-band Hubbard model, sheding new light on systems ranging
from transition metal oxides to organic Mott systems. Even in this
simplest {}``single-site'' version, the method can be easily extended
to multi-band and multy-orbital models, features that are of key importance
in materials such as rare-earth intermetallics (heavy fermion systems)
and the recetnly discovered familty of iron-pnictides \cite{basov2011natphys}.
The technical difficulties in applying the methods to multi-orbital
systems has pleagued early studies, essentially due to lack of a reliable
``quantum impurity solver'' for DMFT equations. Very recent progress
based on ``continuos time quantum Monte Carlo'' methods \cite{troyer11rmp}
has provided a practical and efficient solution in many cases, opening
avenues for applications to many materials. 

Other complications surrounded
efforts to dovetail the DMFT methods for strong correlation with first-principle
density-functional electronic structure methods (DFT). Considerable efforts
have been invested in this program \cite{LDA-DMFT06rmp}, and over
the last few years combined density functional and DMFT studies have bloomed into a veritable
industry, resulting in impressive and accurate first-principles descriptions
of many strongly correlated materials. Finally, effects of inter-site
correlations have also been a focus of much recent work, leading to
various {}``cluster'' generalizations of DMFT,
resulting in substantial new insight \cite{haule07hightc,jarrell11prl}
in systems such as high Tc cuprate superconductors. Different aspect
of these theories have been discussed in detail in several recent
reviews \cite{jarrell05rmp,LDA-DMFT06rmp}. In the following we focus
on applications of DMFT to disordered systems \cite{RoP2005review},
and list the physical phenomena and regimes that have found a natural
description within the DMFT framework, but which remain difficult
to address using complementary approaches. 

The simplest DMFT theories focus on the dynamical effects of local
interactions, while describing the environment of a given site as
an average {}``effective medium''. In presence of disorder, this
formulation reduces \cite{dk-prb94} to the well-known {}``coherent
potential approximation'' (CPA) in the noninteracting limit. This
theory, while providing a reasonable description of disorder-averaged
single-particle spectral functions, is unable to capture Anderson
localization, as well as other imporant disorder effects such as the
formation of inhomogeneous or glassy phases. Recent efforts, however,
have resulted in various extensions of DMFT, which have succeeded
to incorporate various aspects of such disorder-driven phenomena,
as follows.\pagebreak
\begin{itemize}
\item \emph{``Statistical DMFT}'' theories \cite{motand} which calculate
the local site-dependent self-energies for a given fixed realization
of disorder. Here the strong correlation effects are described in
DMFT fashion, while disorder fluctuations are treated in an exact
numerical approach. In a sense, this method can be regarded as a quantum
generalization of the ``TAP equation''
\cite{re:TAP} approach to classical spin glasses. In applicatio to
disordered Huibbard and Anderson lattrice models, it led to the discovery
of disorder-driven non-Fermi liquid behavior \cite{RoP2005review}
and {}``Electronic Griffiths phases'' \cite{mirandavlad1,tanaskovicetal04,andrade09prl}. 
\item {}``\emph{Typical-Medium Theories}'' \cite{tmt} which treat both
the interactions and disorder fluctuations in a DMFT-like self-consistent
fashion. This method is the simplest order-parameter theory of Mott-Anderson
localization. 
\item {}``\emph{Exteded DMFT (EDMFT)}'' theories discribing the effect
of the bosonic collective modes due to inter-site interactions. The
EDMFT approach, which was much utilized in recent work on quantum
criticality in Kondo systems \cite{qimiao00prb}, also proved very
effective in describing quantum spin glass \cite{sachdevreadopper}, and
especially in describing the effects of the long-ranged Coulomb interaction
\cite{chitra00prl}. It provided a quantitatively accurate description
of the pseudogap phenomena in Coulomb systems \cite{pankov05prl},
and clarified the relation between glassy freezing and formation \cite{pastor-prl99}
of the universal Efros-Shklovskii Coulomb gap in presence of disorder. 
\end{itemize}

\subsection{Conclusions and outlook}

Recent years have seen considerable progres in finding fascinating
but often baffling examples of materials that seem to belong to the
metal-insulator transition region. In many of these examples, the
electron localization does not appear to have a conventional {}``band
transition'' character, where some kind of uniform ordering leads
to a band gap opening at the Fermi surface. Instead, mechanisms such
as Mott and Aderson localization are invoqued, processes that simply
do not fit the cherished mold of spontaneous symmtry breaking. Neverthless,
in systems ranging from heavy doped semiconductors, dilute two-dimensional
electron gases, to organic charge-transfer salts, to cuprate-oxide
and iron-pnictide materials, the metal-insualtor transition region
typically assumes the form expected of quantum criticality. 

The traditional approaches to the metal-insulator transition, dating
to the 1980s, tried to circumvent the absence of an obvious order-parameter
description by systematically examining the stability of the metallic
(Fermi liquid) phase to weak disorder. This approach, despite its
formal elegance and conceptual simplicity, did not find many convincing
applications, most likely because its inherent inability to describe
genuine strong correlation effects - the role of {}``Mottness''.
Indeed, much of the efforts of the theoretical condensed matter community
over the last twenty years has focused on the Mott physics in its
many forms. 

One central physics question has emerged from all these studies. This
issue, first emphasized in the pioneering works of Mott and Anderson,
relates to what are the relevant length-scales that dominate the {}``bad-metal''
regime. The perspective provided by the Fermi liquid picture of metals
and the symmetry breaking paradigm of conventional criticality both
suggest that long lengthscales should hold the key. However, as hinted
by the title of Phillip Anderson's 1978 Nobel Prize lecture {}``Local
Moments and Localized States'' \cite{andersonlocrev}, sufficiently
close to insulating states, the real-space representation may offer
a better starting point. 

Over the last twenty years, these local ideas acquired a precise and
systematic language with the rise of DMFT approaches. This provided
a natural \emph{dynamical} order-parameter description for strongly
correlated systems with and without disorder. The progress made by
the various forms and extensions of DMFT is very encouraging, since
they proved capable of incorporating all the basic mechanisms for electron
localization. Many of these phenomena, such as the emergence of strongly
inhomogneous phases, or the description of glassy dynamics, proved
to be beyond the scope of conventional Fermi-liqui based theories
of interaction-localization. In all fairness, though, the problem
remains remains far from being resolved, and much more focused effort
will be neccessary to combine all the facets of these fasciating theories
in a comprehensive and well established picture of the metal-insulator
transition region.

\section{Acknowledgments}

This work was supported by the NSF through grant DMR-1005751, and
by the National High Magnetic Field Laboratory. 

\bibliographystyle{OUPnamed_notitle}
\bibliography{vlad11}

\end{document}